\documentclass[a4paper, onecolumn]{quantumarticle}
\pdfoutput=1
\usepackage{pgfplots}
\pgfplotsset{compat=newest}
\usepackage{graphicx} 
\usepackage{amsmath}
\DeclareMathOperator{\sgn}{sgn}
\usepackage{amssymb}
\usepackage{bm}
\usepackage[version=4]{mhchem}
\usepackage{ amsfonts}
\usepackage{physics}
\usepackage{float}
\usepackage{caption}
\usepackage{tikz}
\usepackage{todonotes}
\usepackage{mathtools,leftindex,tensor,mhchem}
\usepackage{authblk}
\usepackage{abraces}
\usepackage{dsfont}
\usepackage{changepage}
\usepackage{mathrsfs}

\usepackage{booktabs}

\usepackage[numbers, sort&compress]{natbib}

\usepackage{xurl}

\usepackage{setspace}

\usepackage{amsthm}
\newtheorem{definition}{Definition}

\usepackage[breaklinks=true]{hyperref}
\usepackage{cleveref}

\title{Fermion Doubling in Quantum Cellular Automata}
\author[1]{Dogukan Bakircioglu}
\email{dogukan.bakircioglu@universite-paris-saclay.fr}
\author[1]{Pablo Arnault}
\email{pablo.arnault@universite-paris-saclay.fr}
\author[1]{Pablo Arrighi}
\email{pablo.arrighi@universite-paris-saclay.fr}

\affil[1]{\small Université Paris-Saclay, INRIA, CNRS, ENS Paris-Saclay, LMF, 91190 Gif-sur-Yvette, France \vspace{0.4cm}}
\numberwithin{equation}{section}
\usepackage{appendix}
\usepackage{nccmath,amsmath,lipsum}
\usepackage{tabularx}

\newcommand{\defeq}{\vcentcolon=}

\begin{document}

	\maketitle
\begin{abstract}
\noindent   
A Quantum Cellular Automaton (QCA) is essentially an operator driving the
evolution of particles on a lattice, through local unitaries. Because
$\Delta_t=\Delta_x = \epsilon$, QCAs constitute a privileged framework to cast the digital quantum simulation of relativistic quantum particles and their interactions with gauge fields, e.g.,
$(3+1)$D Quantum Electrodynamics (QED). But before they can be adopted, simulation schemes for high-energy physics need prove themselves against specific numerical issues, of which the most infamous is Fermion Doubling (FD). FD is well understood in particular in the real-time, discrete-space \emph{but} continuous-time settings of Hamiltonian Lattice Gauge Theories (LGTs), as the appearance of spurious solutions for all  $\Delta_x=\epsilon\neq 0$. We rigorously extend this analysis to the real-time, discrete-space \emph{and} discrete-time schemes that QCAs are. We demonstrate the existence of FD issues in QCAs for $\Delta_t =\Delta_x = \epsilon \neq 0$. By applying a covering map on the Brillouin zone, we provide a flavor-staggering-only way of fixing FD that does not break the chiral symmetry of the massless scheme. We explain how this method coexists with the Nielsen-Ninomiya no-go theorem, and give an example of neutrino-like QCA showing that our model allows to put chiral fermions interacting via the weak interaction on a spacetime lattice, without running into any FD problem.
\end{abstract}	





	\section{Introduction}

\noindent 
{\bfseries Quantum Cellular Automata.}
Quantum Cellular Automata (QCAs) are shift-invariant, causal and unitary operators over spatial lattices of quantum systems~\cite{ArrighiOverview,FarrellyReview}. A QCA can always be realized as a unitary quantum circuit, infinitely repeating across space and time~\cite{Arrighi1DQCA,ArrighiUCAUSAL}. In the line of what was envisioned by Feynman right at the birth of Quantum Computing~\cite{FeynmanQC, FeynmanQCA}, it has been shown that these QCAs can be used for the digital Quantum Simulation of Quantum Field Theories~(QFTs)~\cite{ABF20, EDMMplus2023}. For this purpose, one tries and discretizes the QFT in steps of $\Delta_{t}=\Delta_{x}=\epsilon \in \mathbb R_+$. We define the success of such an attempt as the wires of the circuit of the QCA ``matching'' the speed of light of the simulated QFT. 
Such a discrete-spacetime dynamics, in the limit of a large number of time steps, and more specifically in an appropriate continuum limit that one can consider (namely, $\Delta_t = \Delta_x = \epsilon \rightarrow 0$), faithfully reproduces~\cite{ArrighiDirac} the propagation of the simulated massive particles.
Various types of discrete-spacetime Lorentz covariance have even been suggested for these \mbox{models~\cite{ArrighiLORENTZ, ArrighiLorentzCovariance, PaviaLORENTZ, Debbasch2019a}}. This is in contrast with standard Hamiltonian-based Quantum Simulation~\cite{Banuls2020-vg,Martinez2016-kj}, which is always grounded in one way or another on continuous-time Hamiltonian models, which are fundamentally non-relativistic, and which remain so even if digitalized into a QCA, e.g., by means of a trotterization~\cite{doi:10.1126/science.273.5278.1073, APAF18, MA19}. \\

 

\noindent
{{\bfseries Fermion Doubling.}} The classical simulation of QFTs has a long history. Discrete-spacetime formulations of QFTs, a.k.a.\ Lattice Gauge Theories (LGTs), were formally introduced using the Lagrangian formalism by Wilson in Ref.\ \cite{Wilson74}. A discrete-space but continuous-time reformulation of Wilson's LGTs, in the Hamiltonian formalism this time, was then provided by Kogut and Susskind in Ref.\ \cite{KogSuss75a}. They, however, noticed a strange behavior in the dispersion relation of the fermionic lattice theory. Namely, they observed spurious, non-physical solutions persisting whenever $ \Delta_{x} \neq 0$ however small, which would inevitably pollute the numerics: this problem is referred to as Fermion Doubling (FD)---see also Ref.\ \cite{Susskind_1977} for more details seminally given on the problem. In Ref.\ \cite{KogSuss75a}, Kogut and Susskind introduce their Hamiltonian LGT with a lattice scheme that actually already fixes FD, by placing the two different chiralities of the Dirac spinor on adjacent but different lattice sites: this solution was referred to as ``staggered fermions''. The staggering method was then extended to all possible internal components of Dirac fields, including spin, flavors, or ``colors''~\cite{RotheBook}. On the other hand, Wilson also proposed to fix FD---which also appears in his Lagrangian, discrete-spacetime LGTs for $\Delta_t=\Delta_x=\epsilon \neq 0$---, by adding an $\epsilon$-dependent term with discretized second-order spatial derivatives,
disappearing in the continuum limit: this solution is referred to as ``Wilson fermions''. Unfortunately, both solutions break the chiral symmetry of the respective massless schemes. Afterwards, this FD problem was shown to be a fundamental issue of lattice fermionic systems~\cite{NielsenI1981,Nielsen1981,Friedan:1982nk}, so that any discrete model involving fermions must address this issue before producing numerical results, and QCAs are no exception to that. \\
   


\noindent
{\bfseries The FD problem in QCAs.}
In recent years, several aspects of numerical analysis of QCA-based models for QFT have been investigated~\cite{MoscoPerturbation,ArrighiQEDContinuum,trezzini2024renormalisationquantumcellularautomata}. In this literature, the FD problem has been neglected, with the exceptions of Refs.\ \cite{ABF20, APAF18, Arnault2022, Jolly2023-cy}. However, in all these four articles, the FD analysis is based on those conducted for discrete-space but continuous-time formulations of LGTs, in the Hamiltonian formalism---which falls short in the discrete-spacetime setting. In particular, this leads the authors of Ref.\ \cite{ABF20} to falsely conclude that FD does not occur in their model. The reason why, despite describing models in discrete time, the previously cited four articles based their FD analysis on that of continuous-time/Hamiltonian LGTs rather than on that of Lagrangian/Wilsonian LGTs, which \emph{do} discretize time, is that the framework of QCAs is, despite being in discrete time, in a sense closer to Hamiltonian LGTs than to Lagrangian/Wilsonian LGTs: this is because in both QCAs and Hamiltonian LGTs, we are especially interested in evolving the system in time (e.g., for quantum simulation~\cite{Martinez2016-kj}), by which we mean \emph{real time}, that is, without  Wick rotation. \\



\noindent
{{\bfseries Contributions.}}
In this article, we develop a rigorous, general FD analysis method in order to assess the presence of FD in QCAs as well as in real-time discrete-spacetime models in general, the most common ones being Hamiltonian-based---for such models in imaginary time, see, e.g., Ref.\ \cite{RotheBook}. 
We apply it to the latest QCA models of QFT, namely, the $(1+1)$D QED QCA of Ref.\ \cite{ABF20}, and the $(3+1)$D QED QCA of Ref.\ \cite{EDMMplus2023}. We prove that these QCAs do suffer from a FD problem, although we show that they are less FD-prone than those based on a naive discretization of the Dirac equation---that is, \emph{both} continuous-time/Hamiltonian LGTs and Lagrangian/Wilsonian LGTs. We explain how to fix this FD problem, according to a general method, which can be interpreted geometrically as introducing a covering map on the Brillouin zone. In direct space, our solution amounts to staggering \emph{only} the flavor degree of freedom that we introduce, so that we call it the \emph{flavor-staggering-only} method. Interestingly, such a fix does not break  the chiral symmetry of the massless scheme. 
We fully describe the FD-fixed, flavored-QCA versions of the QCAs of Refs.\ \cite{ABF20,EDMMplus2023}. These feature $ \mathbb{Z}_{2}$ and $ \mathbb{Z}_{2} \cross \mathbb{Z}_{2} \cross \mathbb{Z}_{2}$ flavor symmetries, respectively. The reader should note that our way of introducing the flavor degree of freedom is ``\emph{minimal}'' with respect to our initial, flavorless model, and in particular different from the flavoring needed to solve the FD problem in standard LGT \cite{RotheBook}.  \\
 
\noindent    
{{\bfseries Roadmap.}} In Sec.\ \ref{sec:standardFD}, we provide a way to analyze FD in discrete-space and discrete-time formulations of real-time fermionic systems. Then, in Sec.\ \ref{sec:qedqcafd}, we apply this analysis to the $(1+1)$D QED QCA and to the $(3+1)$D QED QCA cited previously. We realize that both of them suffer from a FD problem, albeit less severely than usual lattice fermion models. 
In Sec.\ \ref{sec:solution}, we fix the FD problem of these models by introducing flavor degrees of freedom respecting the dynamics of the models (unlike Wilson FD-fixing terms), yielding us the flavored versions of these QCAs. In Sec.\ \ref{sec:formaltreatment}, we prove that the two solutions proposed in Sec.\ \ref{sec:solution} admit a geometrical understanding as the application of a covering map on their Brillouin zone. Lastly, in Sec.\ \ref{sec:discussion}, we discuss how our results fit together with the Nielsen-Ninomiya no-go theorem, and show that it \emph{is} possible to have a neutrino-like particle in a flavored QCA such as those we have introduced, i.e., our model \emph{does} allow to put chiral fermions interacting via the weak interaction on a lattice without having to introduce their chiral counterpart, thanks, again, to the flavor we have introduced.

    \section{Background on fermion doubling}\label{sec:standardFD}

	This section aims at conveying a general way of analyzing and resolving FD on lattice QFT models, by means of a paradigmatic example. Here we are interested in discrete-space discrete-time formulations of QFTs, unlike Refs.\ \cite{Susskind_1977,KogSuss75a}, whose framework is in discrete space but continuous time. We will not base ourselves on a  Hamiltonian as they do. We will not base ourselves on a classical Lagrangian either, as in Refs.\ \cite{ Wilson74,SHARATCHANDRA1981205}. Instead, our analysis will depart directly from a discretization of the relevant equation of motion, which is the Dirac equation. We draw inspiration from the Green's-function-based methods of Refs.\ \cite{RotheBook,Smit_2023}.


   \subsection{Continuum and naive discrete-spacetime equations of motions}\label{subsec:eqmotions}

    \noindent    
{{\bfseries Lattice and fields on it.}}
	Consider a $2$D spacetime lattice, $ \Lambda^{2} \subset \mathbb{R}^{2},$ with spacing $ \epsilon \in \mathbb R_+$. In the limit where the lattice spacing becomes smaller and smaller, the lattice approaches $  \mathbb{R}^{2}$.
    Let $t$ and $x$ be, respectively, the time and space coordinates over $\mathbb{R}^{2}$.
    One can express the elements of $ \Lambda^{2}$ by introducing $ t_n \defeq n \epsilon$ and $ x_k \defeq k \epsilon$, with $ n,k \in \mathbb{Z}$. We will however sometimes use the notations $t$ and $x$ instead of $t_n$ and $x_k$, respectively, since there will be no possible confusion regarding the fact that we are speaking of the lattice coordinates and not the continuum's. In QFTs, or rather, more appropriately, in their associated classical field theory, fields are defined to be maps from a manifold, where physical objects ``live'', to, typically, a $d$-dimensional complex vector space $\mathbb{C}^d$, with $d \in \mathbb N^\ast$. Here we consider a classical spin-1/2 field
    on the lattice as a map $\psi : \Lambda^{2} \to \mathbb{C}^{2}$ obeying a discretization of the free Dirac equation. To have a simple writing but still insist on the discrete nature of our model, we will consider $\psi$ actually depending directly on $(n,k)$ rather than on $(t_n,x_k)$.  We introduce the fields $\psi^+$ and $\psi^-$, from $\Lambda^2$ to $\mathbb C$, which are such that $\psi(n,k) \equiv [\psi^+(n,k), \psi^-(n,k)]^\top$, where $\top$ denotes the transposition. \\

        \noindent    
{{\bfseries Continuum Dirac equation.}}
    Recall that in the $(1+1)$D continuum, the Dirac equation for a spin-1/2 field $ {\psi}_{\text{cont.}}(t,x)$ can be written, by an appropriate choice of representation of the gamma/alpha matrices, namely, $\alpha_1 = \sigma_3$ and $\alpha_0 = \sigma_1$ (as in Ref.\ \cite{ABF20}), as
	
	\begin{equation}\label{eq:Dirac}
		(i \mathbb{I}_{2} \partial_{t}   + i\sigma_{3} \partial_{x} - m  \sigma_{1} ) {\psi}_{\text{cont.}}(t,x) = 0 \, ,
	\end{equation}
	where  (i) $ \mathbb{I}_{2}$ is  the $ 2 \cross 2 $ identity matrix,  (ii) $\sigma_{1}$ and $\sigma_{3} $ are the well-known first and third Pauli matrices, respectively, (iii) and $m \in \mathbb R_+$ is the mass. We call $(i \mathbb{I}_{2} \partial_{t}   + i\sigma_{3} \partial_{x} - m  \sigma_{1})$ the Hamiltonian Dirac operator, which is an operator in the sense of the single-particle quantum-mechanical interpretation of the classical spin-1/2 field as a (two-component) wavefunction.  \\

            \noindent    
{{\bfseries Naive spacetime discretization of the Dirac equation.}}
    In order to obtain a discretization of the previous equation, a typical first idea would be to replace partial derivatives by right (or left) finite differences, also called ``lattice derivatives'', but this breaks the Hermicity of the single-particle momentum and energy operators,
    which are respectively $\mathcal P \defeq -i\partial_{x}$ and $\mathcal E \defeq  i\partial_{t}$~\cite{Susskind_1977,RotheBook}. Traditionally, one rather wants these momentum and energy operators to remain Hermitian after discretization, so that their spectrum remains real~\cite{book_Peskin_Schroeder}. This is achieved by replacing partial derivatives by symmetric lattice derivatives:
    \begin{subequations}
	\begin{align}
		{\mathcal P \psi}(t,x)=-i \partial_{x} {\psi}_{\text{cont.}} (t,x) \ &\longrightarrow \ \hspace{0.1cm} \mathcal {P}_{\text{naive}} \,\psi(n,k)\defeq -\frac{i}{2 \epsilon} [ \psi(n, k+1) - \psi(n, k-1)] \\ 
		{\mathcal E \psi}(t,x)=i \partial_{t} {\psi}_{\text{cont.}} (t,x) \hspace{0.33cm} \ &\longrightarrow \ \ \, \mathcal  {E}_{\text{naive}} \, \psi(n,k)\defeq \frac{i}{2 \epsilon} [ \psi(n+1, k) - \psi(n-1, k)] \, .
	\end{align} 
\end{subequations}
   It turns out that this yields, not only Hermitian discrete momentum and energy operators, but also, overall, a Hermitian discretized Hamiltonian Dirac operator, associated to the following discrete-spacetime equation of motion:
	\begin{equation}\label{eq:MotionDirac}
		\frac{i \mathbb{I}_{2}}{2 \epsilon}[\psi(n+1,k) - \psi(n-1,k) ] + \frac{i \sigma_{3}}{2 \epsilon}[\psi(n,k+1) - \psi(n,k-1) ] - m \sigma_{1} \psi(n,k) = 0 \, .
	\end{equation}
    The fact that this overall Hermiticity cannot be obtained using right (or left) lattice derivatives is proven in App.\ \ref{sec:apphermicity}. We refer to this discretization procedure via symmetric lattice derivatives, or, equivalently, to the previous equation, as the ``naive discretization scheme''. \\

            \noindent    
{{\bfseries Fourier-space continuum Dirac equation.}}
    Now, before proceeding to the Fourier analysis of the previous classical-fields discrete-spacetime scheme, let us recall the conclusions of a few facts---detailed in App.\ \ref{app:on-shellness}---regarding the single-particle quantum-mechanical interpretation of the classical spin-1/2 Dirac field, which deliver us an equation that we will afterwards need for our reasoning: if we fully know/knew the bivariate function ${\psi}_{\text{cont.}}(t,x)$, we can/could define a spacetime Fourier transform
    \begin{equation}
    \tilde{\psi}_{\text{cont.}}(E,p) \defeq \int_{-\infty}^{+\infty} \frac{dE}{2\pi} \int_{-\infty}^{+\infty}\frac{dp}{2\pi} \, {\psi}_{\text{cont.}}(t,x) \, e^{iEt - i px} \, , 
    \label{eq:cont_Fourier_transform}
    \end{equation}
    which we know satisfies/would satisfy the equation
    \begin{equation}
    \label{eq:thefinalequation}
    (\mathbb{I}_{2} E  - \sigma_{3} p - m  \sigma_{1} )  \tilde{\psi}_{\text{cont.}}(E,p) = 0 \, .
    \end{equation}

    \subsection{Fourier analysis}

            \noindent    
{{\bfseries Framework for a Fourier analysis in both discrete space and discrete time.}}    
    Sequences on $ \Lambda^{2}$ can be compactly Fourier decomposed due to the periodicity of the lattice under translations of   $\vec{a}_0 \defeq \vec{t} \epsilon $ and $\vec{a}_1 \defeq \vec{x} \epsilon $, where $ \vec{t}$ and $ \vec{x} $ are orthonormal vectors. The reciprocal lattice vectors for this square lattice $\Lambda^2$ are then $ \vec{b}_0 \defeq 2 \pi \vec{t}/\epsilon$ and $ \vec{b}_1 \defeq 2 \pi \vec{x}/\epsilon$. They generate the Brillouin Zone (BZ) $\mathcal{B} \defeq \{(p,E)\in [-\pi/\epsilon,\pi/\epsilon]^2\}$, which is the integration domain of the Fourier decomposition. 
    The spin-1/2 field on the lattice is Fourier decomposed via\footnote{We choose, for further convenience in the computations we carry out, a normalization that does not match, in the continuum limit, that used in Eq.\ \eqref{eq:cont_Fourier_transform}, but to achieve the matching it would be enough to have an overall factor $\epsilon^2$ in the definition of $\tilde \psi(E,p)$ below instead of the overall factor $\epsilon$, and, accordingly, an overall factor of $1$ in the expression of $\psi(t,x)$ below instead of the overall factor $\epsilon$.}
	\begin{subequations}\label{eq:Fourier}
    \begin{align}
           \tilde{\psi}(E,p)  &\defeq \epsilon \sum_{ n= - \infty}^{\infty} \sum_{ k = - \infty}^{\infty}\hspace{1mm}  \psi(n,k)  \, e^{iEn \epsilon-ipk\epsilon} \label{eq:FT}\\
            \psi(n,k) &= \epsilon \int_{- \pi/\epsilon}^{\pi/\epsilon}  \frac{dE}{2 \pi } \int_{- \pi/\epsilon}^{\pi/\epsilon}  \frac{dp}{2 \pi } \hspace{1mm} \tilde{\psi}(E,p) \, e^{-iEn \epsilon+ipk\epsilon} \, , \label{eq:ansatz}
    \end{align}
	\end{subequations}
    where the Fourier transform  $\tilde{\psi}$ is a map from $\mathcal{B}$ to  $ \mathbb{C}^{2}$. 
    Let us mention that introducing the energy integrals in Eq.\ \eqref{eq:cont_Fourier_transform} or Eq.\ \eqref{eq:ansatz}, is nothing but adopting the so-called off-shell perspective of QFT, which is also relevant in mere classical field theory; working off shell will greatly simplify this work, but it is not a necessity. \\

                \noindent    
{{\bfseries Fourier-transformed naive discretization scheme.}}  
    Inserting the Fourier-decomposed $ \psi(n,k)$ of the previous equation into Eq.\ \eqref{eq:MotionDirac}, we obtain (details in App.\ \ref{sec:ftofnaive}):
	\begin{equation}\label{eq:MotionFourier}
			 \int_{- \pi/\epsilon}^{\pi/\epsilon}  \frac{dE}{2 \pi} \int_{- \pi/\epsilon}^{\pi/\epsilon}  \frac{dp}{2 \pi } \hspace{1mm}  e^{-iEn \epsilon+ ipk\epsilon}  {D}_{\mathcal{B}} (E,p) \tilde{\psi}(E,p) = 0 \, ,
	\end{equation}
	where
    \begin{equation}
    \label{eq:DBEp}
		{D}_{\mathcal{B}} (E,p) \defeq
			\mathbb{I}_{2} \sin(E \epsilon)  - \sigma_{3} \sin(p \epsilon)   - m  \epsilon\sigma_{1}  \, .
	\end{equation}
    Under the limit $\epsilon \to 0 $ around the points  $ E\epsilon = 0$ and $ p \epsilon = 0 $, and using $ \sin(x) \sim x $ for $ x$ small, we see that ${D}_{\mathcal{B}} (E,p)/\epsilon$ converges to 
    $\left(\mathbb{I}_{2} E  - \sigma_{3} p   - m \sigma_{1} \right)$,
    which matches the Fourier behavior of the continuum Dirac Eq.\ \eqref{eq:Dirac}, for which we know, due to the single-particle quantum-mechanical interpretation of the classical spin-1/2 Dirac field (see App.\ \ref{app:on-shellness}), that Eq.\ \eqref{eq:thefinalequation} holds. 
    Hence, we will, in this discrete-spacetime context also, only consider the  solutions to Eq.\ \eqref{eq:MotionFourier} which satisfy
    \begin{equation}
    \label{eq:Fourierspacedynamicalscheme}
    {D}_{\mathcal{B}}(E,p)\tilde{\psi}(E,p)=0 \, ,
    \end{equation}
    that is, essentially (we will see in Sec.\ \ref{subsubsec:disp_rel} why we say here ``essentially'') the so-called ``on-shell'' solutions. After all, Eq.\ \eqref{eq:MotionFourier} appears to have the form of the Fourier decomposition of a vanishing function, and one solution is to take all Fourier coefficients to be zero. Other solutions may exist, but they are not considered in a physics context~\cite{Hall2013}, which is related to the fact that in a certain regime we want the single-particle quantum-mechanical interpretation of the classical spin-1/2 field to hold, whether we are in the continuum (see App.\ \ref{app:on-shellness}) or on a spacetime lattice.


    \subsection{Spurious solutions}
    
            \noindent    
{{\bfseries First analysis.}} 
    The match with the continuum mentioned previously occurs around the points lying in the disk\footnote{To be more formal in defining an exact disk, we could consider some maximum fixed radius $R/\epsilon$ with $R \ll 1$. We think this is simple enough for the reader to understand it and complete this if necessary.} $U_{(0,0)} \defeq \{(p,E) \in\mathcal{B} : \hspace{1mm}  d((p,E) , (0,0)) \ll \epsilon^{-1}  \} $, where $ d(x,y)$ is the usual Euclidean distance over $\mathcal{B}$.
    However, because $ \sin( \pi -x) = \sin(x) $ and recall that $\sin(x) \sim x$ for small $x$, the neighborhoods $ U_{(r,s)} \defeq \{( r \pi \epsilon^{-1}+ p , s \pi \epsilon^{-1}+E )\in\mathcal{B} \ :\ (p,E) \in U_{(0 ,0)} \}$, where $ r,s \in \{-1,0,1\}  $, other than $U_{(0,0)}$, behave, in the limit $\epsilon \rightarrow 0$, just like $U_{(0,0)}$ as far as ${D}_{\mathcal{B}}$ is concerned (sometimes up to additional minus signs in front of $p$ or/and $E$). Because of this, each transformation $p \mapsto r \pi \epsilon^{-1}+ p$ or/and $E \mapsto s \pi \epsilon^{-1}+ E$ will be called a ``$\pi$-symmetry''. \\
    
                \noindent    
{{\bfseries Removing redundancies by wrapping up the Brillouin zone.}} 
    In fact, the symmetries of the free dynamical scheme that Eq.\ \eqref{eq:MotionDirac} is\footnote{Symmetries which reflect in the symmetries of the lattice since usually the lattice is chosen to have the symmetries of the free dynamical scheme, and which are here translations by the lattice spacing $\epsilon$ in both the space and time direction. Another situation is that the lattice is fixed from the start, i.e., cannot be chosen (e.g., some solid crystal of atoms), but there exists some dynamical scheme on that lattice, evolving either on the full lattice, or on a sublattice.}, translate into a $2\pi$-periodicity of the Fourier-space dynamical scheme, Eq.\ \eqref{eq:Fourierspacedynamicalscheme}. Hence, since the Brillouin zone is meant to analyze this scheme, this $2\pi$-periodicity makes it meaningful to wrap up $ \mathcal{B}$ as the torus $\mathbb{T}^{2} \defeq \mathbb{S}^{1} \cross \mathbb{S}^{1} $---where $\mathbb{S}^{1}$ is the sphere of dimension $1$, that is, the circle---, as suggested in Fig.\ \ref{fig:degeneracieslgt} with the disks having matching colors.

	\begin{center}
\begin{tikzpicture}
      \begin{scope}[xshift= 50mm]
			\draw[thick,->] (2,2) -- (4.5,2);
			\draw[thick,->] (2,2) -- (2,-0.5);
			\draw[thick,->] (2,2) -- (-0.5,2);
			\draw[thick,->] (2,2) -- (2,4.5);
	
			\draw[dashed, black, thick] (4,4) --(4,0);
			\draw[dashed, orange, thick] (4,4) --(0,4);
			\draw[dashed,black,thick] (0,4) --(0,0);
			\draw[dashed,orange,thick] (0,0) --(4,0);

			\draw[red,dashed, ultra thick] (2,2) circle[radius = 3.5 mm];
              \draw node at (2.5,2.5) { $U_{(0,0)} $};
			\filldraw[blue](4,4) circle[radius=3mm];
            	\draw node at (5,4) { $U_{(1,1)} $};
			\filldraw[blue](4,0) circle[radius=3mm] ;
            \draw node at (5,0) { $U_{(1,-1)} $};
			\filldraw[blue](0,0) circle[radius=3mm] ;
            \draw node at (-1,0) { $U_{(-1,-1)} $};
			\filldraw[blue](0,4) circle[radius=3mm] ;
			\draw node at (-1,4) { $U_{(-1,1)} $};
			\filldraw[yellow](4,2) circle[radius=3mm] ;
               \draw node at (4.8,1.6) { $U_{(1,0)} $};
        
			\filldraw[green](2,4) circle[radius=3mm] ;
             \draw node at (1.4,3.5) { $U_{(0,1)} $};
			\filldraw[green](2,0) circle[radius=3mm] ;
              \draw node at (1.4,0.5) { $U_{(0,-1)} $};
			\filldraw[yellow](0,2) circle[radius=3mm] ;
            \draw node at (-0.8,1.6) { $U_{(-1,0)} $};

			\draw node at (4.7,2) {$ p$};
			\draw node at (2,4.7) {$ E$};
			\draw node at (4.3,2.5) { $ \frac{\pi}{\epsilon}$};
    			\draw node at (2.5,4.24) { $ \frac{\pi}{\epsilon}$};
			\draw node at (-0.24,2.5) { $ -\frac{\pi}{\epsilon}$};
			\draw node at (2.5,-0.24) { $- \frac{\pi}{\epsilon}$};
		\end{scope}

	\end{tikzpicture}
		\captionof{figure}{ \emph{Fermion-doubling neighborhoods for the symmetric-finite-differences scheme (yellow, green and blue), and (see Sec.\ \ref{subsubsec:QWFD}) for the $(1+1)$D Dirac discrete-time quantum walk (blue only).} The neighborhoods having the same color are in fact identical due to the toric geometry. \label{fig:doublers}} 

        \label{fig:degeneracieslgt}
	\end{center}

	\vspace{2mm}

	After this wrapping up, we see that there are three spurious $U$-disks, defined as disks containing an energy and/or a momentum that in the continuum limit are sent to infinity because infinitesimally close to the edges of the Brillouin zone, but that, as far as ${D}_{\mathcal{B}}(E,p)$ is concerned, behave just like the disk $U_{(0,0)}$, which is the only one yielding finite and thus acceptable values of momentum and energy in the continuum limit. Intuitively, this is problematic, because this suggests that no matter how small $\epsilon$ is, these infinite-energy and/or infinite-momentum Fourier modes (which we may simply call ``infinite modes'' in what follows) will continue to behave just like legitimate solutions and pollute the numerics---only to vanish at exactly $\epsilon=0$ when they become ill-defined, but of course this is relevant only in a pure mathematical sense, since when running numerical simulations we never arrive to $\epsilon = 0$ strictly, and so we never get rid of the problem. But, to rigorously decide if these infinite modes actually cause physical problems, let us look at the dispersion relation and the propagator of the discretized theory.
    
    \subsection{Dispersion relation and propagator}


\subsubsection{Dispersion relation}
\label{subsubsec:disp_rel}

Consider the physical, i.e., the ${D}_{\mathcal{B}}(E,p)\tilde{\psi}(E,p)=0$ solutions to Eq.\ \eqref{eq:MotionFourier}.
At coordinates $(E,p)$ where $\tilde{\psi}(E,p)\neq 0$, this forces 
\begin{equation}
\label{eq:determinant}
\det{{D}_{\mathcal{B}}(E,p)} = 0 \, ,    
\end{equation}
where the determinant refers to the internal-degree-of-freedom operator/matrix structure---which is anyways the only remaining one in Fourier space---, and can easily be computed from Eq.\ \eqref{eq:DBEp}, which delivers
\begin{equation}
    \det{{D}_{\mathcal{B}}(E,p)}=\left[\sin^2(E\epsilon )-\sin^2(p \epsilon)\right]/\epsilon^2 - m^2 \, .
\end{equation}
The condition of Eq.\ \eqref{eq:determinant} is called the ``dispersion relation'', and it is sometimes temporarily relaxed, in which case $\tilde{\psi}(E,p)$ is said to be ``off shell''. For true, i.e., ``on-shell'' solutions, the support of $\tilde{\psi}(E,p)$, and hence its entire norm, must therefore lie in the region of the BZ where (we have combined the two previous equations)
\begin{equation}
        \left[\sin^2(E\epsilon )-\sin^2(p \epsilon)\right]/\epsilon^2 -m^2=0 \, .\label{eq:det}
\end{equation}
So, even though $\tilde{\psi}(E,p)$ has as a domain the bidimensional surface $\mathcal{B}$, only the lines satisfying Eq.\ \eqref{eq:det} matter, namely, the lines satisfying the dispersion relation.

\subsubsection{Propagator/Green's function}

{{\bfseries Introduction.}} 
 The quantity $\det{{D}_{\mathcal{B}}(E,p)}$ does not just serve to restrict the support of the on-shell solutions. It plays another fundamental role that we are now going to explain. In high-energy-physics experiments, for example, certain physical quantities are of paramount importance, and the so-called propagator is one of them. The propagator, word usually used in a quantum context, is the quantum probability amplitude for a particle moving from $ (n',k')$ to $ (n,k)$---you can think of, e.g., the quantum first-quantization interpretation of our classical field $\psi(n,k)$ as a quantum wavefunction, or of the quantum \emph{field} quantization of that classical field, both giving you a notion of propagator. Now, both because we here work in discrete spacetime---rather than in continuous spacetime as in standard quantum mechanics or QFT---, and because the points we want to make can be made at the level of mere classical fields, we will use the concept of Green's function---which can be introduced also in discrete spacetime---rather than that of propagator, the former being more general, and anyways essentially reducing to that of propagator in standard quantum mechanics or QFT. In order to go directly to the points we want to make, we will not recall generalities on Green's function, but will just apply well-known definitions and results directly to our problem. \\

\noindent
    {{\bfseries Green's function equation.}} 
    Green's function for our problem, which is solving Eq.\ \eqref{eq:MotionDirac}, is the function $G(n,k ; n',k')$ satisfying the following finite-differences equation\footnote{It can be easily verified that this definition implies that we choose to define Green's function as strictly equal to the propagator---arising if we endow our classical field theory with some quantum interpretation---, and in particular there is no factor of $i \hbar$ between both objects as one sometimes find.},
    \par\nobreak
    {\small    
    \begin{align}
             &\frac{i \mathbb{I}_{2}}{2 \epsilon}[G(n+1,k ; n',k') - G(n-1,k ; n',k') ] + \frac{ \sigma_{3}}{2 \epsilon}[G(n,k+1 ; n',k') -G(n,k-1 ; n',k')]
            - m \sigma_{1}G(n,k ; n',k') \nonumber\\ 
            & \hspace{10.3 cm}= \delta(n-n') \delta(k-k')\, , \label{eq:Greensfunction}
   \end{align} }%
   where---for simplicity, and because we have already done that for sequences on $\Lambda^2$---we have used the continuous-variables notation $\delta(x)$ although we are here just dealing with the Kronecker deltas used for discrete variables. Now, generically speaking, the main interest of Green's function is that, if one manages to determine it, i.e., to solve the previous equation, then one can evolve any initial condition $\psi(n',k')$ known at time $n'$, up to any other time $n$, via the following equation (this is a standard result of Green's function theory),
  \begin{equation}
 \psi(n,k) =   \sum_{k'}  G(n,k; n',k') \psi(n',k') \, .
  \end{equation}

\vspace{2mm}
\noindent
    {{\bfseries Solution in Fourier space.}} 
   As before for the field itself, we are going to look for a solution $G(n,k;n',k')$ of Eq.\ \eqref{eq:Greensfunction} under the form of the following ansatz, which is strictly exactly the same as the one used before for the field itself if we take into account the fact that the space and time translation invariances of the model (i.e., of Eq.\ \eqref{eq:Greensfunction}) imply that $G(n,k;n',k')$ is actually a function solely of $n-n'$ and $k-k'$,
    \begin{equation}\label{eq:propagator}
	G(n,k ; n'k') =  \int_{- \pi/\epsilon}^{\pi/\epsilon}  \frac{dE}{2 \pi } \int_{- \pi/\epsilon}^{\pi/\epsilon}  \frac{dp}{2 \pi } \hspace{1mm}  
			e^{-iE(n-n') \epsilon + ip(k-k')\epsilon}  G_{\mathcal{B}}(E,p) \, .
    \end{equation}
    Generally speaking, Green's function is defined as the right inverse of the equation-of-motion linear operator, in the sense of distributions~\cite{economou2006green}, as expressed by Eq.\ \eqref{eq:Greensfunction} above. Now, by inserting the ansatz of the previous equation into Eq.\ \eqref{eq:Greensfunction}, we realize (after a computation very similar to that yielding Eq.\ \eqref{eq:MotionFourier}) that it is sufficient to ask, for Eq.\ \eqref{eq:Greensfunction} to be satisfied, that the following Fourier-space equation is satisfied,
    \begin{equation}
    {D}_{\mathcal{B}} (E,p)  G_{\mathcal{B}}(E,p) = \mathbb{I}_{2} \, . 
    \end{equation}
    Finally, to solve the previous equation, we can just use Cramer's rule~\cite{Smit_2023} to find the inverse of the $2\times 2$ matrix ${D}_{\mathcal{B}}(E,p)$ given in Eq.\ \eqref{eq:DBEp},---a formula which in this low-dimensional case is very easy to work out---, and this eventually yields: 
	\begin{equation}\label{eq:green}
		G_{\mathcal{B}}(E,p) =  \frac{  \mathbb{I}_{2} \sin(E \epsilon) 
        +\sigma_{3} \sin(p \epsilon) + m  \epsilon\sigma_{1} }{\det\{{D}_{\mathcal{B}}(E,p)\}} .
	\end{equation}

\vspace{2mm}
\noindent
    {{\bfseries The origin of the FD problem.}} Having in mind Eqs.\ \eqref{eq:propagator} and \eqref{eq:green}, it is natural to think that the most significant contribution to the integral of the first equation may come from the neighborhoods of the zeroes of the denominator $\det\{{D}_{\mathcal{B}}(E,p)\}$, since these are likely to be poles of $G_{\mathcal{B}}(E,p)$. In fact, it is a standard result that in the continuum limit where $\epsilon$ goes to zero, these neighborhoods in $\mathcal{B}$ will be, strictly, the only contributors to the integral, as it can be seen when the integral is turned into a complex contour integration~\cite{book_Peskin_Schroeder}. In other words, in the continuum limit the pairs $(E,p)$ that make the denominator zero are the only ones that allow propagation, i.e., the only physically relevant pairs. More precisely, such pairs determine the solution space of the Dirac equation: any solution must a be a quantum superposition\footnote{Here, a Fourier integral over the appropriate $E$s and $p$s, by putting in the integrand a Dirac delta factor centered on the dispersion-relation expression, which restablishes on-shellness.} of the spacetime wavefunctions that correspond to these $(E,p)$ pairs. \\

\noindent
    {{\bfseries Explaining the FD problem in more depth.}} 
Let us reformulate and develop the ideas of the previous paragraph and give a final explanation of the FD problem, via the neighborhoods $U_{(r,s)}$ that we introduced previously and via the dispersion relation. Inspecting Eq.\ \eqref{eq:green}, we see that as $\epsilon$ gets close to zero,  $G_{\mathcal{B}}(E,p)$ also gets close to zero except for the points in the neighborhoods\footnote{Expanding both the numerator and the denominator of $G_{\mathcal{B}}(E,p)$ in some neighborhood $U_{(r,s)}$, we see that the leading order of the numerator is of order $\epsilon$, whereas the next-to-leading order of the denominator, which is the one that matters, is of order $\epsilon^2$, which makes $G_{\mathcal{B}}(E,p)$ ``blow up'' in such a neighborhood. The leading order of the denominator is of order $1$, but its coefficient is the dispersion-relation expression $E^2 -p^2 -m^2$, which vanishes once we take the integral over the energy $E$ to put us on-shell---and we are in the continuum limit, so that indeed we finally get the continuum on-shell relation $E^2 = p^2+m^2$.} $U_{(r,s)}$. As a consequence, all of the $U_{(r,s)}$ contribute to the propagator just like $U_{(0,0)}$ does, leading to over-counting. In the exact continuum limit where $\epsilon=0$, the Fourier space $\mathcal{B}$ becomes $\mathbb{R}^2$, so that the points in the boundary of $ \mathcal{B}$ are no longer valid Fourier modes: except for $U_{(0,0)}$, the $U_{(r,s)}$ are no longer defined in the continuum limit. But, mathematically, we could still define some analogs of them by considering neighborhoods of the $p$-axis and of the $E$-axis at infinity, that is, in the case we are on-shell, neighborhoods of the lines defined by $\pm E = \pm\sqrt{p^2+m^2}$, which for $E$ and $p$ going to infinity are the lines $\pm E = \pm p$. These neighborhoods would again still be defined in the exact continuum limit, where they would again lead to over-counting.


\subsection{Resolution}

\noindent 
{{\bfseries Introduction.}} In order to remove FD from the previous lattice formulation of fermions, one can seek to obtain in $ \det\{D_{\mathcal{B}}(E,p)\}$ terms that depend on  $\sin({p \epsilon / 2})$ and $\sin({E \epsilon / 2})$, since $\sin(x/2)$ is not symmetric under the transformation $x \to \pi - x$. This is the road that has been followed by several methods, falling into two families. In the first family, the Dirac equation is modified either additively in the case of Wilson fermions~\cite{Wilson:1975id}, or multiplicatively in the case of Ginsparg-Wilson fermions~\cite{PhysRevD.25.2649}, both leading to a breaking of the chiral symmetry of the initial massless scheme. In the second family, called ``staggered fermions'', the lattice vectors are modified by 1/ choosing the Weyl representation of the Dirac equation---and, more generally, in higher dimensions, fully diagonalizing the internal-degree-of-freedom part of the transport term~---, then 2/ {\em flavoring}, i.e., putting in the model ne necessary number of flavors (for the following step to in the end work), and, finally, 3/ {\em staggering}, i.e., laying out, in simple cases, the $\psi^+$ and $\psi^-$ chiral components, and, in more involved cases, all the internal components of the ``multi-flavor wavefuntion'' (typically: chirality, spin, ``color'', flavor), at different lattice sites~\cite{Susskind_1977}, more precisely, each internal component populates a single and unique sublattice\footnote{The necessary number of flavors for this ``staggered fermions'' method to work depends on the dimension of the system; in (1+1)D spacetime there is actually no need for flavoring, but in higher dimensions there is.}. In the third step, it is known that staggering the chiral components breaks the chiral symmetry of the massless scheme. Now, the chiral-symmetry breaking of the massless lattice model is not always wanted, but it can be a choice motivated by considerations of lattice quantum chromodynamics~\cite{PhysRevD.15.1111}. We speak of \emph{choice}, because indeed nothing forces the lattice model to always break the chiral symmetry of the massless scheme: Indeed, the infamous no-go theorem about chiral-symmetry breaking on the lattice~\cite{NielsenI1981} only concerns \emph{chirally charged particles}---as discussed further in Sec.\ \ref{sec:discussion}. Let us show that for Dirac particles, \emph{which are not chirally charged}, this does not have to be the case---i.e., this breaking of chiral symmetry is not necessarily brought about by the latticization of the model---, by staggaering the flavor only but not the chirality. \\

\noindent 
{{\bfseries Decribing our flavor-staggering-only solution.}}
In the second family of methods, one essentially seeks to implement a map $\rho:\mathcal{B}\to\mathcal{B'}$ such that $(p,E) \mapsto (p',E')\defeq(p/2, E/2)$, so that $ \det\{D_{\mathcal{B}'}(E,p)\}$ depends on $(\sin(p')$, $\sin(E'))$: since $\mathcal{B'}$ is half of $\mathcal{B}$, the degeneracies due to the $\pi$-symmetries will be avoided. But, what sort of map is this $\rho$? Recall that $\mathcal{B} $ is the reciprocal space of $ \Lambda^{2}$, from which it follows that $\rho$ has a corresponding map at the level of the lattice, which we call $ \rho^{*} :  \Lambda^{2} \to  \Lambda^{2}_{(0,0)}$, where $\Lambda^{2}_{(0,0)}$ is the sublattice whose generating vectors are  $\{ \vec{a}_{0}', \vec{a}_{1}' \}$ with $\vec{a}_{\mu}' \defeq 2 \vec{a}_{\mu}$, $\mu=0,1$. Notice that $\Lambda^{2}$ has actually four such sublattices, i.e. $\Lambda^{2} =  \Lambda^{2}_{(0,0)} \cup  \Lambda^{2}_{(1,0)} \cup   \Lambda^{2}_{(0,1)} \cup  \Lambda^{2}_{(1,1)}$, where $\Lambda^{2}_{(i,j)}\defeq \Lambda^{2}_{(0,0)} + (i\vec{a}_0,j\vec{a}_1)$, $i,j=0,1$. 

Now, the solution we propose to solve the FD problem is to simply ``flavor'' these sublattices, i.e., to consider that they host different particle species. That way, we are effectively breaking the $\{a_0 , a_1 \}$ lattice symmetry down to a $ \{ \vec{a}_{0}', \vec{a}_{1}'\}$ lattice symmetry. Technically, the one-particle sector of the QFT induced by our model, one-particle sector which can be thought of as represented by our classical-fields model, therefore got split into four distinguished sectors of different flavors which we denote as: $\psi_{f} : \Lambda^{2}_{f} \to \mathbb{C}^{2} $, where $ f \in \{ (0,0) , (1,0) ,(0,1) , (1,1) \} $. 
This $f$ degree of freedom appears to be associated to a global lattice-symmetry argument which the dynamics does not see, which justifies the use of the word ``flavor'', as opposed to the word ``color'' which is normally used for gauge-symmetry-induced degrees of freedom that are coupled to the dynamics. Back to Fourier space, notice also that whilst $\mathcal{B}$ had four neighborhoods contributing to the propagator, $\mathcal{B'}$ has only one such neighborhood, but now we have four copies of it, one for each flavor. The present resolution of the FD did not change the size of the solution space, it just induced a continuum which matches this solution space. \\

\noindent 
{{\bfseries Final comments.}}
Notice again that we have not changed the dynamics, which is implicit in the concept of ``flavor''. What we have done  is changing the type of questions that can be asked, both in the discrete and in the continuous model. All questions asked now need to include flavor information. Endowed with this extra flavor degree of freedom, both the discrete and the continuous models can be related to each other consistently, i.e., the discrete model can be used to simulate the continuous one without errors.

We believe that this {\em flavor-staggering-only} resolution of FD is very general. Here, we started from a discrete-space discrete-time equation of motion. But we could have started from a discrete-space continuous-time one. Either way, one should count the degeneracies of $ \det\{D_{\mathcal{B}}(E,p)\}$, which do not vanish in the continuum limit, and define a map $\rho$ on the BZ to leave them out of reach. This induces a first sublattice in direct space, whose number of distinct translates corresponds to the number of degeneracies. Finally, each sublattice is attributed a flavor.

Here, the lattice equation of motion was obtained by discretizing, in a Hermiticity-preserving manner, the continuum equation of motion arising from a well-known first-quantization\footnote{It actually turns out that the quantized fields anyways also satisfy the exact same equation.} quantum Hamiltonian---having no standard classical counterpart because of the purely quantum internal-degree-of-freedom part. That first-quantization Hamiltonian equation of motion can also be seen, mathematically speaking, as purely classical, if the wavefunction is seen as a classical field, but in that case, the relevant physical object is not any first-quantization Hamiltonian, but either a Lagrangian field density, or a Hamiltonian field density. In the first case, one would discretize directly the field Euler-Lagrange equations, and in the second case, one would discretize Hamilton's field equations for the \emph{field} Hamiltonian (which is not the previous fist-quantization quantum Hamiltonian, although it is related to it). In the next section, we will see that a similar discretization method---but preserving, this time, a \emph{higher number} of relevant properties of the continuum dynamics\footnote{Such as the chiral nature of the transport, or the unitarity. One can prove that most often, unitarity cannot be preserved by a naive space and time discretization of the model, we show this in an article in preparation, unpublished yet.}---applies for a discrete-time (and discrete-space) dynamics prescribed directly in terms of a unitary operator rather than on a Hamiltonian or a Lagrangian, simply because we will again base ourselves on the relevant equation of motion.
	
    \section{Fermion doubling in QCAs}\label{sec:qedqcafd}
	
	We investigate Fermion Doubling in QCA models. We do it for the QCA models that appeared in Refs.\ \cite{ABF20} and~\cite{EDMMplus2023}. These models give an algorithm to simulate QED, using local unitary quantum gates that evolve the states. Actually, we only need to focus on the non-interacting one-particle sector of these models. Indeed, FD is a problem that can in the end be analyzed and treated within the sole non-interacting one-particle sector of a fermionic system, even if generates actual problems only in the interacting model, whether it is one-particle (background interaction field, usually not very interesting) or multi-particle (which is usually the regime we are interested in). Let us clarify this point. If one considers non-interacting models, of one or several particles, the FD problem can simply be avoided by means of well-behaving assumptions of initial conditions. This is because if the initial wave function of a particle is both smooth enough and, in Fourier space, as little populated on the FD neighborhoods as possible, it will remain so, and this can be used to establish convergence~\cite{ArrighiDirac}. Once interactions are ``turned on'', however, this initial situation may be jeopardized, and very-high-frequency modes may appear, producing non-physical solutions--- again, even in the way a single particle propagates. To come back to our initial claim, the analysis of FD really takes us back to the non-interacting one-particle sector to answer questions such as: {\em what goes wrong numerically, if high-frequency modes are enabled? How can we fix those, without relying on well-behaving assumptions on the initial condition?}

	\subsection{FD problem in a $(1+1)$D QED QCA} \label{subsec:1qedqcafd}

	A $(1+1)$D QCA simulating QED has been described in Ref.\ \cite{ABF20}. Since FD is, as we said above, a problem that can in the end analyzed and treated in the simple situation of free Dirac fermions, we can just turn off the interaction terms of Ref.\ \cite{ABF20}, falling back into a model that we call $(1+1)$D free Dirac QCA. In fact, we can even, again as we said, focus on the one-particle sector, a model we call $(1+1)$D Dirac discrete-time quantum walk (QW).

    \subsubsection{A standard $(1+1)$D Dirac-QW model}

\noindent
    {{\bfseries State space.}}     
    An arbitrary state of this single-particle model in the Schrödinger picture, at time $ t = n \epsilon $,  can be written as
	
	\begin{equation}\label{eq:psiQW}
		\ket{\psi(n)} \equiv \sum_{k} \psi^{+}(n,k) \ket{k_{+}} + \psi^{-}(n,k) \ket{k_{-}},
	\end{equation}
	where $\ket{k_{\pm}} \defeq \ket k \otimes \ket \pm$ represents a basis state of position $ x = k \epsilon$ and chirality $\pm$, and belongs, as $\ket{ \psi(n)}$, to the state composite Hilbert space $\mathcal H_{\text{position}} \otimes  \mathcal H_{\text{chirality}} $, with obvious notations. The functions $ \psi^{\pm}(n,k)$ are from $ \Lambda$ to $ \mathbb{C}$. \\

    \noindent
    {{\bfseries Dynamics in terms of a unitary evolution operator.}}
    When the  rule for the unitary dynamics of the $(1+1)$D Dirac QW of Ref.\ \cite{Arrighi1DQCA} are applied to $  \ket{\psi(n)}$, we obtain
    \begin{subequations}
    \label{eqs:Dirac}
	\begin{align}
		\psi^{+}(n+1 , k) &=  \cos(m \epsilon)\, \psi^{+}(n,k-1) - i \sin(m \epsilon)\,\psi^{-}(n ,k) \label{eq:1}\\
		\psi^{-}(n+1 , k) &= \cos(m \epsilon)\,\psi^{-}(n,k+1) -  i\sin(m \epsilon)\,\psi^{+}(n ,k) \label{eq:2} \, ,
	\end{align}
    \end{subequations}
	where $ m $ is again the mass of the fermion in the continuum and $\epsilon$ the spacetime discretization step. Defining $ \psi(n,k)  \defeq (\psi^{+}(n,k),  \psi^{-}(n,k))^{\top} 
    : \Lambda^{2} \to \mathbb{C}^{2}$, with $\top$ denoting the transposition, and take the continuum limit, Eqs.\ \eqref{eqs:Dirac} converge to the Dirac equation~\cite{ ABF20} with $\psi(n,k)$ becoming the Dirac spinor $\psi_{\text{cont.}}(t,x)$. We can write Eqs.\ \eqref{eqs:Dirac} in matrix form as
	
	\begin{equation}\label{eq:neweqmotion}
    \psi(n+1,k) =  (\mathcal{U}^{1}_{QW} \psi)(n,k)   \, ,
	\end{equation}
	where we have introduced the one-particle one-step evolution operator
    \begin{equation}
       \mathcal{U}^{1}_{QW}  \defeq \begin{bmatrix}
				\cos(m \epsilon) \mathcal{S} &  - i \sin(m \epsilon) \\
				- i \sin(m \epsilon) & \cos(m \epsilon) \mathcal{S}^{\dagger}
			\end{bmatrix} \, ,
    \end{equation}
    where $\mathcal{S}$ is the spatial-translation operator to the right (in position representation of quantum mechanics), that acts on wavefunctions as $ (\mathcal{S} \psi)(n,k) \equiv  \psi(n,k-1) $, so that  $(\mathcal{S}^{\dagger} \psi)(n,k) =  \psi(n,k+1) $. \\

        \noindent
    {{\bfseries Time operator and equation-of-motion operator.}}
    In the same manner, we can define a forward time-translation operator $ \mathcal{T}$, whose action is $ (\mathcal{T} \psi)(n,k) \equiv \psi(n-1, k)$, so that $ (\mathcal{T}^{\dagger} \psi)(n,k) = \psi(n+1, k)$. With this time-translation operators at hand, we can rewrite Eq.\ \eqref{eq:neweqmotion} as
	\begin{equation}\label{eq:1dQCA}
			(\mathcal{M}_{QW}^{1} \psi)(n,k) = 0 \, ,
	\end{equation}
    where we have introduced the equation-of-motion operator
    \begin{equation}
    \label{eq:eomoperator}
      \mathcal{M}_{QW}^{1} \defeq 
			\begin{bmatrix}
				\cos(m \epsilon) \mathcal{S} - \mathcal{T}^{\dagger} &  - i \sin(m \epsilon) \\
				- i \sin(m \epsilon) & \cos(m \epsilon) \mathcal{S}^{\dagger} - \mathcal{T}^{\dagger}
			\end{bmatrix} \, .
    \end{equation}
    

    \subsubsection{What the two-step description suggests regarding FD}

    Now, we can actually make contact with the naive discretization scheme of Sec.\ \ref{subsec:eqmotions} by substracting to the operator of Eq.\ \eqref{eq:eomoperator} its Hermitian conjugate\footnote{This means, essentially, taking the anti-Hermitian part of that operator.} (and by multiplying by $i$ in order to preserve overall Hermicity), so as to make appear symmetric finite differences, as we are going to show. We indeed have that
    \par\nobreak
    {\small
    \begin{subequations}
    \label{eq:antihermqca-several}
	\begin{align}\label{eq:antihermqca}
		- \frac{i }{2\epsilon}(\mathcal{M}_{QW}^{1 } - \mathcal{M}_{QW}^{1 \dagger}) &= \frac{1}{\epsilon}\begin{bmatrix}
		-i\cos(m \epsilon) (\mathcal{S} -\mathcal{S}^{\dagger})/2 +i(\mathcal{T}^{\dagger}-\mathcal{T})/2 & - \sin(m \epsilon) \\
			 - \sin(m \epsilon) &  -i\cos(m \epsilon)(\mathcal{S}^{\dagger} -\mathcal{S})/2 + i(\mathcal{T}^{\dagger}-\mathcal{T})/2
		\end{bmatrix} \\
        &= \cos(m \epsilon)\frac{ i \sigma_{3}}{ 2 \epsilon}    (\mathcal{S}^{\dagger} -\mathcal{S})  + \frac{ i\mathbb{I}_{2}}{ 2 \epsilon} (\mathcal{T}^{\dagger}-\mathcal{T})  - \sigma_{1} \frac{\sin(m \epsilon)}{\epsilon} \, .
	\end{align}
    \end{subequations}}%
Applying now the operator of the previous, last expression, to $\psi$, taking the whole at $(n,k)$, and making it vanish because of Eq.\ \eqref{eq:1dQCA}, we obtain
    \par\nobreak
    {\small
	\begin{equation}
    \label{eq:finaleqsimnaive}
	    	\frac{ i\mathbb{I}_{2}}{ 2 \epsilon} (\psi(n+1,k) - \psi(n-1,k)) +\cos(m \epsilon)\frac{ i \sigma_{3}}{ 2 \epsilon}    (\psi(n,k+1) -\psi(n,k-1))\\  - \sin(m \epsilon) \frac{\sigma_{1}}{\epsilon}  \psi(n,k)= 0 \, .
	\end{equation}}%
	The scheme of the previous equation is almost what we had in the naive discretization of the Dirac equation in Sec.\ \ref{subsec:eqmotions}---in the massless case $m=0 $,  it should be noticed that the previous equation simply exactly coincides with Eq.\ \eqref{eq:MotionDirac}.
    However, remember that the operator of Eqs.\ \eqref{eq:antihermqca-several} is the anti-Hermitian part of the operator of Eq.\ \eqref{eq:eomoperator}, not the original operator itself, and because this anti-Hermitianization adds twice the original operator (up to minus signs), this raises hopes that at the level of the original operator there are less FD degeneracies, i.e., this original operator could be less FD-prone than Eqs.\ \eqref{eq:finaleqsimnaive} and \eqref{eq:MotionDirac}. We are going to see that this is indeed what happens. 

    \subsubsection{Analyzing FD in the (1+1)D QED QCA}
    \label{subsubsec:QWFD}

    \noindent
    {{\bfseries Introduction.}}
Let us now analyze for good FD for the $(1+1)$D Dirac QW. In the case of the naive discretization via finite differences, we had looked at the continuum limit of the Fourier-space Green's function, and realized that it was non-vanishing only at the zeroes of its denominator, which is the determinant of the Fourier-space equation-of-motion operator. One can convince oneself that this a general behavior for any dynamical scheme and any discretization of the latter. In any case, one can rigorously prove that this is indeed the case for the particular discretization of the Dirac equation that the $(1+1)$D Dirac QW is, namely, Eq.\ \eqref{eq:1dQCA}, with the equation-of-motion operator being this time $\mathcal{M}^1_{QW}$, given by Eq.\ \eqref{eq:eomoperator}. So let us focus on the determinant of the Fourier-space version of $\mathcal{M}^1_{QW}$. \\

    \noindent
    {{\bfseries Fourier-space description.}}
We go to Fourier space by using Eqs.\ \eqref{eq:Fourier}, which in particular imply
	\begin{subequations}
	\begin{align}
		(\mathcal{S} \psi ) (n,k) &= \frac{\epsilon}{4 \pi^{2}}\int_{- \pi/\epsilon}^{\pi/\epsilon}  dE \int_{- \pi/\epsilon}^{\pi/\epsilon} dp \hspace{1mm} \tilde{\psi}(E,p) e^{-iEn \epsilon + ipk\epsilon} e^{-ip\epsilon}\\
		(\mathcal{S}^{\dagger} \psi) (n,k) &= \frac{\epsilon}{4 \pi^{2}}  \int_{- \pi/\epsilon}^{\pi/\epsilon}  dE \int_{- \pi/\epsilon}^{\pi/\epsilon}  dp \hspace{1mm} \tilde{\psi}(E,p) e^{-iEn \epsilon+ipk\epsilon} e^{ip\epsilon}\\
		(\mathcal{T}^{\dagger}\psi)(n,k) &=  \frac{\epsilon}{4 \pi^{2}} \int_{- \pi/\epsilon}^{\pi/\epsilon}  dE \int_{- \pi/\epsilon}^{\pi/\epsilon} dp \hspace{1mm} \tilde{\psi}(E,p) e^{-iEn \epsilon+ipk\epsilon} e^{-iE\epsilon}.
	\end{align}
    \end{subequations}
	Note that the integration domain is again $(p,E)\in\mathcal{B}$, the reciprocal space of the lattice coordinates $ \Lambda^2$. 
    Inserting the previous equations into Eq.\ \eqref{eq:1dQCA}, we find
	\begin{equation}\label{eq:neweqmotion-fourier}
		\begin{split}
			\frac{1}{4 \pi^{2}} \int_{- \pi/\epsilon}^{\pi/\epsilon}  dE \int_{- \pi/\epsilon}^{\pi/\epsilon} dp \hspace{1mm}  e^{-iEn \epsilon+ipk\epsilon} M^1_{QW; \mathcal B}(E,p)
				\tilde{\psi}(E,p)
				= 0 \, ,
		\end{split}
	\end{equation}
    where we have introduced
    \begin{equation}
        M^1_{QW; \mathcal B}(E,p) \defeq \begin{bmatrix}
				\cos(m \epsilon)  e^{-ip\epsilon} -  e^{-iE\epsilon} &  - i \sin(m \epsilon) \\
				- i \sin(m \epsilon) & \cos(m \epsilon) e^{ip\epsilon} - e^{-iE\epsilon}
			\end{bmatrix} \, ,
    \end{equation}
which, because of Eq.\ \eqref{eq:neweqmotion-fourier}, we recognize as being the Fourier-space equation-of-motion operator of our model. \\

    \noindent
    {{\bfseries The Fourier-space determinant.}}
The determinant of the matrix of the previous equation, that we denote by ${\mathscr{D}}^{1}(E,p)$, can be computed easily, and the final expression can be written
    
\begin{subequations}
	\begin{align}
		{\mathscr{D}}^{1}( E,p)&= 1 + e^{- 2iE\epsilon} -2 e^{-iE\epsilon} \cos(p \epsilon)\cos(m \epsilon) \\
        &= 2e^{-iE\epsilon}[\cos(E\epsilon) - \cos(p\epsilon) \cos(m\epsilon)] \, .
	\end{align}
\end{subequations}
	Taylor expanding the previous expression for small $ \epsilon$, we get
	\begin{equation}
		\frac{{\mathscr{D}}^{1}( E,p)}{\epsilon^2} = -E^{2} + p^{2} + m^{2}+ {O}(\epsilon) \, ,
	\end{equation}
	so that in the exact continuum limit $\epsilon=0$, by setting the previous expression to zero (which as we know gives fully on-shell solutions), we recognize the dispersion relation of the continuum Dirac theory. \\

        \noindent
    {{\bfseries Explaining the FD problem.}}
    Now, let us get to the zeroes of ${\mathscr{D}}^{1}( E,p)$. One thing we can already notice is that, that unlike for the naive discretization, the modes $( E, p \pm \pi/\epsilon)$ and $ ( E \pm \pi/\epsilon, p )$ for $ E$ and $ p$ finite,  do not pollute the continuum limit of the $(1+1)$D Dirac QW, because as $\epsilon\to 0$ we have that
	\begin{equation}
		\begin{split}
			{\mathscr{D}}^{1}( E, p \pm \pi/\epsilon) &\neq {\mathscr{D}}^{1}( E, p)+ O(\epsilon^3)\\
			{\mathscr{D}}^{1}( E  \pm \pi/\epsilon, p) &\neq {\mathscr{D}}^{1}( E, p) + O(\epsilon^3) \, .
		\end{split}
	\end{equation}
	However for the modes $ ( E \pm \pi/\epsilon, p \pm \pi/\epsilon)$ we do have
	\begin{equation}
		\begin{split}
			\mathscr{D}^{1}( E \pm \pi/\epsilon, p \pm \pi/\epsilon)=-\mathscr{D}^{1}( E, p).
		\end{split}
	\end{equation}
    Hence, the $(1+1)$D Dirac QW has spurious solutions due to the specific $\pi$-symmetries $(p,E) \mapsto (  p \pm \pi/\epsilon, E \pm \pi/\epsilon)$ symmetry, which we call \emph{spatiotemporal doublers}, since the symmetry affects \emph{both} the momentum \emph{and} the energy, in contrast for example with so-called (purely) \emph{spatial doublers}, ubiquitous in continuous-time LGT, or with the (purely) \emph{temporal doublers} we have seen before in the naive discretization, which are those for which the $\pi$-symmetry affects the energy solely. \\

            \noindent
    {{\bfseries Conclusion: there is still one doubler.}}
    As a conclusion, we see that the $(1+1)$D Dirac QW is indeed less FD-prone than the naive discretization. High-momentum but low-energy modes, that is, potential spatial doublers, which appear for example in continuous-time LGT,  are actually no longer problematic for our QW (i.e., these modes are not doublers in this case), and this has led to the mistaken claim that this QW is FD-free~\cite{ABF20}. Unfortunately, this is not quite the case: clearly the four neighbourhoods $U_{(\pm 1,\pm 1)}$ contribute to the propagator just like $U_{(0,0)}$ does, leading again to over-counting, which will pollute the numerics.  Fig.\ \ref{fig:doublers} portraits the degeneracies.
    The dashed-\textcolor{red}{red} circled neighbourhood $U_{(0,0)}$ is the usual solution of the Dirac theory. Recall that this Dirac theory was polluted by three extra neighbourhoods $ \{U_{(r,s)}: r,s=-1,0,1\} $ in the naive discretization (yellow, green, and blue). With the $(1+1)$D Dirac QW, it is only polluted by the  \textcolor{blue}{blue} neighbourhoods $ U_{( \pm 1, \pm 1)}$, which, remember, can actually be seen as the same, single neighborhood, due to the toric geometry of the BZ.

	\subsection{FD problem in a $(3+1)$D QED QCA}\label{subsec:3qedqcafd}

\subsubsection{Standard $(3+1)$D Dirac-QCA and Dirac-QW models}

{\bfseries State space.} Again, a $(3+1)$D QCA simulating QED has been proposed in Ref.\ \cite{EDMMplus2023}, but since FD is a problem that can in the end be analyzed and treated within the sole non-interacting one-particle sector of a fermionic system, we turn-off the interaction term, focus on the one-particle sector, and fall back onto a $(3+1)$D Dirac QW.     
    A $(3+1)$D Dirac QW is a set of rules that describes the unitary and local discrete-time dynamics of a multicomponent wavefunction in a $3$D spatial lattice, and which has as (at least one possible) continuum limit the $(3+1)$D Dirac equation. Naturally, we will consider the specific $(3+1)$D Dirac QW used in Ref.\ \cite{EDMMplus2023}. This QW decomposes into four substeps: three steps implementing propagation along each space dimension, and one step for the mass term. 
    The model of course replaces continuous spacetime with a four-dimensional orthogonal lattice, $\Lambda^{4}$,  still with lattice spacing $\epsilon$.  In the multiparticle model, each lattice site is occupied by $4$ qubits, which correspond to the occupation numbers for the four possible components of the Dirac spinor that we discretize. In the one particle sector, so coming back to the QW, and using the Schr\"odinger picture, at a given time step $n$, the overall state, which we denote by $\ket{\psi^{3} ({n})} $, is thus of the form 
	\begin{align}\label{eq:3dwavefunction}
	   \ket{\psi^{3} ({n})} = \sum_{k,l,m} &\Big[ \psi^{(1)}(n,k,l,m) \ket{1 0 0 0 }^{k,l,m} + \psi^{(2)}(n,k,l,m) \ket{0 1 0 0 }^{k,l,m} \nonumber\\ & \ \ + \psi^{(3)}(n,k,l,m) \ket{0 0 1 0 }^{k,l,m} + \psi^{(4)}(n,k,l,m) \ket{0 0 0 1 }^{k,l,m} \Big] \, ,
	\end{align}
	where each $\psi^{(i)}(n,k,l,m) \in \mathbb C$, $i=1,...,4$, and where the kets in the sum are taken from the multiparticle formalism but restricted to the one-particle subspace, so that one could introduce a single-particle position Hilbert space and a 4D internal Hilbert space with basis $\{\ket 1, \ket 2, \ket 3, \ket 4 \}$, and write $\ket{1000}^{k,l,m} \equiv \ket{k,l,m}\ket{1}$, $\ket{0100}^{k,l,m} \equiv \ket{k,l,m}\ket{2}$, $\ket{0010}^{k,l,m} \equiv \ket{k,l,m} \ket{3}$, and $\ket{0001}^{k,l,m} \equiv \ket{k,l,m} \ket{4}$. \\

    \noindent
{\bfseries Dynamics.} Letting $ \psi(n,k,l,m) \equiv \psi^3(n,k,l,m) := (\psi^{(1)}(n,k,l,m),$ $ \psi^{(2)}(n,k,l,m),$ $ \psi^{(3)}(n,k,l,m),$ $\psi^{(4)}(n,k,l,m))^{\top}$, 
the function $\psi : \Lambda^{4} \to  \mathbb{C}^{4}$ is the discretized Dirac multicomponent wavefunction. When applying the unitary prescribed by the model of Ref.\ \cite{EDMMplus2023} in the single-particle sector, we obtain four coupled equations, one for the dynamics of each $\psi^{(i)}(n,k,l,m)$, that we can condense into the following operator equation of motion:
	\begin{equation}\label{eq:MQW 3}
		(\mathcal{M}_{QW}^{3} \psi)(n,k,l,m) = 0,
	\end{equation}
	where 
    \begin{equation}
     \mathcal{M}_{QW}^{3} \defeq \mathcal{U}_{QW}^{3} - \mathcal{T}^{\dagger}    \, ,
    \end{equation}
    with $\mathcal{T}$ again the forward time-translation operator, and where $\mathcal{U}_{QW}^{3} $  is the unitary time-evolution operator of the $ (3+1)$D Dirac QW, as defined by the product of four substeps mentioned previously,
   \begin{equation} \label{eq:u3qw}
       \mathcal{U}^{3}_{QW} \defeq   \mathcal{U}^{3}_{m} \mathcal{U}^{3}_{{x}} \mathcal{U}^{3}_{{y}}  \mathcal{U}^{3}_{{z}},
   \end{equation}
   with 
   \begin{subequations}
   \begin{align}
           \mathcal{U}^{3}_{{z}} &\defeq \exp{-i \epsilon\mathcal{P}_{z} \sigma_{3} \otimes \sigma_{3}} \\  \mathcal{U}^{3}_{{y}} &\defeq \exp{- i\epsilon \mathcal{P}_{y} \sigma_{3} \otimes \sigma_{2} } \\
             \mathcal{U}^{3}_{{x}} &\defeq \exp{-i \epsilon \mathcal{P}_{x} \sigma_{3} \otimes \sigma_{1} } \\  \mathcal{U}^{3}_{m} &\defeq \exp{- i \epsilon m \sigma_{2} \otimes \mathbb{I}_{2} } \, ,
   \end{align} 
   \end{subequations}
   where $\mathcal P_x$, $\mathcal P_y$ and $\mathcal P_z$ are the (position-representation) momentum operators of the \mbox{continuum---but} obviously reduced to the Brillouin zone when viewed in momentum representation. We call the substep operators $  \mathcal{U}^{3}_{{x}},  \mathcal{U}^{3}_{{y}} $ and $  \mathcal{U}^{3}_{{z}}  $, \emph{transport terms}, whereas $  \mathcal{U}^{3}_{m}  $  is the \emph{mass term}.\\ 

\subsubsection{Analyzing FD in the $(3+1)$D QED QCA}

\noindent
{\bfseries Going to Fourier space.} Let us proceed to analyze FD. First, let us prepare the stage by introducing many relevant objects. We start by expressing $    \mathcal{U}^{3}_{QW}$ in terms of the (position-representation) translation operators defined via
\begin{subequations}
\label{eq:translation operators}
	\begin{align}
	    (\mathcal S_{x}\psi)(n,k,l,m) &\equiv \psi(n,k-1,l,m) \\
			(\mathcal  S_{y}\psi)(n,k,l,m) &\equiv \psi(n,k,l-1,m)\\
			(\mathcal  S_{z}\psi)(n,k,l,m) &\equiv \psi(n,k,l,m-1) \, .
	\end{align}
\end{subequations}
  This gives the following operator-valued $4 \times 4$ matrix,
\par\nobreak  
	\begin{multline}\label{eq:UQW 3d}
		\mathcal{U}^{3}_{QW}=\frac{1}{4} \mathcal  S_{x}^{\dagger}\mathcal  S_{y}^{\dagger}\mathcal  S_{z}^{\dagger}\left[
		\begin{matrix}
			(1+i) \left(\mathcal  S_{x}^2 \left(1-i \mathcal  S_{y}^2\right)+\mathcal  S_{y}^2-i\right) c& (1+i) \mathcal S_{z}^2 \left(i \left(\mathcal S_{x}^2+i\right) \mathcal  S_{y}^2+\mathcal  S_{x}^2-i\right) c \\
			(-1+i) \left(i \mathcal S_{x}^2 \left(\mathcal  S_{y}^2+i\right)+\mathcal  S_{y}^2-i\right) c& (1+i) \mathcal  S_{z}^2 \left(\mathcal  S_{x}^2 \left(\mathcal  S_{y}^2-i\right)-i \mathcal  S_{y}^2+1\right) c \\
			(1+i) \left(\mathcal  S_{x}^2 \left(1-i \mathcal  S_{y}^2\right)+\mathcal  S_{y}^2-i\right) s& (1+i) \mathcal  S_{z}^2 \left(i \left(\mathcal  S_{x}^2+i\right) \mathcal  S_{y}^2+\mathcal  S_{x}^2-i\right) s \\
			(-1+i) \left(i \mathcal  S_{x}^2 \left(\mathcal  S_{y}^2+i\right)+\mathcal  S_{y}^2-i\right) s& (1+i) \mathcal  S_{z}^2 \left(\mathcal  S_{x}^2 \left(\mathcal  S_{y}^2-i\right)-i \mathcal  S_{y}^2+1\right) s\\
		\end{matrix} \right.\\ \left. 
		\begin{matrix}
			(-1+i) \mathcal S_{z}^2 \left(\mathcal S_{x}^2 \left(\mathcal S_{y}^2+i\right)+i \mathcal S_{y}^2+1\right) s& (1+i) \left(i \left(\mathcal S_{x}^2+i\right) \mathcal S_{y}^2+\mathcal S_{x}^2-i\right) s  \\
			(-1+i) \mathcal S_{z}^2 \left(i \mathcal S_{x}^2 \left(\mathcal S_{y}^2+i\right)+\mathcal S_{y}^2-i\right) s& (-1+i) \left(\mathcal S_{x}^2 \left(1+i \mathcal S_{y}^2\right)+\mathcal S_{y}^2+i\right) s\\
			(1+i) \mathcal S_{z}^2 \left(\mathcal S_{x}^2 \left(1-i \mathcal S_{y}^2\right)+\mathcal S_{y}^2-i\right) c& (1+i) \left(\mathcal S_{x}^2 \left(-1-i \mathcal S_{y}^2\right)+\mathcal S_{y}^2+i\right) c  \\
			(1+i) \mathcal S_{z}^2 \left(\mathcal S_{x}^2 \left(\mathcal S_{y}^2+i\right)-i \mathcal S_{y}^2-1\right) c& (1-i) \left(\mathcal S_{x}^2 \left(1+i \mathcal S_{y}^2\right)+\mathcal S_{y}^2+i\right) c \\
		\end{matrix}
		\right] \, ,
	\end{multline}%
	where $ s \defeq \sin(m \epsilon)$, $c \defeq \cos(m \epsilon)$. Also, in this $(3+1)$D context the Fourier-transform ansatz is defined via:
\par\nobreak
\begin{subequations}
	\begin{align}
          \tilde{\psi}( E,p_{x},p_{y},p_{z} ) &\equiv \epsilon^{2}\sum_{n,k,l,m = - \infty}^{\infty} e^{i(E n -p_{x}k -p_{y} l -p_{z} m) \epsilon} \psi( n,k,l,m)\\
        \psi( n,k,l,m) &\equiv \epsilon^{2} \int_{-\frac{\pi}{\epsilon}}^{\frac{\pi}{\epsilon}} \int_{-\frac{\pi}{\epsilon}}^{\frac{\pi}{\epsilon}} \int_{-\frac{\pi}{\epsilon}}^{\frac{\pi}{\epsilon}} \int_{-\frac{\pi}{\epsilon}}^{\frac{\pi}{\epsilon}} \frac{dE \hspace{0.5mm} dp_{x} \hspace{0.5mm}dp_{y} \hspace{0.5mm} dp_{z}  }{(2 \pi)^{4}} e^{-i(E n -p_{x}k -p_{y} l -p_{z} m) \epsilon}\tilde{\psi}( E,p_{x},p_{y},p_{z} ).
	\end{align}
\end{subequations}%
The integration domain is now four-dimensional and can be denoted for example by some $ \mathcal{B} \cross \mathcal{B} \equiv \mathcal{B}^{2}$. Finally, the translations operators shown in Eqs.\ \eqref{eq:translation operators} can be expressed in Fourier space via the following equations, respectively, with the addition of the time translation,
\begin{subequations}
	\begin{align}
        (\tilde{\mathcal S}_{x}\tilde\psi)(E, p_x,p_y,p_z) &\equiv e^{-i p_{x} \epsilon} \tilde\psi(E, p_x,p_y,p_z) \\
        (\tilde{\mathcal S}_{y}\tilde\psi)(E, p_x,p_y,p_z) &\equiv e^{-i p_{y} \epsilon}\tilde\psi(E, p_x,p_y,p_z) \\ 
        (\tilde{\mathcal S}_{z}\tilde\psi)(E, p_x,p_y,p_z) &\equiv e^{-i p_{z} \epsilon}\tilde\psi(E, p_x,p_y,p_z) \\
        (\tilde{\mathcal T}\tilde\psi)(E, p_x,p_y,p_z) &\equiv e^{i E \epsilon}\tilde\psi(E, p_x,p_y,p_z) \, .
	\end{align}
\end{subequations}

\vspace{2mm}
\noindent
    {\bfseries FD analysis.}
	Now, again, we analyze FD by examining whether the determinant of the Fourier-space version of $  \mathcal{M}_{QW}^{3}$, that we denote by $\mathscr{D}^{3} (E, p_{x},p_{y},p_{z})$, leads to spurious solutions in the continuum limit. A cumbersome but straightforward computation that one can do with a formal-computation software gives
    \par\nobreak
	\begin{multline}
	\mathscr{D}^{3} (E, p_{x},p_{y},p_{z} ) =	\frac{1}{8} e^{-2 i \epsilon (2 E+p_{x}+p_{y}+p_{z})} \bigg[e^{2 i \epsilon E} \cos (2 \epsilon m) \bigg\{8 e^{2 i \epsilon (p_{x}+p_{y}+p_{z})}\\
		+e^{4 i \epsilon (p_{x}+p_{y}+p_{z})}+e^{4 i \epsilon (p_{x}+p_{y})}+e^{4 i \epsilon (p_{x}+p_{z})}+e^{4 i \epsilon p_{x}}+e^{4 i \epsilon (p_{y}+p_{z})}+e^{4 i \epsilon p_{y}}+e^{4 i \epsilon p_{z}}+1\bigg\} \\ 
		+4 \bigg\{-8 \left(1+e^{2 i \epsilon E}\right) \cos (\epsilon m) \cos (\epsilon p_{x}) \cos (\epsilon p_{y}) \cos (\epsilon p_{z}) e^{i \epsilon (E+2 (p_{x}+p_{y}+p_{z}))}\\ 
		+2 e^{2 i \epsilon (E+p_{x}+p_{y}+p_{z})}+2 e^{2 i \epsilon (2 E+p_{x}+p_{y}+p_{z})}+e^{2 i \epsilon (E+2 p_{x}+p_{y}+p_{z})}+e^{2 i \epsilon (E+p_{x}+2 p_{y}+p_{z})}\\ 
		+e^{2 i \epsilon (E+p_{x}+p_{y}+2 p_{z})}+e^{2 i \epsilon (E+p_{x}+p_{y})}+e^{2 i \epsilon (E+p_{x}+p_{z})}+e^{2 i \epsilon (E+p_{y}+p_{z})}+2 e^{2 i \epsilon (p_{x}+p_{y}+p_{z})}\bigg\}\bigg] \, .\label{eq:det3D}
	\end{multline} %
	Taylor-expanding the previous expression in $\epsilon (E, p_x , p_y , p_z)$ delivers
    \par\nobreak
	\begin{equation}
		\mathscr{D}^{3} (E, p_{x},p_{y},p_{z} )  = \epsilon^{4} (E^{2} - p_{x}^{2} -p_{y}^{2} - p_{z}^{2} - m^{2} )^2+  O(\epsilon^5) \, ,
	\end{equation}%
	which coincides with the continuum theory. These modes, whose energy and momentum values $E,p_x,p_y,p_z$, remain finite in the continuum limit, belong to the neighbourhood $  U_{\bold{0}} \defeq \{ q \in \mathcal{B}^{2}\ :\ d(q,(0 , 0, 0, 0)) \ll \epsilon^{-1} \} $. However, as far as the formula for  $ \mathscr{D}^{3} (E, p_{x},p_{y},p_{z} ) $ is concerned, the modes that are in the following neighbourhoods behave just the same:  
    \begin{equation}
            U_f \defeq  \{ (f_{t}\pi \epsilon^{-1}+ E , f_{x}\pi \epsilon^{-1}+p_{x}, f_{y}\pi \epsilon^{-1}+p_{y} ,f_{z}\pi \epsilon^{-1}+p_{z}) \in\mathcal{B}^{2}  \\:\ (E, p_{x},p_{y} ,p_{z}) \in U_{\bold{0}} \} \, ,
    \end{equation}
     with 
     \par\nobreak
     {\small
\begin{equation}
   f \defeq (f_t,f_x,f_y,f_z) \in \{ ( \pm  ,\pm ,0, 0), (\pm , 0,  \pm , 0) , (\pm,0,0, \pm), (0,\pm, \pm,0), \\ (0,\pm,0,\pm), (0,0,\pm,\pm), (\pm,\pm,\pm,\pm) \}.
\end{equation}}%
    Even though the number of possible $f$s is\footnote{This number can be computed via: $2^2*\binom{4}{2}+2^4 = 40$.} $40 $, they really correspond to $7$ distinct neighbourhoods due to the boundary conditions of $ \mathcal{B}^{2}$ (hyper-toric geometry), as shown in Fig.\ \ref{fig:degeneracies3D} below. These $7$ extra neighbourhoods will again pollute the numerics.

	\begin{center} 
		\begin{tikzpicture}[scale=1]

        \begin{scope}[xshift=-20mm]
			\coordinate ($p_x$) at (2,0,0);
			\coordinate ($p_y$) at (0,2,0);
			\coordinate ($p_z$) at (0,0,2);
			\draw[thick,->] (0,0,0) -- (2.5,0,0);
			\draw[thick,->] (0,0,0) -- (-2.5,0,0);
			\draw[thick,->] (0,0,0) -- (0,3,0);
			\draw[thick,->] (0,0,0) -- (0,-3,0);
			\draw[thick,->] (0,0,0) -- (0,0,3);
			\draw[thick,->] (0,0,0) -- (0,0,-3);
			\draw node at (2.7,0,0) {$p_x$};
			\draw node at (0,3.2,0) {$p_y$};
			\draw node at (0,0,3.3) {$p_z$};
			
			\draw node at (-3,3,0) {$E=0$};

			\draw[dashed , red] (2,0,2)--(2,0,-2);
           \draw[ red] (2,0,-2)--(-2,0,-2);
          \draw[dashed , red] (-2,0,-2)--(-2,0,2);
         \draw[ red] (-2,0,2) --(2,0,2);

			\filldraw[brown] (-2,0,-2) circle (2pt) ;

			\filldraw[brown] (2,0,2) circle (2pt) ;
			
			\filldraw[brown] (-2,0,2) circle (2pt) ;

			\filldraw[brown] (2,0,-2) circle (2pt) ;

			\draw node at (2.2,0.3,0) { $ \frac{\pi}{\epsilon}$};
			\draw[ultra thick] (2,-0.1,0)--(2, 0.1 ,0);
			\draw[ultra thick] (2,0,-0.1)--(2, 0 ,0.1);
			\draw node at (0.2,2.3,0) { $ \frac{\pi}{\epsilon}$};
			\draw[ultra thick] (-0.1,2,0)--(0.1,2,0);
			\draw[ultra thick] (0,2,-0.1)--(0,2,0.1);
			\draw node at (0.1,0.4,2) { $ \frac{\pi}{\epsilon}$};
			\draw[ultra thick] (0,-0.1,2)--(0, 0.1 ,2);
			\draw[ultra thick] (-0.1,0,2)--(0.1, 0 ,2);

			\draw node at (-2.2,0.3,0) { $ -\frac{\pi}{\epsilon}$};
			\draw[ultra thick] (-2,-0.1,0)--(-2, 0.1 ,0);
			\draw[ultra thick] (-2,0,-0.1)--(-2, 0 ,0.1);
			\draw node at (0.2,-2.3,0) { $ -\frac{\pi}{\epsilon}$};
			\draw[ultra thick] (-0.1,-2,0)--(0.1,-2,0);
			\draw[ultra thick] (0,-2,-0.1)--(0,-2,0.1);
			\draw node at (0.3,0.7,-2) { $ -\frac{\pi}{\epsilon}$};
			\draw[ultra thick] (0,-0.1,-2)--(0, 0.1 ,-2);
			\draw[ultra thick] (-0.1,0,-2)--(0.1, 0 ,-2);

			\draw[dashed, blue] (2,2,0)--(2,-2,0);
           \draw[blue](2,-2,0) --(-2,-2,0);
        \draw[ dashed,blue]   (-2,-2,0) --(-2,2,0);
        \draw[ blue]  (-2,2,0)  --(2,2,0);

			\filldraw[cyan] (-2,-2,0) circle (2pt) ;

			\filldraw[cyan] (2,2,0) circle (2pt) ;
			
			\filldraw[cyan] (-2,2,0) circle (2pt) ;

			\filldraw[cyan] (2,-2,0) circle (2pt) ;

			\draw[dashed, green] (0,2,2)--(0,-2,2);
          \draw[ green]  (0,-2,2) --(0,-2,-2);
         \draw[ dashed, green]  (0,-2,-2) --(0,2,-2);
          \draw[ green] (0,2,-2)--(0,2,2);
			
			\filldraw[yellow] (0,-2,-2) circle (2pt) ;

			\filldraw[yellow] (0,2,2) circle (2pt) ;

			\filldraw[yellow] (0,2,-2) circle (2pt) ;
			
			\filldraw[yellow] (0,-2,2) circle (2pt) ;
			
			\draw[red,dashed, ultra thick] (0,0,0) circle [radius = 4mm];

			\end{scope}

			\draw node at (3,3,0) {$E= \pm \frac{\pi}{\epsilon}$};
			\draw[thick,->] (6,0,0) -- (8.5,0,0);
			\draw[thick,->] (6,0,0) -- (3.5,0,0);
			\draw[thick,->] (6,0,0) -- (6,2.5,0);
			\draw[thick,->] (6,0,0) -- (6,-2.5,0);
			\draw[thick,->] (6,0,0) -- (6,0,3);
			\draw[thick,->] (6,0,0) -- (6,0,-3);
			
			\draw node at (9.2,0,0) {$p_x$};
			\draw node at (6,3.2,0) {$p_y$};
			\draw node at (6,0.5,4) {$p_z$};

			\draw node at (8,0.5,0) { $ \frac{\pi}{\epsilon}$};
			\draw node at (4,0.5,0) { $ -\frac{\pi}{\epsilon}$};
			\draw node at (6.5,2,0) { $ \frac{\pi}{\epsilon}$};
			\draw node at (6.5,-2,0) { $ -\frac{\pi}{\epsilon}$};
			\draw node at (6,0.3,2) { $ \frac{\pi}{\epsilon}$};
			\draw node at (6.01,-0.25,-2) { $ -\frac{\pi}{\epsilon}$};

			\filldraw[orange] (8,0,0) circle (2pt) ;
			\filldraw[orange] (4,0,0) circle (2pt) ;
			\filldraw[green] (6,2,0) circle (2pt) ;
			\filldraw[green] (6,-2,0) circle (2pt) ;
			\filldraw[blue] (6,0,2) circle (2pt) ;
			\filldraw[blue] (6,0,-2) circle (2pt) ;

			\filldraw[pink] (8,2,2) circle (2pt) ;
			\filldraw[pink] (4,2,2) circle (2pt) ;
			\filldraw[pink] (8,2,-2) circle (2pt) ;
			\filldraw[pink] (4,2,-2) circle (2pt) ;
			\draw[dashed, cyan] (8,2,2)--(8,2,-2)--(4,2,-2)--(4,2,2)--(8,2,2);

			\filldraw[pink] (8,-2,2) circle (2pt) ;
			\filldraw[pink] (4,-2,2) circle (2pt) ;
			\filldraw[pink] (8,-2,-2) circle (2pt) ;
			\filldraw[pink] (4,-2,-2) circle (2pt) ;
			\draw[dashed, cyan] (8,-2,2)--(8,-2,-2)--(4,-2,-2)--(4,-2,2)--(8,-2,2);

			\draw[ dashed, cyan ] (8,2,2)--(8,-2,2)--(4,-2,2)--(4,2,2)--(8,2,2);
			\draw[  dashed, cyan] (4,2,-2)--(4,-2,-2);
			\draw[ dashed, cyan] (8,2,-2)--(8,-2,-2);
			
		\end{tikzpicture}
		\captionof{figure}{ \emph{Fermion doubling for the $(3+1)$D Dirac discrete-time quantum walk.} We draw the neighbourhoods on the hyper-surfaces where $E=0$ (left) and $ E= 
	\pm \pi/\epsilon$ (right).The neighborhood having the same colour are identical due to the hyper-toric geometry.}
    \label{fig:degeneracies3D}
	\end{center}
	\section{Solution  of fermion doubling in QCAs} \label{sec:solution}

	Having demonstrated the existence of FD problems in  $(1+1)$D and $(3+1)$D QED QCAs, we now propose {\em flavor-staggering-only} resolutions of the problems, along the lines followed in Sec.\ \ref{sec:standardFD}. Recall that the method is to change the BZ in a way that avoids the FD neighbourhoods. The avoidance of the FD neighborhoods in the BZ produces a new, desired BZ, associated to new reciprocal-space lattice vectors and the corresponding reciprocal-space lattice, which has of course a counterpart in direct space associated to new direct-space lattice vectors. This new direct-space lattice is a sublattice of the original lattice $\Lambda^4$. Each one-site translate of that reference sublattice needs a flavor.
	
	\subsection{Solution of FD for the $(1+1)$D QED QCA}\label{subsec:1qedqcafdfix}

\subsubsection{FD problem and solution in Fourier space}

            \noindent
    {{\bfseries A general procedure to solve the problem.}}
	Recall that FD occurs because of the periodicity of the functions involved in the determinant of the Fourier-space equation-of-motion operator, determinant denoted in the $(1+1)$D QED QCA model by $\mathscr{D}^{1}(E,p)$, which is also called \emph{dispersion-relation expression}. For this model, the exact symmetries are
	\begin{equation}\label{Eq: symmetries of d}
			\mathscr{D}^{1}(E,p) =  -\mathscr{D}^{1}(- \sgn(E)\frac{\pi}{\epsilon}+E, - \sgn(p)\frac{\pi}{\epsilon}+p) \, ,
	\end{equation}
	where $ (p, E) \in {\cal B}\defeq[-\pi/\epsilon , \pi/ \epsilon] \cross[-\pi/\epsilon , \pi/ \epsilon] $. 
    We are going to avoid these symmetries simply by reducing the size of the domain. To achieve this, we do what is illustrated in in Fig.\ \ref{fig:shrinking1d}, that is: we (left figure) 1/ cut the BZ ${\cal B}$ into regions that relate to one another by a translation in the BZ; we (left figure again) 2/ color them the same if $\mathscr{D}^{1}$ acts the same upon them; and finally, we (right figure) 3/ make sure that ``pasting'' each border region having the same color as a certain central region on top of the latter, indeed makes the FD-neighbourhood associated to that border region behave exactly the same as the associated central neighborhood. \\
    
	\begin{center}
		\begin{tikzpicture}
			\draw[thick,->] (0,2) -- (2.2,2);
			\draw[thick,->] (0,2) -- (0,-0.5);
			\draw[thick,->] (0,2) -- (-2.2,2);
			\draw[thick,->] (0,2) -- (0,4.5);
			
			\draw[dashed] (2,4) --(2,0);
			\draw[dashed] (2,4) --(0,4);
			\draw[dashed] (-2,4) --(-2,0);
			\draw[dashed] (-2,0) --(2,0);

			\draw node at (2.4,2) {$ p$};
			\draw node at (0,4.7) {$ E$};
			\draw node at (2.1,2.4) { $ \frac{\pi}{\epsilon}$};
			\draw node at (0.4,4.24) { $ \frac{\pi}{\epsilon}$};
			\draw node at (-2.24,2.4) { $ -\frac{\pi}{\epsilon}$};
			\draw node at (0.4,-0.24) { $- \frac{\pi}{\epsilon}$};
			
			\filldraw[cyan, opacity=0.5] (2,2)--(2,0)--(0,0)--(2,2);
			\filldraw[lime, opacity=0.5] (0,4)--(2,2)--(2,4)--(0,4);
			\filldraw[orange, opacity=0.5] (-2,2)--(-2,0)--(0,0)--(-2,2);
			\filldraw[purple, opacity=0.5] (0,4)--(-2,2)--(-2,4)--(0,4);

			\filldraw[lime, opacity=0.5] (-2,2)--(0,0)--(0,2)--(-2,2);
			\filldraw[purple, opacity=0.5] (0,0)--(2,2)--(0,2)--(0,0);but if it is 1 instead then sheet is in the lifted surface.
			\filldraw[orange, opacity=0.5] (0,4)--(0,2)--(2,2)--(0,4);
			\filldraw[cyan, opacity=0.5] (0,4)--(0,2)--(-2,2)--(0,4);
			\filldraw[blue](2,4) circle[radius=3mm] ;
			\filldraw[blue](2,0) circle[radius=3mm] ;
			\filldraw[blue](-2,0) circle[radius=3mm] ;
			\filldraw[blue](-2,4) circle[radius=3mm] ;
			\draw[red,dashed, ultra thick] (0,2) circle[radius = 4mm];

			\draw[thick,->] (5,2) -- (7.2,2);
			\draw[thick,->] (5,2) -- (5,-0.5);
			\draw[thick,->] (5,2) -- (2.8,2);
			\draw[thick,->] (5,2) -- (5,4.5);
			
			\draw[dashed] (7,4) -- (7,0);
			\draw[dashed] (3,4) --(7,4);
			\draw[dashed] (3,4) --(3,0);
			\draw[dashed] (3,0) --(7,0);

			\draw node at (7.4,2) {$ p$};
			\draw node at (5,4.7) {$ E$};
			\draw node at (7.1,2.4) { $ \frac{\pi}{\epsilon}$};
			\draw node at (5.4,4.24) { $ \frac{\pi}{\epsilon}$};
			\draw node at (2.76,2.4) { $ -\frac{\pi}{\epsilon}$};
			\draw node at (5.4,-0.24) { $- \frac{\pi}{\epsilon}$};

			\filldraw[lime] (3,2)--(5,0)--(5,2)--(3,2);
			\filldraw[purple] (5,0)--(7,2)--(5,2)--(5,0);
			\filldraw[orange] (5,4)--(5,2)--(7,2)--(5,4);
			\filldraw[cyan] (5,4)--(5,2)--(3,2)--(5,4);
			
			\filldraw[blue](5,2) circle[radius=3mm] ;
			\draw[red,dashed, ultra thick] (5,2) circle[radius = 4mm];
			
		\end{tikzpicture} 
		\captionof{figure}{
        \emph{Schematic representation of the dispersion-relation expressions of the $(1+1)$D Dirac QW (left) and its flavored version (right).} 
        On the left, due to the periodicity of $\mathscr{D}^{1}(E,p)$, the FD neighborhoods will pollute the continuum limit. On the right is the two-sheeted (BZ vision), or flavored (direct-space vision) solution to this problem.  }
         \label{fig:shrinking1d}
	\end{center}

            \noindent
    {{\bfseries The solution we choose: the rhombus BZ.}}    
	More precisely, Fig.\ \ref{fig:shrinking1d} is one possible solution implementing the three-step procedure described in the previous paragraph, having $8$ regions and $4$ colors. 
    As we cut the $4$ border triangles, and paste them on top of the central rhombus according to matching colors, we obtain the diagram on the right, whose denser opacity indicates the presence of the two sheets.
This procedure can be generalized and  formalized in terms of a Riemannian-surface structure: we postpone this till Sec.\ \ref{sec:formaltreatment}.
    Still informally for now, we observe that the procedure solves FD because there are no longer spurious solutions, but this is at the cost of having two sheets, each contributing to the continuum limit. This will correspond to having  \textcolor{red}{red}-flavored and \textcolor{blue}{blue}-flavored  fermions.

 \subsubsection{Connecting the ``new Fourier space'' to a ``new direct space''}

             \noindent
    {{\bfseries The new, rhombus (or diamond) lattice, or rotated square lattice.}}
Let us explain this formally. If we want to stay with a simple, i.e., Cartesian BZ, restricting the original BZ down to the rhombus must imply changing its reciprocal vectors as something like
\begin{equation}
\label{rel:1}
\begin{split}
     \vec b_{0}= \frac{\pi}{\epsilon} \vec{E} \ \  &\longrightarrow \ \  \vec b'_{0}\defeq \frac{\pi}{2\epsilon} (\vec{E} - \vec{p})\\
       \vec b_{1}= \frac{\pi}{\epsilon} \vec{p} \, \ \ &\longrightarrow  \ \  \vec b'_{1} \defeq \frac{\pi}{2\epsilon} (\vec{E} +\vec{p}) \, .
\end{split}  
\end{equation}
In terms of the direct-space lattice vectors, this means performing the change
\begin{equation}
\begin{split}
     \vec{a}_0=\epsilon \vec{t}  \ \  &\longrightarrow \ \ \vec{a}'_0=\epsilon (\vec{t} -\vec{x}) \\
      \vec{a}_1=\epsilon \vec{x} \, \ &\longrightarrow  \ \  \vec{a}'_1=\epsilon( \vec{t} +\vec{x} ) \, ,
\end{split}  
\end{equation}
as explained in App.\ \ref{sec:applattice}.
But these two new lattice vectors only generate half the points of $ \Lambda^{2}$, namely the $45\deg$-rotated orthogonal lattice  $ \Gamma_{\textcolor{red}{r}}^{2}$.  We call  $ \Gamma_{\textcolor{blue}{b}}^{2}$ the lattice given by $ \Gamma_{\textcolor{red}{r}}^{2}$. Notice that for $ \vec v_r \in  \Gamma_{\textcolor{red}{r}}^{2} $ and $ \vec v_b \in  \Gamma_{\textcolor{blue}{b}}^{2} $,   we have that $ \vec v_r  = \vec v_b + n_1 \vec{x} + n_2 \vec{t}$ when $ n_1+ n_2 $ is an odd integer.
Both $\Gamma_{\textcolor{red}{r}}^{2}$ and $\Gamma_{\textcolor{blue}{b}}^{2}$ admit the same rhombus-shaped BZ $\tilde{\mathcal{B}}$, with basis vectors
$\vec b'_{0}$ and $\vec b'_{1}$ given in Transformations \eqref{rel:1}. Thus, each of these two sublattices generate one of the two sheets of Fig.\ \ref{fig:shrinking1d} right. \\
  
	\begin{figure}
\begin{center}
\hspace{-0.8cm}
		\begin{tikzpicture}
			\draw node at (0.4,-0.4) {$ \vec{a}_1$};
			\draw node at (-0.4,0.4) {$ \vec{a}_0 $};
			\draw[thick,->] (0,0) -- (0.7, 0);
			\draw[thick,->] (0,0) -- (0, 0.7);
			
			\draw[thick,->] (4,1) -- (3.3, 1.7);
			\draw[thick,->] (4,1) -- (4.7, 1.7);
			\draw node at (3.5,1.3) {$ {\color{red} \vec{a}'_0}$};
			\draw node at (4.6,1.3) {$ {\color{red} \vec{a}'_1}$};

			\draw[thick,->] (8,0) -- (7.3, 0.7);
			\draw[thick,->] (8,0) -- (8.7, 0.7);
			\draw node at (7.5,0.3) {${\color{blue} \vec{a}'_0}$};
			\draw node at (8.6,0.3) {$ {\color{blue} \vec{a}'_1}$};
   
			\filldraw[blue](0,0) circle[radius = 1.5mm];
			\filldraw[red](1,0) circle[radius = 1.5mm];
			\filldraw[blue](2,0) circle[radius = 1.5mm];
			\filldraw[red](3,0) circle[radius = 1.5mm];
			\filldraw[blue](4,0) circle[radius = 1.5mm];
			\filldraw[red](5,0) circle[radius = 1.5mm];
			\filldraw[blue](6,0) circle[radius = 1.5mm];
			\filldraw[red](7,0) circle[radius = 1.5mm];
			\filldraw[blue](8,0) circle[radius = 1.5mm];
			\filldraw[red](9,0) circle[radius = 1.5mm];
			\filldraw[red](0,1) circle[radius = 1.5mm];
			\filldraw[blue](1,1) circle[radius = 1.5mm];
			\filldraw[red](2,1) circle[radius = 1.5mm];
			\filldraw[blue](3,1) circle[radius = 1.5mm];
			\filldraw[red](4,1) circle[radius = 1.5mm];
			\filldraw[blue](5,1) circle[radius = 1.5mm];
			\filldraw[red](6,1) circle[radius = 1.5mm];
			\filldraw[blue](7,1) circle[radius = 1.5mm];
			\filldraw[red](8,1) circle[radius = 1.5mm];
			\filldraw[blue](9,1) circle[radius = 1.5mm];
			\filldraw[blue](0,2) circle[radius = 1.5mm];
			\filldraw[red](1,2) circle[radius = 1.5mm];
			\filldraw[blue](2,2) circle[radius = 1.5mm];
			\filldraw[red](3,2) circle[radius = 1.5mm];
			\filldraw[blue](4,2) circle[radius = 1.5mm];
			\filldraw[red](5,2) circle[radius = 1.5mm];
			\filldraw[blue](6,2) circle[radius = 1.5mm];
			\filldraw[red](7,2) circle[radius = 1.5mm];
			\filldraw[blue](8,2) circle[radius = 1.5mm];
			\filldraw[red](9,2) circle[radius = 1.5mm];
		\end{tikzpicture}
		\caption{\em Sublattices of the $(1+1)$D flavored Dirac QCA.  \label{fig:f-lattice1d} }
\end{center}
	\end{figure}

\noindent
{{\bfseries The flavor Hilbert space.}}
Let us now distinguish these sublattices at the level of the state space, whilst keeping the $(1+1)$D Dirac-QCA dynamics as close as possible to the original one. 
To do this we just add two qubits at each site $(n,k)$ (we tensor them with the other two qubits at this location coding for presence or absence of right or left movers), and set them (in the single-particle perspective to come regarding how we used the names ``blue'' and ``red'' here) to   \textcolor{red}{red} ``$=$'' $\ket{10}^{n,k}$ (resp.\ \textcolor{blue}{blue}  ``$=$'' $\ket{01}^{n,k}$)  for even (resp.\ odd) $ n+k $. In the single-particle sector of our QCA, i.e., in our QW, this means we have a new internal degree of freedom with basis states $\ket{\text{red}} \equiv \ket r \equiv  \ket1$ and $\ket{\text{blue}}\equiv\ket b \equiv \ket0$, which is tensored with the position degree of freedom and with the other internal degree of freedom, namely, chirality.
We call this new degree of freedom ``flavor'', because in particle physics the term ``flavor'' is widely used to name a degree of freedom associated to an \emph{internal global symmetry} that is not gauged.
Flavor degrees of freedom are internal degrees of freedom which traditionally do not intervene in a free dynamics, i.e., they are not coupled to the external (i.e., spacetime) degrees of freedom when the dynamics is free---unlike other internal degrees of freedom like chirality or spin. 
The global symmetry corresponding to our flavor degree of freedom will here be, as we will see in more detail in Sec.\ \ref{sec:formaltreatment}, the standard $2\times 2$ matrix representation of the group $ \mathbb{Z}_{2}\defeq\{-1,1\}$, namely, $ \rho(\mathbb{Z}_{2}) \defeq\{\mathbb I_2,\sigma_x\}$, where $\sigma_x$ is the element of the group that exchanges the two flavors, which is the only non-trivial operation in this group. 

 \subsubsection{Solution in direct space: the $(1+1)$D flavored-QW dynamics}

\noindent
{{\bfseries Flavored state.}}
In terms of the single-particle sector, i.e., of the $(1+1)$D Dirac QW, here is how we have to modify the dynamics to include this flavor degree of freedom. We start with a state of our new, flavored QW (FQW), at time $n=0$, having the following very particular form,
\begin{equation}
\label{eq:FQWstate1}
    \ket{\psi^f(n)} \defeq \ket{\chi_{i=0}(n)} \defeq \sum_{k} \left(\chi^{+}(n,k) \ket{k}_{+}  + \chi^{-}(n,k) \ket{k}_{-}\right) \otimes \sigma_{x}^{n+k} \ket{i=0} \, .
\end{equation}
One can introduce $0$-flavored and $1$-flavored wavefunctions, $ \psi_{0}^{\pm}$ and $ \psi_{1}^{\pm}$, respectively, in the following way,
\begin{subequations}
\label{eqs:flavor-content-wavefunctions1}
\begin{align}
      \chi_{\text{F};i}^{\pm}(n,k) &\equiv  \frac{1}{2}\left[\psi^{\pm}_{i}(n,k) \left((-1)^{n+k}+1\right)+\psi^{\pm}_{(i+1) \, \text{mod} \, 2}(n,k) \left((-1)^{n+k+1}+1\right)\right]\\
    &=  \left((-1)^{n+k}+1, (-1)^{n+k+1}+1\right) \, . \begin{pmatrix}
        \psi^{\pm}_{i}(n,k)\\
        \psi^{\pm}_{(i+1) \, \text{mod} \, 2}(n,k)
    \end{pmatrix} \, .
\end{align}
\end{subequations}

\vspace{2mm}
\noindent
{{\bfseries Temporal ket basis, and abstract temporal  translation operator.}} We introduce a Hilbert space $\mathcal H_{\text{time}}$ with basis $(\ket{n})_{n\in\mathbb Z}$, as well as a quantum state $\ket{\psi}$ belonging to the Hilbert space $\mathcal H_{\text{time}} \otimes \mathcal H_{\text{position}} \otimes \mathcal H_{\text{chirality}}$, such that $\ket{\psi(n)} \equiv \langle n | \psi \rangle$. (For the flavored state $\ket{\psi^f}$, just include an extra tensorization by the flavor Hilbert space in the definition of the total Hilbert space.) This enables us to define the abstract version of $\mathcal T$, acting on $\mathcal H_{\text{time}}$, that we denote by $\hat T$:
\begin{equation}
\label{eq:time_trans}
 \hat T \defeq \sum_{n} \ket{n+1} \bra{n} \, .   
\end{equation}

\vspace{2mm}
\noindent
{{\bfseries Flavored dynamics.}}
We now turn to the one-step evolution operator that evolves $ \ket{\psi^f(n)}$. This operator is built as the following modification  of our $(1+1)$D Dirac QW of Sec.\ \ref{subsec:1qedqcafd}:
	\begin{subequations}
    \label{eqs:theeqs}
		\begin{align}	\hat{S}^{f}&\defeq\hat{S}\otimes\sigma_{x}\\
        \hat{T}^{f}&\defeq\hat{T}\otimes\sigma_{x}\\
        \hat{U}^{1}_{FQW}&\defeq 
        \begin{bmatrix}
				\cos(m \epsilon) \hat{S}^f &  - i \sin(m \epsilon)\\
				- i \sin(m \epsilon) & \cos(m \epsilon) \hat{S}^{f\dagger}
			\end{bmatrix} \label{eq:UFQW}\\
        \hat{M}_{FQW}^{1}  &\defeq \hat{U}^{1}_{FQW} - \hat T^{f \dag} \\
		&= \begin{bmatrix}
			[ \cos(m \epsilon)  \hat{S} -\hat{T}^\dag ]\otimes \sigma_{x}& - i \sin(m \epsilon)\\
			- i \sin(m \epsilon)  &[\cos(m \epsilon)  \hat{S}^{\dagger}-\hat{T}^\dag ] \otimes \sigma_{x}
		\end{bmatrix} \, ,
        \end{align}
	\end{subequations}
where $\hat S$ is the usual abstract spatial translation operator, which we have defined in Eq.\ \eqref{eq:space_trans}.
  
The reader can check that this FQW consistently preserves the form chosen here for the initial state, Eq.\ \eqref{eq:FQWstate1}, that is, we can show by induction that for any $n$, we have
 	\begin{equation}
    \label{eq:outstate}
		\ket{\psi^f(n)} = \ket{\chi_{i=0}(n)} = \sum_{k} \left(\chi^{+}(n,k) \ket{k}_{+}  + \chi^{-}(n,k) \ket{k}_{-}\right) \otimes \sigma_{x}^{n+k} \ket{i=0} \, .
	\end{equation} 
    
In App.\ \ref{app:cont_lim_DQW}, we show that our FQW scheme indeed delivers, in the continuum limit, a flavored Dirac equation, that is, the same Dirac equation for each wavefunction $\psi_{\text{cont.}, i}$, $i=0,1$, which is the continuum limit of $\psi_{i}$.

 \subsubsection{Final check in Fourier space}

In order to check that FD is indeed resolved, let us look again at the relevant determinant, which is this time that of the Fourier-space version of $\hat{M}_{FQW}^{1}$. A simple computation shows that this determinant takes the following form:
	\begin{equation}
		\mathscr{D}^{1}_{f}(E,p)=   4 e^{-2 i \epsilon E} [\cos (E \epsilon )-\cos (m \epsilon )  \cos (p \epsilon)]^2 \, .
	\end{equation}
	Notice that $\mathscr{D}^{1}_{f}(E,p)  = [(\mathscr{D}^{1}(E,p)]^{2}  $. It thus looks like the zeros of $\mathscr{D}^{1}_{f}(E,p)$ are the same as those of $\mathscr{D}^{1}(E,p)$, but that is not the case, because its domain of definition is smaller: the rhombus-shaped BZ of the FQW excludes the spurious solutions.
    Still, the square captures the fact that the continuum limit is, in full generality, recovered not once, but twice. Indeed: we have traded our single fermion, that had spurious solutions, for two flavors of fermions with no spurious solutions. Finally, let us recall that the flavoring procedure can of course be applied on the multiparticle scheme, i.e., on the QCA, so as to obtain a flavored QCA.

\subsection{Solution of FD for the  $(3+1)$D QED QCA}\label{subsec:3qedqcafdfix}

\subsubsection{The FD problem in Fourier space}
 
Again we go back to the determinant of the Fourier-space version
of $\mathcal{M}_{QW}^3$, as given in Eq.\ \eqref{eq:det3D}, and identify the symmetries that cause the spurious solutions: 
	\begin{subequations}
		\begin{align}
			\mathscr{D}^{3} ( E,  p_{x},p_{y},p_{z} ) &=  \mathscr{D}^{3} ( - \sgn(E) \frac{\pi}{\epsilon} +E,  -\sgn(p_{x})\frac{\pi}{\epsilon} + p_{x},p_{y},p_{z} )\\
			\mathscr{D}^{3} ( E, p_{x}, p_{y},p_{z} ) &=  \mathscr{D}^{3} ( - \sgn(E)\frac{\pi}{\epsilon} + E,   p_{x}, - \sgn(p_{y})\frac{\pi}{\epsilon}+ p_{y},p_{z} )\\
			\mathscr{D}^{3} ( E,p_{x},p_{y}, p_{z} ) &=  \mathscr{D}^{3} ( - \sgn(E)\frac{\pi}{\epsilon} + E,   p_{x},p_{y},-\sgn(p_{z})\frac{\pi}{\epsilon}+  p_{z} ) \, .
		\end{align}
	\end{subequations}
Note that these equations are analogs of Eq.\ \eqref{Eq: symmetries of d} for each spatial dimension. This is why the symmetries of $\mathscr{D}^{3} (E, p_{x},p_{y},p_{z} ) $ can be visualized with three copies of Fig.\ \ref{fig:shrinking1d}, as represented in Fig.~\ref{fig:shrinking3d} {(left)}. In this figure we condense the three graphs into one by laying them on top of one another. Note that through the combinations of the above symmetries, which we call \emph{basis symmetries}, there actually exist $2^3 = 8$ possible classes of symmetries in $ 8$ distinct hyperplanes, which are two-dimensional planes, respectively. We have opted to take only three of them as a basis of all the 8 symmetries for the sake of simplicity.
 Notice that although the symmetry structure in each hyperplane is the same, their FD neighborhoods remain distinct and hence need bear different colors, as we had done for Fig.\ \ref{fig:degeneracies3D}. We lined them up according to the order of the hyperplanes, i.e., \textcolor{orange}{orange} comes first as it is associated to the hyperplane $E$\hspace{0.05cm}---\hspace{0.05cm}$p_x$ (look at the left graph of Fig.~\ref{fig:degeneracies3D}), then \textcolor{green}{green} as it is associated to $E$\hspace{0.05cm}---\hspace{0.05cm}$p_y$, and finally \textcolor{blue}{blue} as it is associated to $E$\hspace{0.05cm}---\hspace{0.05cm}$p_z$.

	\begin{center}
		\begin{tikzpicture}
			\draw[thick,->] (0,2) -- (2.2,2);
			\draw[thick,->] (0,2) -- (0,-0.5);
			\draw[thick,->] (0,2) -- (-2.2,2);
			\draw[thick,->] (0,2) -- (0,4.5);
			
			\draw[dashed] (2,4) --(2,0);
			\draw[dashed] (2,4) --(0,4);
			\draw[dashed] (-2,4) --(-2,0);
			\draw[dashed] (-2,0) --(2,0);

			\draw node at (3,2) {$ p_{x} ,p_{y},p_{z}$};
			\draw node at (0,4.7) {$ E$};
			\draw node at (2.1,2.4) { $ \frac{\pi}{\epsilon}$};
			\draw node at (0.4,4.24) { $ \frac{\pi}{\epsilon}$};
			\draw node at (-2.24,2.4) { $ -\frac{\pi}{\epsilon}$};
			\draw node at (0.4,-0.24) { $- \frac{\pi}{\epsilon}$};
			
			\filldraw[green, opacity=0.5] (2,2)--(2,0)--(0,0)--(2,2);
			\filldraw[yellow, opacity=0.5] (0,4)--(2,2)--(2,4)--(0,4);
			\filldraw[teal, opacity=0.5] (-2,2)--(-2,0)--(0,0)--(-2,2);
			\filldraw[violet, opacity=0.5] (0,4)--(-2,2)--(-2,4)--(0,4);

			\filldraw[lime, opacity=0.5] (-2,2)--(0,0)--(0,2)--(-2,2);
			\filldraw[violet, opacity=0.5] (0,0)--(2,2)--(0,2)--(0,0);
			\filldraw[teal, opacity=0.5] (0,4)--(0,2)--(2,2)--(0,4);
			\filldraw[green, opacity=0.5] (0,4)--(0,2)--(-2,2)--(0,4);
			\filldraw[orange](1.5,4) circle[radius=1.5mm] ;
			\filldraw[green](1.7,4) circle[radius=1.5mm] ;
			\filldraw[blue](1.9,4) circle[radius=1.5mm] ;
			\filldraw[orange](1.5,0) circle[radius=1.5mm] ;
			\filldraw[green](1.7,0) circle[radius=1.5mm] ;
			\filldraw[blue](1.9,0) circle[radius=1.5mm] ;
			\filldraw[orange](-2.1,0) circle[radius=1.5mm] ;
			\filldraw[green](-1.9,0) circle[radius=1.5mm] ;
			\filldraw[blue](-1.7,0) circle[radius=1.5mm] ;
			\filldraw[orange](-2.1,4) circle[radius=1.5mm] ;
			\filldraw[green](-1.9,4) circle[radius=1.5mm] ;
			\filldraw[blue](-1.7,4) circle[radius=1.5mm] ;
			\draw[red,dashed, ultra thick] (0,2) circle[radius = 4mm];

			\begin{scope}[shift={(2,0)}]
				\draw[thick,->] (5,2) -- (7.2,2);
				\draw[thick,->] (5,2) -- (5,-0.5);
				\draw[thick,->] (5,2) -- (2.8,2);
				\draw[thick,->] (5,2) -- (5,4.5);
				
				\draw[dashed] (7,4) -- (7,0);
				\draw[dashed] (3,4) --(7,4);
				\draw[dashed] (3,4) --(3,0);
				\draw[dashed] (3,0) --(7,0);

				\draw node at (7.9,2) {$ p_{x} ,p_{y},p_{z} $};
				\draw node at (5,4.7) {$ E$};
				\draw node at (7.1,2.4) { $ \frac{\pi}{\epsilon}$};
				\draw node at (5.4,4.24) { $ \frac{\pi}{\epsilon}$};
				\draw node at (2.76,2.4) { $ -\frac{\pi}{\epsilon}$};
				\draw node at (5.4,-0.24) { $- \frac{\pi}{\epsilon}$};

				\filldraw[yellow] (3,2)--(5,0)--(5,2)--(3,2);
				\filldraw[violet] (5,0)--(7,2)--(5,2)--(5,0);
				\filldraw[teal] (5,4)--(5,2)--(7,2)--(5,4);
				\filldraw[green] (5,4)--(5,2)--(3,2)--(5,4);
				
				\filldraw[orange](4.8,2) circle[radius=1.5mm] ;
				\filldraw[green](5,2) circle[radius=1.5mm] ;
				\filldraw[blue](5.2,2) circle[radius=1.5mm] ;
				\draw[red,dashed, ultra thick] (5,2) circle[radius = 4mm];

			\end{scope}

		\end{tikzpicture} 
		\captionof{figure}{{\em Scheùatic representation of the dispersion-relation expressions of the $(3+1)$D Dirac QW (left) and its flavored version (right).} 
        On the left, the FD neighborhoods are projected on the hyperplanes $E$\hspace{0.05cm}\textrm{---}\hspace{0.05cm}$p_x$, $E$\hspace{0.05cm}\textrm{---}\hspace{0.05cm}$p_y$, and $E$\hspace{0.05cm}\textrm{---}\hspace{0.05cm}$p_z$. On the right, the eight-sheeted solution.  }               
	\label{fig:shrinking3d}
	\end{center}

\subsubsection{The FD solution in Fourier space}

By observing Fig.\ \ref{fig:shrinking3d} (left), we realize that the shrinking procedure of Sec.\ \ref{subsec:1qedqcafdfix} can be repeated for each hyperplane independently, as illustrated by Fig.\ \ref{fig:shrinking3d} (right). 
The formal treatment of the new BZ and the map that allows us to consider the sheeted structure is discussed in Sec.\ \ref{sec:formaltreatment} below.
Still, notice that these cutting-and-pasting procedures for three hyperplanes, induce cutting and pasting at the level of the other 2D hyperplanes, $p_x$\hspace{0.05cm}\textrm{---}\hspace{0.05cm}$p_y$, $p_x$\hspace{0.05cm}\textrm{---}\hspace{0.05cm}$p_z$, and $p_z$\hspace{0.05cm}\textrm{---}\hspace{0.05cm}$p_y$, and similarly for the 3D hyperplanes, etc.. In the end, the geometry of $ \mathcal{B}^{2}$ is again shrunk into a new BZ as hoped.  
Notice also that cutting and pasting along $E$\hspace{0.05cm}\textrm{---}\hspace{0.05cm}$p_x$ yields two sheets and lays the  \textcolor{orange}{orange} FD neighbourhood on top of the \textcolor{red}{red} correct continuum limit; next, cutting and pasting along $E$\hspace{0.05cm}\textrm{---}\hspace{0.05cm}$p_y$ yields four sheets and lays the \textcolor{green}{green} FD neighborhood on top of the correct continuum limit; and finally, cutting and pasting along $E$\hspace{0.05cm}\textrm{---}\hspace{0.05cm}$p_z$ yields the $8$-sheeted solution with all FD neighborhoods on top of the correct continuum limit. This is because combinations of cutting-and-pasting procedures along these hyperplanes induce the same procedures on the hyperplanes where other FD neighborhoods sit, see Fig.\ \ref{fig:degeneracies3D}.  For example, \textcolor{brown}{brown} neighborhoods sit in the {$p_x$\hspace{0.05cm}\textrm{---}\hspace{0.05cm}$p_y$} hyperplane, as it is seen in  Fig.\ \ref{fig:degeneracies3D}, and cutting along  $E$\hspace{0.05cm}\textrm{---}\hspace{0.05cm}$p_x$ and $E$\hspace{0.05cm}\textrm{---}\hspace{0.05cm}$p_z$ results in these  \textcolor{brown}{brown} neighborhoods being present in the origin point of the sheet that contains only the \textcolor{brown}{brown} neighborhoods.


\subsubsection{Connecting the ``new Fourier space'' to a ``new direct space''}

\noindent
{{\bfseries The idea.}}
These $8$ sheets should be interpreted as $8$ flavors related to a global $ \mathbb{Z}_{2} \cross \mathbb{Z}_{2} \cross \mathbb{Z}_{2} $ symmetry. We can represent these flavor degrees of freedom with a $3$-qubit Hilbert space, generated by the basis $ \{\ket{000}, \ket{100}, \ket{010}, \ket{001}, \ket{110},\ket{101}, \ket{011}, \ket{111}\}$. The first basis qubit corresponds to the choice of sheet obtained by the $E$\hspace{0.05cm}\textrm{---}\hspace{0.05cm}$p_x$-hyperplane cutting-and-pasting procedure, i.e., $\ket 0$ for that carrying the \textcolor{red}{red} original continuum limit, and $\ket{1}$ for that carrying the \textcolor{orange}{orange} fermion doublers. The second basis qubit corresponds to the choice of the sheet obtained by the $E$\hspace{0.05cm}\textrm{---}\hspace{0.05cm}$p_y$-hyperplane cutting-and-pasting procedure, etc.. In direct space, where each flavor maps to one sublattice, it will be convenient to associate $\ket{000}$ to the original $\Gamma^4_{000}$ \textcolor{red}{red} sublattice, and $\ket{100}$ to its $x$-shifted $\Gamma^4_{100}$ \textcolor{orange}{orange} sublattice, so that, e.g., the flavored $x$-shift must be defined as $\hat S_{x}^{f} := \hat S_{x} \otimes (\sigma_{x} \otimes \mathbb{I}_2 \otimes \mathbb{I}_2)$. The other spatial directions are treated similarly. \\

\noindent
{{\bfseries The final connection.}}
Following this logic, one can establish a one-to-one correspondence between the sheet of the new BZ, the qubit encoding of its corresponding flavor, and the corresponding direct-space sublattice at time $n=0$. We find:
 \begin{equation}
 \label{eq:3dflavourcolour}
    \begin{split}
          \text{\textcolor{red}{red}} \equiv \ket{000}, \hspace{2mm}\text{\textcolor{orange}{orange}} \equiv \ket{100}, \hspace{2mm}\text{\textcolor{blue}{blue}} \equiv \ket{010}, \hspace{2mm}\text{\textcolor{green}{green}} \equiv \ket{001}\\
               \text{\textcolor{yellow}{yellow}} \equiv \ket{011}, \hspace{2mm}\text{\textcolor{brown}{brown}} \equiv\ket{101}, \hspace{2mm}\text{\textcolor{cyan}{cyan}} \equiv\ket{110}, \hspace{2mm}\text{\textcolor{pink}{pink}} \equiv \ket{111},
    \end{split}
 \end{equation}
 with the understanding that, e.g., the \textcolor{cyan}{cyan} sublattice $\Gamma^4_{110}$ is the $(0,\epsilon,\epsilon,0)$-translate of the original \textcolor{red}{red} sublattice, etc.. This way, we have deduced Fig.\ \ref{fig:f-lattice3d} (left). \\

 \noindent
{{\bfseries Flavored time translation.}}
 In direct spacetime, after one time step and hence three substeps, a particle will have moved along the three directions and hence the flavored time translation needs to be defined as  $\hat T^{f} := \hat T \otimes (\sigma_{x} \otimes \sigma_{x} \otimes \sigma_{x})$.  From this we deduce Fig.\ \ref{fig:f-lattice3d} (left) and (right). \\

 \noindent
{{\bfseries The final sublattices covering.}} 
Altogether, the $8$ direct-spacetime lattices arising from the flavor symmetry $\mathbb{Z}_{2} \cross \mathbb{Z}_{2} \cross \mathbb{Z}_{2}$ form a covering of the $4$D orthogonal lattice $ \Lambda^{4}$, which can be expressed as the following equality,
	\begin{equation}
		\Lambda^{4} = \bigcup_{j \in  \mathbb{Z}_{2} \cross \mathbb{Z}_{2} \cross \mathbb{Z}_{2}  }  \Gamma^{4}_{j} \, .
	\end{equation}  
The exact correspondence between the generating vectors of these sublattices and the reciprocal vectors of the new BZ is detailed in App.\ \ref{subsec:4DBZ}.

	\begin{center}
		\begin{tikzpicture}

\draw[thick,->] (-2.7,-1,0) -- (-2,-1,0);
\draw node at (-1.8,-1,0) {$x$};
\draw[thick,->] (-2.7,-1,0) -- (-2.7,-0.3,0);
\draw node at (-2.7,-0.1,0) {$y$};
\draw[thick,->] (-2.7,-1,0) -- (-2.7,-1,0.7);
\draw node at (-2.7,-1,1) {$z$};

			\filldraw[red] (0,0,0) circle (3pt) ;
			\filldraw[orange] (1,0,0) circle (3pt) ;
			\filldraw[blue] (0,1,0) circle (3pt) ;
			\filldraw[green] (0,0,1) circle (3pt) ;
			\filldraw[yellow] (0,1,1) circle (3pt) ;
			\filldraw[brown] (1,0,1) circle (3pt) ;
			\filldraw[cyan] (1,1,0) circle (3pt) ;
			\filldraw[pink] (1,1,1) circle (3pt) ;

			\filldraw[blue] (0,-1,0) circle (3pt) ;
			\draw[dashed] (0,-1,0)--(0,0,0);
			\filldraw[cyan] (1,-1,0) circle (3pt) ;
			\draw[dashed] (1,-1,0)--(1,0,0);
			\filldraw[yellow] (0,-1,1) circle (3pt) ;
			\draw[dashed] (0,-1,1)--(0,0,1);
			\filldraw[pink] (1,-1,1) circle (3pt) ;
			\draw[dashed] (1,-1,1)--(1,0,1);
			\draw[dashed] (0,-1,0)--(1,-1,0)--(1,-1,1)--(0,-1,1)--(0,-1,0);

			\draw[dashed] (0,0,0)--(1,0,0)--(1,1,0)--(1,1,1)--(1,0,1)--(0,0,1)--(0,1,1)--(0,1,0)--(0,0,0)--(0,0,1);
			\draw[dashed] (1,0,0)--(1,0,1);
			\draw[dashed] (0,1,1)--(1,1,1);
			\draw[dashed] (0,1,0)--(1,1,0);

			\filldraw[red] (2,0,0) circle (3pt) ;
			\filldraw[green] (2,0,1) circle (3pt) ;
			\filldraw[blue] (2,1,0) circle (3pt) ;
			\filldraw[yellow] (2,1,1) circle (3pt) ;
			\draw[dashed] (1,0,0)--(2,0,0);
			\draw[dashed] (1,0,1)--(2,0,1);
			\draw[dashed] (1,1,1)--(2,1,1);
			\draw[dashed] (1,1,0)--(2,1,0);
			\draw[dashed] (2,1,1)--(2,1,0)--(2,0,0)--(2,0,1)--(2,1,1) ;

			\filldraw[green] (0,0,-0.8) circle (3pt) ;
			\draw[dashed] (0,0,0)--(0,0,-0.8);
			\filldraw[brown] (1,0,-0.8) circle (3pt) ;
			\draw[dashed] (1,0,0)--(1,0,-0.8);
            \filldraw[yellow] (0,1,-0.8) circle (3pt) ;
			\draw[dashed] (0,1,0)--(0,1,-0.8);
            \draw[dashed] (0,0,-0.8)--(0,1,-0.8);
              \filldraw[pink] (1,1,-0.8) circle (3pt) ;
			\draw[dashed] (1,1,0)--(1,1,-0.8);
            \draw[dashed] (1,0,-0.8)--(1,1,-0.8);
             \draw[dashed] (0,1,-0.8)--(1,1,-0.8);
             \draw[dashed](0,0,-0.8)--(1,0,-0.8);

			\draw node at (0.2,-2,0) {$t= 0 $};

			\begin{scope}[xshift=40mm]
				\filldraw[red] (1,1,1) circle (3pt) ;
				\filldraw[orange] (0,1,1) circle (3pt) ;
				\filldraw[blue] (1,0,1) circle (3pt) ;
				\filldraw[green] (1,1,0) circle (3pt) ;
				\filldraw[yellow] (1,0,0) circle (3pt) ;
				\filldraw[brown] (0,1,0) circle (3pt) ;
				\filldraw[cyan] (0,0,1) circle (3pt) ;
				\filldraw[pink] (0,0,0) circle (3pt) ;
				\draw[dashed] (0,0,0)--(1,0,0)--(1,1,0)--(1,1,1)--(1,0,1)--(0,0,1)--(0,1,1)--(0,1,0)--(0,0,0)--(0,0,1);
				\draw[dashed] (1,0,0)--(1,0,1);
				\draw[dashed] (0,1,1)--(1,1,1);
				\draw[dashed] (0,1,0)--(1,1,0);

				\filldraw[brown] (0,-1,0) circle (3pt) ;
				\draw[dashed] (0,-1,0)--(0,0,0);
				\filldraw[green] (1,-1,0) circle (3pt) ;
				\draw[dashed] (1,-1,0)--(1,0,0);
				\filldraw[orange] (0,-1,1) circle (3pt) ;
				\draw[dashed] (0,-1,1)--(0,0,1);
				\filldraw[red] (1,-1,1) circle (3pt) ;
				\draw[dashed] (1,-1,1)--(1,0,1);
				\draw[dashed] (0,-1,0)--(1,-1,0)--(1,-1,1)--(0,-1,1)--(0,-1,0);

				\filldraw[pink] (2,0,0) circle (3pt) ;
				\filldraw[cyan] (2,0,1) circle (3pt) ;
				\filldraw[brown] (2,1,0) circle (3pt) ;
				\filldraw[orange] (2,1,1) circle (3pt) ;
				\draw[dashed] (1,0,0)--(2,0,0);
				\draw[dashed] (1,0,1)--(2,0,1);
				\draw[dashed] (1,1,1)--(2,1,1);
				\draw[dashed] (1,1,0)--(2,1,0);
				\draw[dashed] (2,1,1)--(2,1,0)--(2,0,0)--(2,0,1)--(2,1,1) ;
	
		\filldraw[cyan] (0,0,-0.8) circle (3pt) ;
			\draw[dashed] (0,0,0)--(0,0,-0.8);
			\filldraw[blue] (1,0,-0.8) circle (3pt) ;
			\draw[dashed] (1,0,0)--(1,0,-0.8);
            \filldraw[orange] (0,1,-0.8) circle (3pt) ;
			\draw[dashed] (0,1,0)--(0,1,-0.8);
            \draw[dashed] (0,0,-0.8)--(0,1,-0.8);
              \filldraw[red] (1,1,-0.8) circle (3pt) ;
			\draw[dashed] (1,1,0)--(1,1,-0.8);
            \draw[dashed] (1,0,-0.8)--(1,1,-0.8);
             \draw[dashed] (0,1,-0.8)--(1,1,-0.8);
             \draw[dashed](0,0,-0.8)--(1,0,-0.8);
	
				\draw node at (0.2,-2,0) {$t=\epsilon$};
	
				\end{scope}

		\end{tikzpicture}
		\captionof{figure}{{\em Sublattices of the $(3+1)$D Dirac flavored QCA.}} 
 \label{fig:f-lattice3d}
	\end{center}

\subsubsection{Solution in direct space: the $(3+1)$D flavored-QW dynamics}

 \noindent
{{\bfseries Flavored state.}}
For this flavored lattice, at time $ t = n \epsilon$, a state of the $(3+1 )$D Dirac FQW, i.e., the one-particle sector of the $(3+1 )$D Dirac FQCA, $\ket{\psi^f(n)}$, is chosen to be of the form

	\begin{align}
		 \ket{\psi^f(n)} &\equiv \sum_{k,l,m} \bigg[ \psi^{(1)}(n,k,l,m) \ket{1 0 0 0 }^{k,l,m} + \psi^{(2)}(n,k,l,m) \ket{0 1 0 0 }^{k,l,m} + \psi^{(3)}(n,k,l,m) \ket{0 0 1 0 }^{k,l,m} \nonumber \\  & \ \ \ \ \ \ \ \ + \psi^{(4)}(n,k,l,m) \ket{0 0 0 1 }^{k,l,m}  \bigg] \otimes \bigg[\Big( \sigma_{x}^{n+k}\otimes  \sigma_{x}^{n+l} \otimes  \sigma_{x}^{n+m} \Big) \ket{000} \bigg] \\
         &\equiv \sum_{k,l,m} \bigg[ \Big( \psi^{(1)}(n,k,l,m) \ket{1 } + \psi^{(2)}(n,k,l,m) \ket{2} + \psi^{(3)}(n,k,l,m) \ket{3} \nonumber \\  & \ \ \ \ \ \ \ \ + \psi^{(4)}(n,k,l,m) \ket{4}\Big) \otimes \ket{k,l,m} \bigg] \otimes \bigg[\Big( \sigma_{x}^{n+k}\otimes  \sigma_{x}^{n+l} \otimes  \sigma_{x}^{n+m} \Big) \ket{000} \bigg] \, .
	\end{align}


\vspace{2mm}
 \noindent
{{\bfseries Flavored dynamics.}}
As mentioned above, the translation operators in space and time act also on the flavor degrees of freedom, in association with Eqs.\ \eqref{eq:3dflavourcolour} and Fig.\ \ref{fig:f-lattice3d}: 

\begin{subequations}
	\begin{align}
		\hat S_{x}^{f} &\defeq \hat S_{x} \otimes (\sigma_{x} \otimes \mathbb{I}_2 \otimes \mathbb{I}_2) \\
		\hat S_{y}^{f} &\defeq \hat S_{y}   \otimes (\mathbb{I}_2  \otimes  \sigma_{x} \otimes \mathbb{I}_2)\\
		\hat S_{z}^{f} &\defeq \hat S_{z}   \otimes (\mathbb{I}_2  \otimes \mathbb{I}_2  \otimes  \sigma_{x}) \\
		\hat{T}^{f} &\defeq \hat{T} \otimes (\sigma_{x} \otimes \sigma_{x} \otimes  \sigma_{x}) \, .
	\end{align}
\end{subequations}
To turn $ \hat{M}_{QW}^{3}$ into $\hat{M}_{FQW}^{3}$, one just needs, as in the $(1+1)$D case, to exchange the translation operators with those  of the four previous equations. Recall that $\Hat{M}_{FQW}^{3}$ is the operator which generates the equations of motions for the $(3+1)$D FQW, and, according to what we have just said, together with one or two lines of computations, it is actually straightforward to realize that it has the following form\footnote{Note that this is not true in the case of the $(1+1)$D model, and is due to the fact that we have chosen, in this $(3+1)$D context, a (1+1)D block for each space direction which is slightly different from our $(1+1)$D model---~which contains an on-site term whereas the $(1+1)$D blocks of our $(3+1)$D model do not.},
\begin{equation}
  \label{eq:mfqw}
    \hat{M}_{FQW}^{3} = \hat{M}_{QW}^{3} \otimes (\sigma_{x} \otimes \sigma_{x} \otimes  \sigma_{x}) \, .
\end{equation}
Its determinant ``in Fourier space'', $\mathscr{D}^{3}_f$, turns out be equal to
\begin{equation}
\mathscr{D}^{3}_f( E,  p_{x},p_{y},p_{z} ) = [\mathscr{D}^{3} ( E,  p_{x},p_{y},p_{z} )]^{8} \, ,
\end{equation}
analogously to what we obtained for our $(1+1)$D model. In the end, we have $8$ independent correct solutions, each of a different flavor, coinciding with the right continuum limit---instead of one single correct solution and $7$ spurious ones.

\section{Flavoring is a covering map} \label{sec:formaltreatment}

Let us take a step back. To detect and fix FD, we 1/ started with some discrete equation of motion that produces the Dirac equation in the continuum, 2/ expressed it in Fourier space, 3/~analyzed its determinant, 4/ observed that the determinant had solutions that were spurious, i.e., that would not vanish in the continuum limit, 5/ cut the domain of definition of the determinant, a.k.a. the BZ, into regions of similar behavior, and ``pasted them on top of each other'' in order to remove the spurious solutions, 6/ realized that these regions corresponded to sublattices in direct space, 7/~flavored each sublattice, and, finally, 8/ realized that, this way, each spurious solution had become a correct solution of a different flavor. Now, given an initial BZ such as that considered, i.e., with its FD neighborhoods around the spurious solutions, it is not something completely straightforward to formalize the cutting-and-pasting procedure that we have to apply in order to remove FD---namely, step 5/ in the above summary. This demands some thinking, which this section aims to deliver. We find that the best way is to visualize the initial BZ is as a toric complex plane, and we come up with a covering map whose image is also a toric complex plane, but a different one. In this section, we will give the precise formalization of this idea, both for the $(1+1)$D Dirac QW and the $(3+1)$D Dirac QW.

\subsection{$(1+1)$D Dirac QW}


\subsubsection{Introduction}

\vspace{2mm}
We formalize, with the use of covering maps, the cutting-and-pasting procedures that we have done in the previous section. Here is our strategy: first, we turn $\mathcal{B}$ and $\mathcal{B'}$ into Riemann surfaces with complexification of their coordinates; second, we express $\mathcal{B}$ as a complex torus which is the quotient of the complex plane by a square lattice\footnote{This square lattice is the Fourier-space reciprocal of the real-space lattice,  since indeed it is known than the former is \emph{also} a square lattice (this holds for square lattices only, a priori).}, and likewise we express $\mathcal{B'}$ as a  complex torus which is the quotient of the complex plane by an oblique lattice (which is also a square lattice in the right coordinates); third, we define a map from the square lattice to the oblique lattice; and fourth, we prove that this map induces a covering map from $\mathcal{B}$ to $\mathcal{B'}$. 

We take the definition of a covering map of Ref.\ \cite{gilligan2012lectures}.
\begin{definition}If $ X $ and $ Y $ are topological spaces, a mapping $ p : Y \to X$ is called a covering map if every point $ x \in X$ has an open neighborhood $ U$, such that its preimage $ p^{-1}(U)$ can be represented as
\begin{equation}
    p^{-1}(U) = \bigcup_{j \in J} V_{j} \, ,
\end{equation}
where the $ V_{j}$s are disjoint open subsets of $ Y$, $ j \in J$ with $ J$ some discrete set. Moreover, all the mappings that are restricted to $ V_{j}$, $ p|_{V_j} : V_{j} \to U$, must be homeomorphisms. 
\end{definition}
\noindent
Notice that the last sentence is not relevant in the present context, as it only specified for topologies that may have singularities. 

\subsubsection{Definition of the will-be covering map $\varphi$}

Let us start by the first step, namely, the complexification of the coordinates of BZ, via $ (E,p) \cong z = \epsilon(E+i p)/2\pi$. We will, from now on, not introduce new notations for many of the complexified objects, we will just identify the original ones to them. Secondly, we find that the BZ of a square lattice can be constructed by a quotient topology, $\mathcal{B}\cong \mathbb{C}/ \Lambda$, where\footnote{Before we had used the notation $k$ for $m$, but it the same concept. Actually, we have used for $\Lambda$ the notation we use for the real-, or direct-space lattice (including the labels of the coordinates), but the $\Lambda$ we refer to here is the Fourier-space version of the original $\Lambda$, but we have kept the notation $\Lambda$ because mathematically the two are the same, and this section is rather mathematical. We will do the same for $\Gamma$ afterwards.} $ \Lambda \cong \{ n e^{ 2i \pi}+ m e^{i \pi/2}: n, m \in \mathbb{Z} \}$,  meaning that $z_1 \sim z_2 $ if $ z_1-z_2 \in \Lambda$.  In Sec.\ \ref{sec:solution}, we worked out a way of departing from $ \mathcal{B}$ and arriving to the rhombus-shaped $\mathcal{B}'$, which is also constructed by a quotient topology, namely, $\mathcal{B}'\cong \mathbb{C}/ \Gamma^{2}$, with\footnote{In App.\ \ref{app:cont_lim_DQW}, we have used the notations $\rho$ and $r$ for the $m$ and $n$ that just follow, respectively, since we are dealing with the rhombus, i.e., square-rotated lattice, but these are the same concepts.}  $  \Gamma^{2} \cong \{ (    n e^{ -i \pi/4} + m e^{i \pi/4}  )/\sqrt{2}: n, m \in \mathbb{Z} \}$ and under the same complexification, as it is clear from Fig.\ \ref{fig:shrinking1d}.

Notice that the lattice $\Gamma^{2}$ characterizes $\mathcal{B}'$, and hence we define a map $ \varphi: \mathcal{B} \to \mathcal{B}'$ characterized by the map $\upsilon: \Lambda \to \Gamma^{2}$ defined as a multiplication by $ j \defeq e^{-i \pi/4}/\sqrt{2}$, i.e., as $\Gamma^{2} \defeq j \Lambda$. Again, this already implicitly defines $\varphi(z)$, and now we are going to extract from this implicit definition and explicit one. 
We are first going to define a suitable map $\varphi^{*}: \mathbb{C} \to \mathbb{C}$. Any complex number $Z$ can be uniquely decomposed as $Z = z + \lambda $ with $ z \in \mathcal{B} $ and $ \lambda \in \Lambda$. 
We hence define $\varphi^{*}( Z)$ as
\begin{equation}
    \varphi^{*}( z+ \lambda)\defeq z+ j\lambda \, .
\end{equation}
Next, we define 
\begin{equation}
\varphi( z ) \defeq \varphi^{*}(z+\lambda)/\Gamma^{2}  \, ,
\end{equation}
which gives us
 \begin{subequations}
   \begin{align}
         \varphi(z) &= ( z +  j\lambda)/ \Gamma^{2} \\
         &= z/\Gamma^{2} \, ,\quad\textrm{ since }j\lambda \in \Gamma^{2},\label{eq:first} \\
         &= z -  \gamma \label{eq:secondd} \, ,
            \end{align}
 \end{subequations}
for some $\gamma \in \Gamma^2$ (Eq.\ \eqref{eq:secondd} ensues from the mere definition of what $z/\Gamma^2$ is),  i.e.\ (see the definition of $\Gamma^2$ above), for some $n$ and $m$ such that
\begin{equation}
\gamma =  n j+ \frac{m}{2 j} \, .
\end{equation}
Below, we are going to prove which $m$ and $n$ are acceptable, and we are actually going to choose them in a particular manner so as to pick up a particular $\varphi$ of our convenience.

\subsubsection{Proof that $\varphi$ is a covering map}

To show that $\varphi$ is indeed a covering map, we need to show that, for  every open $ U  \subset \mathcal{B}'$, $ \varphi^{-1}(U)$ is a disjoint union of open subsets of $ \mathcal{B}$.

 As it is the case in general for maps of the type of $\varphi$ that we have just explicitly defined~\cite{diamond2006first}, it is clear that $\varphi$ is not invertible, e.g., $\varphi(0)=\varphi(j)= 0$,  and $ \varphi(\frac{e^{-i \pi/4}}{\sqrt{2}}) = \varphi(\frac{e^{5 i \pi/4}}{\sqrt{2}}) =\frac{e^{-i \pi/4}}{\sqrt{2}}$. 
 But, it is easy to realize that for $ \varphi(0) \in U \subset \mathcal{B}'$, we can find disjoint open neighborhoods of $0 \in \mathcal B$ and $j \in \mathcal B$, namely, $V_0$ and $V_1$, respectively, such that $ \varphi^{-1}(U) = V_{1} \cup V_{2}$, and this is an example of a $U$ for which $ \varphi^{-1}(U)$ is a union of disjoint subsets of $ \mathcal{B}$. 


 Nevertheless, the fact that  $\varphi$ is a \underline{\emph{two-covering}}, \emph{in general}, i.e., for any $z \in \mathcal B$, needs to be proven. 
First, we notice that $\mathcal{B'} \subset  \mathcal{B}$, so that we can \emph{choose} that the restriction of $\varphi$ to $ \mathcal{B'}$ is the identity.
For every $z\in U\subseteq  \mathcal{B'}$, we thus start with  $z=\varphi(z)$ (which is known to be true with the choice we just made), and then $z=\varphi(z') \neq z'$, and $z=\varphi(z'') \neq z''$, where $ z \in V_{1}$, $ z' \in V_{2}$, $ z'' \in V_{3}$, and $V_{1},V_{2},V_{3}$ are open neighborhoods of $\mathcal{B}$. We are going to show (i) that we can choose a $\varphi$ which indeed corresponds to our graphical intuition, and (ii) that we can find a $\varphi$ such that that $z' = z''$.

Now let us see if the choice  $z' \defeq z+ 2 h j + g/j$, with $h,g \in \{-1,0,1\}$, which is simply directly suggested by our graphical construction in Fig.\ \ref{fig:shrinking1d}, is compatible with our formal definition of $\varphi$ above, which is the main thing we have to check here. Notice that we identify two choices of $(h,g)$ that would give the same $z'$ due to the toric geometry of the BZ. Inserting this choice for $z'$ into Eq.\ \eqref{eq:secondd} taken for $z=z'$, we get
 \begin{equation}\label{eq:covering1d}
     \varphi(z') = z + (2h - m)j + \frac{2g - n}{2j} \, ,
  \end{equation}
so it is sufficient and actually necessary to choose $m \defeq 2h$ and $n \defeq 2g$ to find $ \varphi(z') = z  $, which is what we want. This proves that $ \varphi(V_{2}) =\varphi(V_{1})$.

Moreover, if we pick $z'' \defeq z+ 2 h' j + g'/j \in V_3$ (this form is the only possibility for $z''$ if again we want to stick to pour graphical construction) such that $z'' \neq z$, then it is easy to realize graphically that we must have $z''=z'$ because otherwise $ z'' \notin \mathcal{B}$, or $ z'' = z$. Hence, $ V_{3}= V_{2} $, and we have proved that $\varphi$ is a two-covering map. 

The fact that $\varphi$ is a two-covering between the BZs $\mathcal{B}$ and $\mathcal{B}'$ gives the definitive formal geometrical understanding of FD as a $\mathbb{Z}_2$ symmetry, as announced previously.

\subsection{$(3+1)$D Dirac QW}

We would like to use the same reasoning as in the previous section for the $(3+1)$D case, and show that the cutting-and-pasting procedure of Fig.\ \ref{fig:shrinking3d} is an eight-covering map $\varphi_{2}$ from  $\mathcal{{B}}^{2}$ to $(\mathcal{B}^{2})'$. However, here the four-dimensional $(\mathcal{B}^{2})'$ is \emph{not} a rhombus nor a Cartesian product of rhombuses; rather, it only admits a rhombus shape after a projection on a two-dimensional hyperplane is applied.  But, we \emph{have actually} worked out a way of representing $(\mathcal{B}^{2})'$ as a Cartesian product of three $ \mathcal{B}'$s in App.\ \ref{sec:applowerBZ}. Such a presentation makes use of the Bragg equation, which allowed us to introduce degenerative representations of the BZ.
Namely, we found that $(\mathcal{B}^{2})'=\Pi (\mathcal{B'}^{3})$ with $ \mathcal{B'}^{3}\defeq \mathcal{B}'\cross\mathcal{B}'\cross\mathcal{B}'\cong \mathbb{C}/ \Gamma^{2} \cross \mathbb{C}/ \Gamma^{2} \cross \mathbb{C}/ \Gamma^{2} $ and $ \Pi$ a projection map. In the same manner, we found that $\mathcal{{B}}^{2}=\Pi (\mathcal{B}^{3})$ with $ \mathcal{B}^{3}\defeq \mathcal{B}\cross\mathcal{B}\cross\mathcal{B}\cong\mathbb{C}/ \Lambda \cross \mathbb{C}/ \Lambda \cross \mathbb{C}/ \Lambda $  and $\Pi$ the same projection. This $\Pi$ takes $ (z,w,x) \in \mathbb{C}^3 / \Lambda^{3}$ into
\begin{equation}
    \Pi(z,w,x)= (\Re(z)+ i \Im(w),\Im(w)+ i \Im(x) ).
\end{equation}
Geometrically, $\Pi$ must be understood as projecting on the hyperplane $\Re(z)=\Re(w)=\Re(x)$.

We can now decompose the cutting-and-pasting procedures that we have in Fig.\ \ref{fig:shrinking3d} as follows. First, we embed $\mathcal{{B}}^{2}$ into $ \mathcal{B}^{3}$ as we did in Fig.\ \ref{fig:shrinking3d} (left), i.e., we represent $ \mathcal{B}^{2}$ as a subset of the Cartesian product of three independent two-dimensional surfaces. Second, we define a covering map, $\varphi_{3}$ from  $\mathcal{B}^{3}$ to $\mathcal{B'}^{3}$, just by letting $\varphi_3=\varphi\cross\varphi\cross \varphi$, i.e.,
\begin{equation}
     \varphi_{3}( z, w, x) \defeq  (\varphi( z),\varphi( w),\varphi(x))= ( z - \gamma_{0}, w -  \gamma_{1}, x - \gamma_{2}) \, ,
\end{equation}
where $  \gamma_{0}, \gamma_{1}, \gamma_{2} \in \Gamma^{2}$; since $\varphi$ is a two-covering, then we immediately find that $\varphi_{3}$ is an eight-covering.
Finally, we apply the projection map $ \Pi$, which enables us to define a composite map $\varphi_{2} \defeq \Pi \circ \varphi_{3} \circ \Pi^{-1}$, which maps $\mathcal{B}^{2}$ to $(\mathcal{B}^{2})'$. This $\varphi_{2}$ ``remains'' an eight-covering map (with respect to $\varphi_3$), because the multi-valuedness of $\Pi^{-1}$ gets cancelled by $\Pi$, as detailed in App.~\ref{sec:applowerBZ}, and this completes our proof.


\vspace{2mm}

\vspace{2mm}


\vspace{2mm}

\section{Chiral symmetry and no-go theorem}\label{sec:discussion}
In this section, we show that the QCA models of Refs.\ \cite{ABF20, EDMMplus2023}  are chirally symmetric. We also show that the flavored versions of these QCA models remain chirally symmetric. This may come as surprising to the more QFT-inclined reader, with awareness of the famous no-go theorem by Nielsen and Ninomiya. We take an in-depth look at the relationship between this theorem and our flavored-QCA models. This section may be skipped by the more quantum-information-inclined reader. 

\subsection{Chiral symmetry on Dirac QCAs}
\label{subsec:ChiralSymmetry}

\subsubsection{Reminder: definition of both axial and chiral symmetries}

Although chiral symmetry is a well-known topic of fermionic theories (see Ref.\ \cite{book_Peskin_Schroeder} for a comprehensive treatment), let us remind the reader of its definition in order to make this exposition self-contained.

Let us first recall axial symmetry---or, more precisely, U$(1)$-axial symmetry---and then we will recall chiral symmetry.  When we say that the massless $ (D+1)$D Dirac equation, which can be written $ i \gamma^{\mu} \partial_{\mu} \psi_{\text{cont.}}(x^{\mu})=0$---for $ \mu \in \{0 , 1, \dots, D \}$, and with the shortcut $x^\mu$ for $\underline{x} \defeq (x^0=t,x^1=x,x^2,\dots,x^D)$---, is axially symmetric, we mean that $i \gamma^{\mu} \partial_{\mu} (\exp{i \theta \gamma^{5}} \psi_{\text{cont.}})(x^{\mu}) = 0 $, where $\theta \in \mathbb R$, and where $ \gamma^{5} \defeq i \gamma^{0} \gamma^{1} \dots \gamma^{D}$, so that $ [\gamma^{5}, \gamma^{\mu}]= 0$. This is the basic most-general symmetry associated to $\gamma^5$ that one usually wants to preserve. Now, in most cases, including the case of the present, free massless Dirac equation, axial symmetry is equivalent to  so-called chiral symmetry\footnote{Whilst axial symmetry and chiral symmetry often go hand-in-hand, this is not always the case, e.g., in the presence of weak interaction or Adler-Bell-Jackiw anomaly. In such cases one does not have axial symmetry but one still has chiral symmetry; the massless Dirac operator, $ i \gamma^{\mu} \partial_{\mu}$, commutes with the chiral projection operators (or chiral projectors)---see definitions below in the main text---, and one characterizes chiral symmetry by means of this commutation relation.}, which, on that equation, reads $i \gamma^{\mu} \partial_{\mu}( \gamma^{5} \psi_{\text{cont.}})(x^{\mu}) = 0 $. We will hence focus on chiral symmetry here.

Whatever the space dimension $D$, one can show that the matrix $ \gamma^{5}$ has two eigenvalues, namely, $ 1$ and $-1$, which naturally induce a splitting of the total internal degree of freedom into what we call \emph{chirality}, degree of freedom of Hilbert-space dimension $2$, and the rest (typically, the spin). The two eigenvalues have two corresponding eigenspaces, and we can look for solutions of the Dirac equation using the partition of the total Hilbert space induced by these eigenspaces: we call the wavefunctions associated to each eigenspace ${\psi}^{R}_{\text{cont.}}$ and ${\psi}^{L}_{\text{cont.}}$, respectively. Together, they form the wavefunction, namely $  \psi_{\text{cont.}} = {\psi}^{R}_{\text{cont.}} +{\psi}^{L}_{\text{cont.}}$. Hence $\gamma^{5}\psi_{\text{cont.}}= {\psi}^{R}_{\text{cont.}} - {\psi}^{L}_{\text{cont.}}$. The chiral projectors, each projecting on one eigenspace, are defined as
\begin{subequations}
\begin{align}
     P_{R} &\defeq \frac{1}{2}(\mathbb{I} + \gamma^{5}) \\
     P_{L} &\defeq \frac{1}{2}(\mathbb{I} -\gamma^{5}) \, ,
\end{align}
\end{subequations}
where $\mathbb{I}$ is the identity matrix corresponding to the representation of the Clifford algebra that one chooses to work with. The previous projectors project the wavefunction into its right-handed component, $ {\psi}^{R}_{\text{cont.}}$, or left-handed one, $ {\psi}^{L}_{\text{cont.}}$, and then it is trivial to show that the massless Dirac equation decouples into two equations, one for the left-handed component, and one for right-handed one: this fact is what is referred to as chiral symmetry, namely, the exchange of the chiralities does not change the system of equations, since the latter decouples chirality when the mass is zero. Notice that, when $ \exp{i \theta \gamma^{5}}$ is applied on ${\psi}^{R}_{\text{cont.}}$ and $ {\psi}^{L}_{\text{cont.}}$, the latter catch the phases $ \exp{-i \theta } $ and $ \exp{i \theta }$, respectively.  We conclude that the essence of the axial symmetry is having two similar equations of motions whose respective wavefunctions ``take opposite phase orientations when a phase is applied'', rather than taking the same phase orientation in the case of mere $\text{U}(1)$ symmetry.

\subsubsection{The massless $(1+1)$D and $(3+1)$D Dirac QWs are chirally symmetric}

Consider now the $(1+1)$D QED QCA model in Ref.\ \cite{ABF20}. In the non-interacting case, it has as continuum limit the $(1+1)$D Dirac equation in the following representation:
\begin{subequations}
\begin{align}
    ( i  \sigma_{x} \partial_{t} +  \sigma_{y} \partial_{x}-m) {\psi}_{\text{cont.}}(t,x) = 0 \, . 
\end{align}
\end{subequations}
In this representation, the chiral projectors are
\begin{subequations}
\begin{align}
    {P}_{R}^{1} &\equiv \frac{1}{2}(\mathbb{I}_{2}+  \sigma_{x} \sigma_{y}) \\
    {P}_{L}^{1} &\equiv \frac{1}{2}(\mathbb{I}_{2} -  \sigma_{x} \sigma_{y}) \, .
\end{align}
\end{subequations}
Even though these projectors have been defined in the continuum limit, they work unchanged in the discrete, as they do not depend on the scaling $ \epsilon$. The $(1+1)$D QED QCA is a chirally symmetric theory because, in the massless case, the following commutation relations hold:
\begin{subequations}
\begin{align}
   [ \Hat{M}_{QW}^{1}, {P}_{R}^{1}]&= 0 \\
   [ \Hat{M}_{QW}^{1}, {P}_{L}^{1}]&= 0 \, .
\end{align}
\end{subequations}

Now, the unitary evolution operator of the $(3+1)$D QED QCA is written in the Weyl representation, as can be checked starting from Eq.\ \eqref{eq:u3qw}. We call $ \gamma^{5}_{W}$ the $(3+1)$D ``$\gamma^5$ matrix'' in such a representation, which is
\begin{equation*}
    \gamma^{5}_{W} \defeq \begin{pmatrix}
        \mathbb{I}_{2}&0\\
        0& -\mathbb{I}_{2}
    \end{pmatrix} \, ,
\end{equation*}
and the corresponding chiral projectors are 
\begin{subequations}
\begin{align}
 {P}_{ R}^{3} &\defeq  \begin{pmatrix}
        \mathbb{I}_{2}&0\\
        0&0
    \end{pmatrix}  \\
    {P}_{L}^{3} &\defeq  \begin{pmatrix}
        0&0\\
        0& \mathbb{I}_{2}
    \end{pmatrix} \, .
\end{align}
\end{subequations}
Again by checking that the following commutation relations hold in the massless case,
\begin{subequations}
\begin{align}
     [ \Hat{M}_{QW}^{3}, {P}_{L}^{3}] &= 0 \\ [ \Hat{M}_{QW}^{3}, {P}_{R}^{3}] &= 0 \, ,
\end{align}
\end{subequations}
we establish that the $(3+1)$D QED QCA is chirally symmetric.

\subsubsection{The massless $(1+1)$D and $(3+1)$D Dirac FQWs are still chirally symmetric}
\label{subsubsec:ChiralSymmetryFQW}

It is straightforward to see that if a QCA model is chirally symmetric, then its flavored version  is also chirally symmetric. This is because the chiral projectors do not act on the flavor degree of freedom. Let us display this fact on our flavored version of the $(3+1)$D QED QCA. The chiral projection operators for the flavored theory are simply
\begin{subequations}
\begin{align}
    {P}_{F;R}^{3} &\defeq {P}_{R}^{3} \otimes \mathbb{I}_{8} \\ {P}_{F;L}^{3}  &\defeq {P}_{L}^{3} \otimes \mathbb{I}_{8} \ .
\end{align}
\end{subequations}
Due to Eq.\ \eqref{eq:mfqw}, we find
\begin{subequations}
\begin{align}
     [ \Hat{M}_{FQW}^{3}, {P}_{F;R}^{3}]&= 0 \\ [ \Hat{M}_{FQW}^{3}, {P}_{F;L}^{3}]&= 0 \ .
\end{align}
\end{subequations}
The same works for the flavored $(1+1)$D QED QCA, it is also chirally symmetric. This means that for each flavor component, we get, in the massless case, decoupled left-handed and right-handed fermions.

\subsection{No-go theorem \emph{versus} our flavor-staggering-\emph{only} scheme {\color{red}}}

\subsubsection{Chiral symmetry on the lattice and FD do not always go together, such as in our flavored scheme}
 
{\bfseries Original statement of the no-go theorem.} Nielsen and Ninomiya's 1981 three articles about the constraints arising when putting chiral fermions on a lattice~\cite{NielsenI1981, NielsenII1981, Nielsen1981}, are well-known in lattice QFT. Their no-go theorem is stated for the first time (first article, Ref.\ \cite{NielsenI1981}, p.\ 21) in the following terms: \\

\begin{adjustwidth}{1cm}{0cm}
``It is the purpose of this article to formulate and prove a \emph{no-go theorem}: the appearance of equally many right- and left-handed species (types) of Weyl particles with given quantum numbers is an unavoidable consequence of a lattice theory under some \emph{mild assumptions}.'' \\
\end{adjustwidth}

\noindent
What is said in this passage, is that you need a state space that is able to handle as many left-handed as right-handed fermions at each site and for each every quantum number, i.e., charge (in the very generic sense of potentially conserved quantities).  \\

\noindent
{\bfseries Ubiquitous loosely formulated statement of the no-go theorem.} Now, in the Lattice-QFT community, the question of the potential loss of chiral symmetry in the massless scheme, often gets mixed up with the FD issue (see, e.g., Ref.\ \cite{RotheBook}):\\


\begin{adjustwidth}{1cm}{0cm}
``That the doubling phenomenon must occur in a lattice regularization which respects the usual hermitcity, locality and translational invariance requirements, follows from a theorem by Nielsen and Ninomiya [Nielsen (1981)] which states that, under the above assumptions, one cannot solve the fermion doubling problem without breaking chiral symmetry for vanishing fermion mass.'' \\
\end{adjustwidth}

\noindent
Here the author seems so say that FD fixing is also a source of chirality breaking. \\

\noindent
{\bfseries Avoidance of the loosely formulated no-go theorem with our flavor-staggering-only scheme.} The previous sentence is ony true with FD-fixing techniques that will necessarily break chiral symmetry, such as Wilson extra-term techniques or \emph{chirality}-staggering techniques. Our flavor-staggering-\emph{only} FD-fixing technique, instead, is not chirality breaking (see the proof of this in Sec.\ \ref{subsubsec:ChiralSymmetryFQW}), thanks to the introduction of a second fermion flavor\footnote{Actually, to be fully clear on this matter, one should evaluate whether the preservation of chiral symmetry by our FQWs is \emph{solely} due to the fact that we do not stagger chirality, or if there is \emph{also} an extra specificity of our FQW schemes that indeed allows chiral symmetry to be preserves, rather than this being the case of \emph{any} flavor-staggering-only scheme that one could think of.}\raisebox{-0.5ex}{'}\footnote{Notice that this flavor-staggering-only technique could also be used in traditional LGT schemes with symmetric finite differences, and it should also remove FD while preserving chiral symmetry, at least if not staggering the chirality is the main (or even the only) source of preservation of chiral symmetry in the lattice scheme.}.\\

\noindent

\subsubsection{Managing to put the weak interaction on the lattice}

{\bfseries Nielsen and Ninomiya's negative conclusions about putting the weak interaction on the lattice.} According to Nielsen and Ninomiya, their no-go theorem implies the following important consequence: \\

\begin{adjustwidth}{1cm}{0cm}
``The most important consequence of our no-go theorem is that \emph{the weak interaction cannot be put on the lattice}.''	 \\
\end{adjustwidth}

\noindent
Let us explain why they arrive to this conclusion. In the theory of the weak interaction, there are only left-handed neutrinos (hypercharge $-1$), as confirmed by experiments: right-handed neutrinos have never been observed\footnote{The weak interaction also requires the existence of antineutrinos which are necessarily right-handed (hypercharge~$+1$), and, similarly to the case of neutrinos, left-handed antineutrinos have never been observed~\cite{Glashow:1961tr,Salam:1968rm}.}. But, when one discretizes the Weyl equation for these left-handed neutrinos with the standard symmetric finite differences, the only ones that actually exist, then the original formulation of the no-go theorem (stated above) tells us that we see appear as many right-handed neutrinos, which contradicts experiments\footnote{This fact is very easy to check computationally.}. \\

\noindent
{\bfseries Putting the weak interaction on the lattice with our flavor-staggering-\emph{only} scheme.} As announced above, having introduced the flavor degrees of freedom  on the lattice as we have done it, results in a richer structure in relation to the no-go theorem, a structure that escapes, (i) not only the standard, loose understanding of this no-go theorem---as shown in Sec.\ \ref{subsubsec:ChiralSymmetryFQW}---but also, as we are going to show it now, (ii) the impossibility of putting Weyl fermions interacting via the weak interaction on the lattice.

Take the $(1+1)$D Dirac FQW for $m=0$. It has two flavors.
Let us consider the local U$(1)$-axial symmetry for hypercharge on this FQW. Let us first remind how one can make it arise from a gauge principle. When we check the evolution operator of the free FQW for $m=0$, we realize that it entangles different flavors but not different chiralities. Hence, like in the continuum, we have a global $\text{U}_L(1) \times \text{U}_R(1)$ symmetry (this is just another, more explicit name for the global U$(1)$-axial symmetry). This global symmetry can then be gauged via the gauge principle, which demands the introduction of a gauge field which is that of the weak interaction. The generic real-space state of such a model is
\par\nobreak
{\small
\begin{equation}
{\psi}^{\nu}(n,k) \ket{n,k} \defeq  {\psi}_{R}^{\textcolor{red}{r}}(n,k)\ket{0, +} \ket{n,k} +  {\psi}_{R}^{\textcolor{blue}{b}}(n,k) \ket{1,+} \ket{n,k} + {\psi}_{L}^{\textcolor{red}{r}}(n,k) \ket{0, -} \ket{n,k} +  {\psi}_{L}^{\textcolor{blue}{b}}(n,k) \ket{1,-} \ket{n,k}  \, .
\end{equation}}
Let us now define the $\text{U}_L(1) \times \text{U}_R(1)$ local-symmetry operator for our flavored model, via its application on the basis ket states. This is an operator that can be written $\hat Q^{f} \defeq \hat Q \otimes \mathbb I_{2}$, where $\mathbb I_{2}$ acts on the flavor Hilbert space, and $\hat Q$ is the usual $\text{U}_L(1) \times \text{U}_R(1)$ local-symmetry operator, that is, it acts on the ``chirality tensor spacetime-position'' basis as $\hat Q \ket{+}\ket{n,k} \equiv e^{i g_R \theta_R(n,k)} \ket{+}\ket{n,k}$ and $
\hat Q \ket{-} \ket{n,k}\equiv e^{i g_L \theta_L(n,k)} \ket{-} \ket{n,k}$, where  $g_R$ and $g_L$ are the right- and left-chirality coupling constants of the interaction, and $\theta_R$ and $\theta_L$ the right- and left-chirality local phases---which are functions (or rather, sequences) of the spacetime position $(n,k)$, and arise---in first quantization---from operators $\theta_R(\hat n,\hat k)$ and $\theta_L(\hat n,\hat k)$, respectively---, so that the action of $\hat Q^f$ on the full-Hilbert-space basis is
\begin{subequations}
\begin{align}
 \hat Q^f \ket{0,+} \ket{n,k} &\equiv  e^{i g_R \theta_R(n,k)}\ket{0,+}\ket{n,k} \\
 \hat Q^f \ket{0,-} \ket{n,k} &\equiv  e^{i g_L \theta_L(n,k)}\ket{0,-} \ket{n,k} \\
 \hat Q^f  \ket{1,+} \ket{n,k} &\equiv  e^{i g_R \theta_R(n,k)}\ket{1,+}\ket{n,k}  \\
 \hat Q^f  \ket{1,-} \ket{n,k} &\equiv  e^{i g_L \theta_L(n,k)}\ket{1,-} \ket{n,k}  \, .
\end{align}
\end{subequations} 

Now, in order to construct a neutrino state, we have to make several particular choices for our generic state. First, we need to pick $g_L=-1$ and $g_R = 0$. This choice is not something that has to do with our flavor model. It is also made in the standard quantum field theory of the weak interaction. The choice $g_R=0$ ensures that, whether right-handed neutrinos exist or not, they can anyways not be detected in experiments, because such hypothetical particles can in principle only interact via the weak interaction; more precisely, choosing $g_R=0$ ensures that no measurement apparatus can detect either such particles directly, or, actually rather---since neutrinos are most often not detectable by direct observation---, byproducts of particle reactions via the weak interaction which would involve such right-handed neutrinos. Hence, since we do not need to describe this experimental non-reality that right-handed neutrinos are, we can then just choose 
\begin{equation}
{\psi}_{R}^{\textcolor{red}{r}}(n=0,k) = 0 \, ,  \end{equation}
which, thanks to the choice $g_R = 0$, implies the previous equation also holds for any $n$ \emph{in many extensions of our model we could think of}---$g_R=0$ turns out to be actually \emph{not} needed if we only want ${\psi}_{R}^{\textcolor{red}{r}}(n,k) = 0$ in the mere weak-interaction model once we choose ${\psi}_{R}^{\textcolor{red}{r}}(n=0,k) = 0$, the reader will understand this below with the treatment that we make of the blue flavor.

The second choice we have to make is the following:
\begin{subequations}
\begin{align}
\psi^{\textcolor{blue}{b}}_R(n=0,k) &= 0 \\
\psi^{\textcolor{blue}{b}}_L(n=0,k) &= 0 \, .
\end{align}
\end{subequations}
Because $g_R=0$, it is totally obvious that $\psi^{\textcolor{blue}{b}}_R(n,k) = 0$ for any $n$, since this is true for the free equation (since $m=0$), which is not modified by the presence of the weak-interaction gauge field because $g_R=0$. Something that is a bit less obvious is that $\psi^{\textcolor{blue}{b}}_L(n,k) = 0 $ for all $n$, because $g_L = - 1 \neq 0$. But, by inspecting the equation of motion of $\psi^{\textcolor{blue}{b}}_L$ resulting from an interaction with the weak-interaction gauge field, one in the end realizes that  $\psi^{\textcolor{blue}{b}}_L(n,k) = 0 $ for all $n$ even if $g_L = - 1 \neq 0$.

Now, again, let us recall that this weak-interaction lattice model for neutrinos is correct because, \emph{thanks to our lattice structure allowing for two flavors} (even if one of them, namely, the blue flavor, is actually not populated), red right-handed neutrinos do not appear, i.e., we bypass Nielsen-Ninomiya's theorem---in its very original form.

What one should add to all this is that, while $g_R=0$ guaranties the non-appearance of right-handed neutrinos from any of the two flavors in \emph{many extensions of our model we could think of}, this is not as much the case for the appearance of the blue-flavor left-handed  neutrino, because $g_L = -1 \neq 0$.

\section{Closing remarks}

{\bfseries Technical summary.}
In this article, we analyze and fix Fermion Doubling (FD) for Quantum Cellular Automata (QCAs) used for the modelling of Quantum Field Theories (QFTs).

The analysis works by writing the equations of motions of the one-step evolution of the QCA for a single particle in Fourier space and analyzing its characteristic polynomial. Due to its $\cos$ and $\sin$ functions, this characteristic polynomial has periodicities, leading to spurious solutions that survive even at arbitrarily small $\epsilon$, and then become ill-defined when $\epsilon=0$. They are therefore unphysical as they do not belong to the continuum theory, but they will remain there and pollute the numerics even when we increase the precision. We also pinpoint the problem at the level of the propagator of the QCA.

The $(1+1)$D and $(3+1)$D QED QCAs of Refs.\ \cite{ABF20,EDMMplus2023} \emph{do} suffer from the FD problem. We show that the number of solutions (spurious or not) that are present in these QCAs are half the number of those of discrete-time naive lattice fermions. If one goes to continuous time and only considers lattice fermion Hamiltonians, then one misses the Brouillin-zone (BZ) locations where the QCA doublers actually appear, which explains why they had remained oblivious to prior analysis~\cite{ABF20}.

We develop a cutting-and-pasting procedure on the BZ, which we formalize as the application of  covering maps upon the complexification of the BZ. The several sheets of new BZ correspond to flavoring several sublattices that partition direct space, respectively. Overall the number of solutions is preserved by this FD-fixing procedure, but the previously unphysical solutions have become physical solutions of different flavors. We call this solution {\em flavor-staggering-only}. We implement this in full details for the $(1+1)$D and $(3+1)$D QED QCAs, resulting in FD-free $ \mathbb{Z}_{2}$ and $ \mathbb{Z}_{2} \cross  \mathbb{Z}_{2} \cross \mathbb{Z}_{2}$ flavored versions of them. We prove that they still enjoy chiral symmetry in the massless situation.

We discuss misconceptions that surround the Nielsen-Ninomiya no-go theorem, and why it does not imply that \emph{all} FD-fixing solutions break chirality---only \emph{some} of them do---, this is for example the case of our flavored model. It does demand, however, that the number of left-handed and right-handed particles per charge be the same---this is the original statement of the theorem. We show how restricting our flavored model to a single flavor enables to simulate the propagation of massless Weyl fermions---which are of course of definite chirality---without making appear opposite-chirality spurious modes, that is, the weak interaction \emph{can} actually be put on the lattice with our flavored model. \\

 {\noindent {\bfseries Perspectives.}}
{The Nielsen-Ninomiya no-go theorem is of great importance for our understanding of the nature of spacetime, in particular discrete spacetime: this theorem hence deserves a modern overhaul. We would also like to better understand the physical interpretation of the flavor symmetries that FD-fixing introduces. To some extent they are physical, as they make sure that the discrete models match the continuum. But could they justify the existence of different types of particle flavors, e.g., in order to recover weak interactions in comprehensive speculative extensions of our flavored model in which the second-flavor fermion is likely to be populated by the dynamics?}

\paragraph{Acknowledgments.} This project/publication was made possible through the support of the ID\# 62312 grant from the John Templeton Foundation, as part of \href{https://www.templeton.org/grant/the-quantum-information-structure-of-spacetime-qiss-second-phase}{``The Quantum Information Structure of Spacetime'' Project (QISS)}. The opinions expressed in this project/publication are those of the author(s) and do not necessarily reflect the views of the John Templeton Foundation. This work has also been funded by the French National Research Agency (ANR), (i) by the project TaQC ANR-22-CE47-0012, and (ii) within the framework of ``Plan France 2030'', under the research projects EPIQ ANR-22-PETQ-0007, OQULUS ANR-23-PETQ-0013, HQI-Acquisition ANR-22-PNCQ-0001 and HQI-R\&D ANR-22-PNCQ-0002.

\bibliographystyle{unsrturl}
\bibliography{biblio}

	\appendix

	\section{Lattice derivatives and Hermicity}\label{sec:apphermicity}

	In this appendix, we are interested in translating Hermicity properties from the continuum to the discrete. The stage is a $(1+1)$D spacetime $\mathcal{M}$, with boundaries. In the continuum, the Dirac classical field we consider is $\psi_{\text{c}} \equiv {\psi_{\text{cont.}}}$, defined by $ \psi_{\text{c}} : \mathcal{M} \to \mathbb{C}^{2}$, with boundary conditions 
	\begin{equation}
		\psi_{\text{c}}(t,x) = 0 \hspace{1mm} \ \ \text{at } \partial \mathcal{M} \, .
	\end{equation}
    In these conditions, the operator $ i \partial_{x} $ is Hermitian. This statement can be proven by the using the inner product defined for Dirac fields, which is the integration, over the space parametrization, of the Lorentz scalar $ \psi_{\text{c}}^{\dagger} {\psi_{\text{c}}} $ (the dagger here only indication the Hermitian conjugate in the \emph{internal} Hilbert space, the Hermitianity of $i\partial_x$ that we are studying here has to do with the \emph{position} Hilbert space). Indeed, recall that
	\begin{subequations}
		\begin{align}
			\int dx \ \psi_{\text{c}}^{\dagger}(t,x) [i \partial_{x} {\psi_{\text{c}}}(t,x)] 
			&=    \int dx \left[ i  \partial_{x}\left(\psi_{\text{c}}^{\dagger}(t,x) {\psi_{\text{c}}}(t,x) \right)- \left(i\partial_{x} \psi_{\text{c}}^{\dagger}(t,x)\right) {\psi_{\text{c}}}(t,x)\right]\\
			&= i \left(\psi_{\text{c}}^{\dagger}(t,x)\psi_{\text{c}}(t,x) \right)|_{\partial  \mathcal{M}} - \int dx \  \left( i \partial_{x}\psi_{\text{c}}^{\dagger}(t,x)\right)\psi_{\text{c}}(t,x) \\ 
			&=  \int dx \  \left(i\partial_{x} \psi_{\text{c}}(t,x)\right)^{\dagger}\psi_{\text{c}}(t,x) \, .
		\end{align}
	\end{subequations}
 
	Consider now the question of how to discretize the Dirac equation, treated in Sec.\ \ref{sec:standardFD}, and in particular the question of which ``lattice derivatives'' to use. We are going to use the boundary conditions
	\begin{equation}
 \label{eq:BC}
		\psi(n_i , k) = \psi(n_f, k) =\psi(n, k_i) = \psi(n,k_f) = 0 \, .
	\end{equation}
    Let us check if the spatial forward finite difference preserves Hermicity. This finite difference is defined as
	\begin{equation}
		d_x \psi(n,k) \defeq \frac{i}{\epsilon} \left ( \psi(n,k+1) - \psi(n,k) \right) \, .
	\end{equation}
	We wish that
\par\nobreak
 {\small
	\begin{equation}
			\epsilon \sum_{ n= n_i,  k = k_i}^{n_f, k_f} \psi^{\dagger}(n,k) \left( \frac{i}{\epsilon} ( \psi(n, k+1) - \psi(n, k))\right)
			=  \epsilon  \sum_{ n= n_i,  k= k_i}^{n_f, k_f}  \left( \frac{i}{\epsilon} ( \psi(n, k+1) - \psi(n, k))\right)^{\dagger} \psi(n,k) \, .
	\end{equation}}
	Let us see if the previous equation holds, or not. We are going to manipulate the left-hand side of the previous equation, and to use the boundary conditions, to try to prove the equality:
	\begin{subequations}
		\begin{align}
			&\epsilon \sum_{ n= n_i,  k = k_i}^{n_f, k_f} \psi^{\dagger}(n,k) \left( \frac{i}{ \epsilon} ( \psi(n, k+1) - \psi(n, k))\right) \\
			 &\ \ \ \   =  i  \sum_{ n= n_i,  k'= k_i+1}^{n_f, k_f+1}  \psi^{\dagger}(n,k'-1) \psi(n, k') -  i  \sum_{ n= n_i,  k = k_i}^{n_f, k_f} \psi^{\dagger}(n,k)\psi(n, k)\\
			&\ \ \ \   =   \sum_{ n= n_i,  k = k_i}^{n_f, k_f} (-i (\psi^{\dagger}(n,k) -\psi^{\dagger}(n,k-1) ) \psi(n, k)  +  i\psi^{\dagger}(n,k_{f}) \psi(n, k_{f} +1) - i\psi^{\dagger}(n,k_{i}-1) \psi(n, k_{i})\\
			&\ \ \ \   =  \epsilon \sum_{ n= n_i,  k = k_i}^{n_f, k_f} (\frac{i}{\epsilon} (\psi(n,k) -\psi(n,k-1) )^{\dagger}\psi(n, k)\\
            &\ \ \ \   \neq  \epsilon \sum_{ n= n_i,  k = k_i}^{n_f, k_f} (\frac{i}{\epsilon} (\psi(n,k+1) -\psi(n,k) )^{\dagger}\psi(n, k) \, ,
		\end{align}
	\end{subequations}
	so we see that Hermicity is broken.

Let us then check if, this time, the symmetric finite differences do preserve Hermiticity. We want
    \par\nobreak
	{\small
	\begin{equation}
			\epsilon \sum_{ n= n_i,  k = k_i}^{n_f, k_f} \psi^{\dagger}(n,k) \left( \frac{i}{2 \epsilon} ( \psi(n, k+1) - \psi(n, k-1))\right) 
			=  \epsilon  \sum_{ n= n_i,  k= k_i}^{n_f, k_f}  \left( \frac{i}{2 \epsilon} ( \psi(n, k+1) - \psi(n, k-1))\right)^{\dagger} \psi(n,k). 
	\end{equation}}
	One can rewrite the left-hand side of the previous equation as, 
\par\nobreak
{\small
	\begin{subequations}
		\begin{align}
&\epsilon \sum_{ n= n_i,  k = k_i}^{n_f, k_f} \psi^{\dagger}(n,k) \left( \frac{i}{2 \epsilon} ( \psi(n, k+1) - \psi(n, k-1))\right) \\
& = \frac{i}{2 }  \sum_{ n= n_i,  k' = k_i+1}^{n_f, k_f+1} \psi^{\dagger}(n,k'-1) \psi(n, k') -  \frac{i}{2 }  \sum_{ n= n_i,  k'' = k_i -1}^{n_f, k_f -1 } \psi^{\dagger}(n,k''+1) \psi(n, k'') \\
& =   - \frac{i}{2 }  \sum_{ n= n_i,  k= k_i}^{n_f, k_f}(  \psi^{\dagger}(n,k+1) -\psi^{\dagger}(n,k-1)) \psi(n, k) \\
&\ \ \ \ \ +  \frac{i}{2 } \psi^{\dagger}(n,k_f) \psi(n, k_f +1)  - \frac{i}{2 }  \psi^{\dagger}(n,k_i) \psi(n, k_i -1)  - \frac{i}{2 } \psi^{\dagger}(n,k_i-1) \psi(n, k_i)  + \frac{i}{2 } \psi^{\dagger}(n,k_f+1) \psi(n, k_f)\\
&=   \epsilon\sum_{ n= n_i,  k= k_i}^{n_f, k_f} \left(  \frac{i}{2\epsilon}  (  \psi(n,k+1) -\psi(n,k-1)) \right)^{\dagger}   \psi(n, k) \, .
		\end{align}    
	\end{subequations}}
	When going from the before-last line to the last line of the previous equations, we used the boundary conditions introduced in Eq.\ \eqref{eq:BC}. With the last line, we can conclude that we have proven that the Hermicity of $ i \partial_{x}$ (and hence that of $-i\partial_x$), is preserved by replacing it with symmetric finite differences. An analog proof can be done for $i \partial_{t}$.

\section{About on-shellness in single-particle quantum mechanics}
    \label{app:on-shellness}

    Interpreting the Dirac Eq.\ \eqref{eq:Dirac} as an equation of motion for the wavefunction of a single quantum-mechanical particle, and using an abstract-Hilbert-space notation for the position space, we can rewrite that equation as
    \begin{equation}
    (i \mathbb{I}_{2} \partial_{t}   - i\sigma_{3} \hat p_{\text{cont.}} - m  \sigma_{1} ) |\psi_{\text{cont.}}(t)\rangle = 0  \, ,
    \end{equation}
    where $\hat p_{\text{cont.}}$ and $|\psi_{\text{cont.}}(t)\rangle$ are the abstract versions of $\mathcal P_{\text{cont.}}$  and ${\psi}_{\text{cont.}}(t,x)$, respectively. Apply now $\langle p |$ on the left of the previous equation, where $\ket p$ is an eigenstate of $\hat p_{\text{cont.}}$ with eigenvalue $p$: this delivers
    \begin{equation}
    \label{eq:momDirac}
    (i \mathbb{I}_{2} \partial_{t}   - i\sigma_{3}  p - m  \sigma_{1} ) {\psi}^{\text{mom.}}_{\text{cont.}}(t,p) = 0 \, ,
    \end{equation}
    where ${\psi}^{\text{mom.}}_{\text{cont.}}(t,p)$ is the usual, momentum Fourier transform of ${\psi}_{\text{cont.}}(t,x)$. The previous equation implies that the eigenelements $\phi_{\text{cont.}}^E(t,p)$ of the energy operator are exactly the eigenelements of the Fourier-space Hamiltonian $H_{D}(p) \defeq i\sigma_{3}  p + m  \sigma_{1} $, in other words, this single-particle quantum-mechanical framework tells us that we must be ``on-shell''. Now, solving the eigenvalue equation associated to the energy operator yields, in momentum space, solutions of the form
    \begin{equation}
    \label{eq:eigenstate}
     \phi^E_{\text{cont.}}(t,p) \defeq A(E,p) \, e^{-iEt} \,   
    \end{equation}
    where $A(E,p)$ is a coefficient that simply coincides with the (two-component) ``value'' of $\phi^E_{\text{cont.}}(t,p)$ at $t=0$, and which may depend on $E$. Inserting the previous solution into Eq.\ \eqref{eq:momDirac}, we finally end up with the following equation,
    \begin{equation}
    \label{eq:AEp_eq}
    (\mathbb{I}_{2} E - i \sigma_3 p - m\sigma_1) A(E,p) = 0 \, .    
    \end{equation}
    Now, take any function ${\psi}^{\text{mom.}}_{\text{cont.}}(t,p)$ and decompose it on the family of eigenstates of the energy operator, which are exactly the eigenstates of the Hamiltonian operator as we said, which we know form a basis of the total Hilbert space (so that we can indeed decompose any such function on that basis), as
    \begin{align}
    \label{eq:decomp}
    {\psi}^{\text{mom.}}_{\text{cont.}}(t,p) = \int^{+\infty}_{-\infty} \frac{dE}{2\pi} \, \tilde{\psi}_{\text{cont.}}(E,p) \, e^{-iEt} \, ,    
    \end{align}
    where the $\tilde{\psi}_{\text{cont.}}(E,p)$s are defined as the coefficients of ${\psi}^{\text{mom.}}(t,p)$ on the energy eigenbasis. Before making our last comment on the present discussion, let us recall that, in principle, if one knows/knew ${\psi}^{\text{mom.}}_{\text{cont.}}(t,p)$, one can/could compute the coefficients of the previous decomposition via
    \begin{equation}
    \tilde{\psi}_{\text{cont.}}(E,p) = \int_{-\infty}^{+\infty} dt \, {\psi}^{\text{mom.}}_{\text{cont.}}(t,p) \, e^{iEt} \, ,
    \end{equation}
    which can/could then be seen/interpreted as a temporal Fourier transform. Coming now back to our initial line of thought, notice that each term $\tilde{\psi}_{\text{cont.}}(E,p) e^{-iEt}$ of the decomposition of Eq.\ \eqref{eq:decomp}, is of the form of Eq.\ \eqref{eq:eigenstate}, i.e., is an eigenstate of the energy operator. As such, Eq.\ \eqref{eq:AEp_eq} holds for the coefficients $\tilde{\psi}_{\text{cont.}}(E,p)$, that is, 
    \begin{equation}
    (\mathbb{I}_{2} E - i \sigma_3 p - m\sigma_1) \tilde{\psi}_{\text{cont.}}(E,p) = 0  \, ,
    \end{equation}
    which is the final equation we wanted to arrive to.

\section{Fourier transformation of the naive discretization scheme} \label{sec:ftofnaive}

The Dirac field $\psi(n,k)$ that we introduced in Sec.\ \ref{sec:standardFD}, satisfies the following normalization condition,
\begin{equation}
    \epsilon^{2} \sum_{n = - \infty}^{\infty} \sum_{k = - \infty}^{\infty}\psi(n,k) \psi^{\dagger}(n,k)= 1 \, .
\end{equation}
Hence, we conclude that the dimension of $ \psi(n,k)$ is $ \epsilon^{-1}$. 

The following consistency relations holds:
\begin{subequations}
\begin{align}
\psi(n,k) &= \epsilon \int_{- \pi/\epsilon}^{\pi/\epsilon}  \frac{dE}{2 \pi } \int_{- \pi/\epsilon}^{\pi/\epsilon}  \frac{dp}{2 \pi } \hspace{1mm} \tilde{\psi}(E,p) e^{-iEn \epsilon +ipk\epsilon}\\
  &=    \epsilon^2 \sum_{ n= - \infty}^{\infty} \sum_{ k = - \infty}^{\infty}\hspace{1mm}   \int_{- \pi/\epsilon}^{\pi/\epsilon}  \frac{dE}{2 \pi } \int_{- \pi/\epsilon}^{\pi/\epsilon}  \frac{dp}{2 \pi }  \psi(n,k)  e^{-iE(n-n') \epsilon + ip(k-k')\epsilon}\\
      &=  \sum_{ n= - \infty}^{\infty} \sum_{ k = - \infty}^{\infty}\hspace{1mm} \psi(n',k') \delta_{n,n'} \delta_{k,k'} \\
     &=  \psi(n,k).
\end{align}
\end{subequations}

From the Fourier-integral form of $\psi(n,k) $, Eq.\ \eqref{eq:ansatz}, we can obtain  Eq.\ \eqref{eq:MotionFourier}, here is the full calculation:
 
\begin{subequations}
\begin{align}
    	&\frac{i \mathbb{I}_{2}}{2 \epsilon}(\psi(n+1,k) - \psi(n-1,k) ) + \frac{ \sigma_{3}}{2 \epsilon}(\psi(n,k+1) - \psi(n,k-1) ) - m \sigma_{1} \psi(n,k)\\
 & \ \ \ \ \ = \epsilon \int_{- \pi/\epsilon}^{\pi/\epsilon}  \frac{dE}{2 \pi } \int_{- \pi/\epsilon}^{\pi/\epsilon}  \frac{dp}{2 \pi } \hspace{1mm} \left(
 \frac{i \mathbb{I}_{2}}{2 \epsilon}( e^{-iE(n+1) \epsilon + ipk\epsilon} -   e^{-iE(n-1) \epsilon  + ipk\epsilon }) \tilde{\psi}(E,p) \right.\\ & \ \ \ \ \ \ \ \ \ \  \ \ \ \ \ \ \ \ \ \ \left.
 + \frac{ \sigma_{3}}{2 \epsilon}( e^{-iEn \epsilon + ip(k+1)\epsilon} -   e^{-iEn \epsilon + ip(k-1)\epsilon}) \tilde{\psi}(E,p) - m \sigma_{1} e^{-iEn \epsilon + ipk\epsilon} \tilde{\psi}(E,p) \right) \\
 & \ \ \ \ \ = \epsilon \int_{- \pi/\epsilon}^{\pi/\epsilon}  \frac{dE}{2 \pi } \int_{- \pi/\epsilon}^{\pi/\epsilon}  \frac{dp}{2 \pi } \hspace{1mm} 
 \left(\frac{i \mathbb{I}_{2}}{2 \epsilon} ( e^{-iE \epsilon} - e^{iE \epsilon}  ) + \frac{i  \sigma_{3}}{2 \epsilon} (e^{ip \epsilon} - e^{-ip \epsilon}  )  - m \sigma_{1}  \right) e^{-iEn \epsilon + ipk\epsilon} \tilde{\psi}(E,p) \\ 
 & \ \ \ \ \ = \int_{- \pi/\epsilon}^{\pi/\epsilon}  \frac{dE}{2 \pi } \int_{- \pi/\epsilon}^{\pi/\epsilon}  \frac{dp}{2 \pi } \hspace{1mm}  \left(\mathbb{I}_{2} \sin(E \epsilon) -  \sigma_{3}\sin(p \epsilon)  - m \epsilon \sigma_{1}  \right)\tilde{\psi}(E,p) e^{-iEn \epsilon + ipk\epsilon} \label{eq:last}\\  & \ \ \ \ \ =0 \, .
 \end{align}
\end{subequations}
From the very last equation, which holds simply because the very first expression of the list of equations is equal to zero (that is the equation of motion), we conclude that the integrand of the last expression (Eq.\ \eqref{eq:last}), has to vanish:
\begin{equation}
  (  \mathbb{I}_{2} \sin(E \epsilon) -  \sigma_{3}\sin(p \epsilon)  - m \epsilon \sigma_{1})\tilde{\psi}(E,p) = 0 \, ,
\end{equation}
which concludes our proof.

\section{Continuum limit of the $(1+1)$D flavored Dirac quantum walk}
\label{app:cont_lim_DQW}

\subsection{Introduction}

\subsubsection{Presentation of the original QW}

{\bf State.} The state of the $(1+1)$D Dirac QW  at time $ n$ is
\begin{equation} 
    \ket{\psi(n)}\equiv \sum_{k} \psi^{+}(n,k) \ket{k}_{+}  + \psi^{-}(n,k) \ket{k}_{-},
\end{equation}
where $ \ket{k}$ is  a position eigenstate, $\ket{+}$ and $\ket{-}$ are the chirality eigenstates, and $ \ket{k}_{\pm} \defeq \ket{k} \otimes \ket{ \pm}$.

\noindent
{\bf Evolution operator.} The one-step evolution operator of the $(1+1)$D Dirac QW, which is unitary, is 
{\small
\begin{equation} 
    \hat U \defeq \sum_{k} \cos(m \epsilon) \left( \ket{k-1}_{-} \bra{k}_{-}+ \ket{k+1}_{+} \bra{k}_{+} \right) - i \sin(m \epsilon) \left( \ket{k}_{-} \bra{k}_{+}+ \ket{k}_{+} \bra{k}_{-}\right). 
\end{equation}}

\noindent
{\bf Spatial shift operator.}
We define the spatial shift/translation operator, or translation operator in space, and give its Hermitian conjugate (which immediately follows),
\begin{subequations}
\begin{align}
      \hat S &\defeq  \sum_{k} \ket{k+1} \bra{k} \label{eq:space_trans}\\
     \hat S^{\dagger} &=  \sum_{k} \ket{k-1} \bra{k}.
\end{align}
\end{subequations}
The spatial shift operator acts on the QW state as
\begin{subequations}
   \begin{align}
         \hat S \ket{\psi(n)} &= \sum_{k} \psi^{+}(n,k) \ket{k+1}_{+}  + \psi^{-}(n,k) \ket{k+1}_{-}\\ 
        &=  \sum_{k'} \psi^{+}(n,k'-1) \ket{k'}_{+} + \psi^{-}(n,k'-1) \ket{k'}_{-} \, .
   \end{align}
\end{subequations}

We see that even though $ \hat S$ is not an operator which acts on the wavefunctions $ \psi^{\pm}(n,k)$ directly, it nevertheless induces an action on them. Indeed, this induced action is naturally defined via
\begin{equation}
(\mathcal S \psi^\pm)(n,k) \defeq \bra{k}_{\pm} \hat S \ket{\psi(n)} \, ,
\end{equation}
which finally yields
\begin{subequations}
\begin{align}
    (\mathcal S \psi^{\pm})(n,k) &= \psi^{\pm}(n,k-1) \label{eq:Scal}\\
    (\mathcal S^{\dagger} \psi^{\pm})(n,k) &= \psi^{\pm}(n,k+1) \, . 
\end{align}
\end{subequations}

\noindent
{\bf Temporal shift operator.}
We also define the temporal shift/translation operator, or translation operator in time, via Eq.\ \eqref{eq:time_trans}, having first introduced a time Hilbert space $\mathcal H_{\text{time}}$, so that we finally  have
\begin{equation}
    \bra n \hat {T}^\dag \ket{\psi} \defeq \ket{\psi(n+1)},
\end{equation}
and its induced action on the wavefunction is naturally defined by
\begin{equation}
    (\mathcal T^\dag \psi^\pm)(n,k) \defeq (\bra n \otimes \bra{k}_{\pm}) \hat{T}^\dag \ket{\psi} \, ,
\end{equation}
which yields
\begin{equation}
    (\mathcal T^\dag \psi^\pm)(n,k) = \psi^{\pm}(n+1,k) \, .
\end{equation}

\subsubsection{Presentation of our FQW}

\noindent
{\bf \emph{State.}} The state of our FQW at time $n$ is
\begin{equation} 
\label{eq:FQWstate}
    \ket{\chi_{i}(n)} \defeq \sum_{k} \left(\chi^{+}(n,k) \ket{k}_{+}  + \chi^{-}(n,k) \ket{k}_{-}\right) \otimes \sigma_{x}^{n+k} \ket{i} \, ,
\end{equation}
where $i=0$ or $1$, and so where we have introduced two flavor states, $ \ket{0}$ and~$\ket{1}$. From now on, we will choose $i=0$.

\noindent
{\bf \emph{Evolution operator.}} The evolution operator of the FQW is given by Eq.\ \eqref{eq:UFQW}:
\par\nobreak
\begin{equation} 
\label{eq:evol_op}
    \hat U^{1}_{FQW} \defeq 
        \begin{bmatrix}
				\cos(m \epsilon) (\hat{S} \otimes \sigma_x) &  - i \sin(m \epsilon)\\
				- i \sin(m \epsilon) & \cos(m \epsilon) (\hat{S}^{\dagger} \otimes \sigma_x)
			\end{bmatrix}  \, . 
\end{equation}%

\noindent
{\bf \emph{Equation of motion of the FQW.}} The equation of motion of the FQW is
\begin{equation}
(\hat T^\dagger \otimes \mathbb I_2 \otimes \sigma_x) \ket{\psi^f(n)} =  U^{1}_{FQW} \ket{\psi^f(n)} \, ,
\end{equation}
where the $\mathbb I_2$ of the previous operator acts on the chirality space, and where we have chosen to write this equation of motion for a fully generic state $ \ket{\psi^f(n)}$ on the lattice, rather than the particular one we shall choose in the end, and that we have already written above, Eq.\ \eqref{eq:FQWstate}. This is because this has helped us to carry out the computations we needed to in order to provide the proof of this appendix.

\subsection{Relevant set of equations of motions for the continuum limit to be taken}

Before taking the continuum limit, we must find the right set of equations of motion of which we are going to take the continuum limit. The first thing we do is to work with wavefunctions rather than states on the position Hilbert space, because we want to be able to pack together equations of motion for wavefunctions taken at different spacetime points. This is indeed necessary to derive a proper continuum limit, as the reader may guess, since the red and the blue sublattices do not contain the same wavefunctions, but both are necessary in our model, so we have to pack what happens on one of them to what happens on the other. \\

\noindent
Hence, we will work with $\mathcal T$, $\mathcal S$, and $\psi^f(n,k)$, rather than with $\hat T$, $\hat S$, and $\ket{\psi^f(n)}$. We will also ``write a $\hat T$'' on the right side of the previous equations, rather than ``having $\hat T^\dagger$'' on the left, because this has helped us to visualize better the equations of motion on the two different sublattices. Combining all this, we can rewrite the previous equation as 
\begin{equation}
\psi^f(n,k) =  \left[ (\mathcal T \otimes \mathbb I_2 \otimes \sigma_x) \, \mathcal U^1_{FQW} \, \psi^f \right](n,k) \, ,   
\end{equation}
where $\mathcal U^1_{FQW}$ is exactly as $\hat U^1_{FQW}$ but in which we have replaced $\hat S$ by $\mathcal S$.

After a somewhat long but straightforward computation, the previous equation finally leads to
\begin{equation}
\begin{pmatrix}
    \psi^r_+(n,k) \\
    \psi^b_+(n,k) \\
    \psi^r_-(n,k) \\
    \psi^b_-(n,k)
\end{pmatrix} =
\begin{pmatrix}
    c \, \psi^r_+(n-1,k-1) - is \, \psi^b_-(n-1,k) \\
    c \, \psi^b_+(n-1,k-1) - is \, \psi^r_-(n-1,k) \\
    c \, \psi^r_-(n-1,k+1) - is \, \psi^b_+(n-1,k) \\
    c \, \psi^b_-(n-1,k+1) - is \, \psi^r_+(n-1,k)
\end{pmatrix} \, ,
\end{equation}
where $c \defeq \cos(\epsilon m)$ and $s \defeq \sin(\epsilon m)$. \\

\noindent
Starting from now, we recommend the reader to visualize what is going on on a spacetime lattice. We now have to take into account that, in our model, the previous equation never holds exactly as written, since our model only makes sense with a particular form choice for the wavefunction, where the sites with $n+k$ even are occupied only by a red-flavor wavefunction, and the sites with $n+k$ odd are occupied only by a blue-flavor wavefunction.

After a somewhat long but straightforward computation, with some visual help as mentioned before, we finally arrive to the actual set of equations that is satisfied in our model, by combining what happens on the red lattice with what happens on the blue lattice, and by considering $n+k$ even:
\begin{equation}
\label{eq:FFFinal}
\left(
\begin{array}{l}
    \psi^r_+(n,k) \\
    \psi^b_+(n,k+1) \\
    \psi^r_-(n,k) \\
    \psi^b_-(n,k+1)
\end{array} \right) =
\left(
\begin{array}{ll}
    c \, \psi^r_+(n-1,k-1) & \hspace{-0.2cm} - \, \, is \, \psi^b_-(n-1,k) \\
    c \, \psi^b_+(n-1,k) & \hspace{-0.2cm} - \, \, is \, \psi^r_-(n-1,k+1) \\
    c \, \psi^r_-(n-1,k+1) & \hspace{-0.2cm} - \, \, is \, \psi^b_+(n-1,k) \\
    c \, \psi^b_-(n-1,k+2) & \hspace{-0.2cm} - \, \, is \, \psi^r_+(n-1,k+1)
\end{array}  \right) \, .
\end{equation}
The reader will surely notice that this equation is nothing but the previous one but having replaced $k$ by $k+1$ in the second and fourth lines of the column vectors.

\subsection{Continuum limit}

\subsubsection{New coordinates}

We would like to find the continuum limit of Eq.\ \eqref{eq:FFFinal}. However, due to the rombus structure of the sublattices, this cannot be done without performing a $45°$ rotation and $\sqrt{2}$ dilatation of the coordinate system, so as to use coordinates that treat the rhombus lattice as a rotated square lattice. We thus define the following new coordinates,
\begin{subequations}
\begin{align}
    r &\defeq n-k \\
    \rho &\defeq n+k \, ,
\end{align}
\end{subequations}
which are respectively the new time ($r$) and space ($\rho$) coordinates, and that can be inverted into 
\begin{subequations}
\begin{align}
    n &\defeq \frac{\rho+r}{2} \\
    k &\defeq \frac{\rho-r}{2} \, .
\end{align}
\end{subequations}
Note by looking at the two previous equations that, since $n$ and $k$ are integers, $r$ and $\rho$ have to be either odd or even together, so that their sum and difference are even and thus dividable by $2$. As announced, in these new coordinates (and with the constraint that $r$ and $\rho$ have to be either odd or even together), the lattice is naturally viewed as two sublattices, which are square if we rotate the view by 45° with respect to the original lattice; one of these two sublattices corresponds to $r$ and $\rho$ even together, and the other to $r$ and $\rho$ odd together. Each of these two square sublattices have lattice spacing $ \epsilon' \defeq \sqrt{2} \epsilon$. 

\subsubsection{Equations of motion in the new coordinate system}

By inserting the previous two expressions of $n$ and $k$ as functions of $r$ and $\rho$ into Eq.\ \eqref{eq:FFFinal}, we obtain
\begin{equation}
\label{eq: NEWEQ}
\left(
\begin{array}{l}
    \psi^r_+(\frac{\rho+r}{2},\frac{\rho-r}{2}) \\
    \psi^b_+(\frac{\rho+r}{2},\frac{\rho-r}{2}+1) \\
    \psi^r_-(\frac{\rho+r}{2},\frac{\rho-r}{2}) \\
    \psi^b_-(\frac{\rho+r}{2},\frac{\rho-r}{2}+1)
\end{array} \right) =
\left(
\begin{array}{ll}
    c \, \psi^r_+(\frac{\rho+r}{2}-1,\frac{\rho-r}{2}-1) & \hspace{-0.2cm} - \, \, is \, \psi^b_-(\frac{\rho+r}{2}-1,\frac{\rho-r}{2}) \\
    c \, \psi^b_+(\frac{\rho+r}{2}-1,\frac{\rho-r}{2}) & \hspace{-0.2cm} - \, \, is \, \psi^r_-(\frac{\rho+r}{2}-1,\frac{\rho-r}{2}+1) \\
    c \, \psi^r_-(\frac{\rho+r}{2}-1,\frac{\rho-r}{2}+1) & \hspace{-0.2cm} - \, \, is \, \psi^b_+(\frac{\rho+r}{2}-1,\frac{\rho-r}{2}) \\
    c \, \psi^b_-(\frac{\rho+r}{2}-1,\frac{\rho-r}{2}+2) & \hspace{-0.2cm} - \, \, is \, \psi^r_+(\frac{\rho+r}{2}-1,\frac{\rho-r}{2}+1)
\end{array}  \right) \, .
\end{equation}
The reader should not confuse the superscript $r$, which indicates the red flavor, with the coordinate $r$, which is the new time coordinate. 

We will now group the red- and blue-flavored wavefunction images of the left-hand side of the previous equation all at the same spacetime-lattice site, by introducing the following new wavefunctions,
\begin{subequations}
\begin{align}
{\tilde  \psi^r}_{\pm}(r,\rho) &\defeq \psi_\pm^r(\tfrac{\rho+r}{2},\tfrac{\rho-r}{2}) \\
{\tilde  \psi^b}_{\pm}(r,\rho) &\defeq \psi_\pm^b(\tfrac{(\rho+1)+(r-1)}{2},\tfrac{(\rho+1)-(r-1)}{2}) \, .
\end{align}
\end{subequations}
The ``tilde'' should not be confused with a Fourier transform, which it does not represent here, it is just a notation we use for the new wavefunction, which is going to be very temporary anyways.
By using the two previous equations, we can rewrite Eq.\ \eqref{eq: NEWEQ} as
\begin{equation}
\label{eq: NEWEQ2}
\left(
\begin{array}{l}
    {\tilde \psi^r}_{+}(r,\rho) \\
    {\tilde  \psi^b}_{+}(r,\rho) \\
     {\tilde  \psi^r}_{-}(r,\rho) \\
    {\tilde  \psi^b}_{-}(r,\rho)
\end{array} \right) =
\left(
\begin{array}{ll}
    c \, {\tilde  \psi^r}_{+}(r,\rho-2) & \hspace{-0.2cm} - \, \, is \, {\tilde  \psi^b}_{-}(r,\rho-2) \\
    c \, {\tilde  \psi^b}_{+}(r,\rho-2) & \hspace{-0.2cm} - \, \, is \, {\tilde  \psi^r}_{-}(r-2,\rho) \\
    c \, {\tilde  \psi^r}_{-}(r-2,\rho) & \hspace{-0.2cm} - \, \, is \, {\tilde  \psi^b}_{+}(r, \rho-2,) \\
      c \, {\tilde  \psi^b}_{-}(r-2,\rho) & \hspace{-0.2cm} - \, \, is \, {\tilde  \psi^r}_{+}(r-2,\rho)
\end{array}  \right) \, .
\end{equation}

We see that the walkers only explore the new coordinate system every two sites, and this is the one that corresponds to the actual sublattices structure without introducing extra sites, so we introduce $r'$ and $\rho'$ so that
\begin{subequations}
\begin{align}
r &\equiv2\rho'  \\
 \rho &\equiv 2r'  \, ,
\end{align}
\end{subequations}
by having assumed that we chose for the previous grouping the sublattice where $r$ and $\rho$ were even together---which is more natural since the origin can then indeed be $(r=0,\rho=0)$; we also introduce the following new wavefunctions,
\begin{subequations}
\begin{align}
    {\psi^r}{\raisebox{0.7ex}{\hspace{0.08mm}$'$}}_{\! \! \! \! \! \pm}(r',\rho') &\defeq {\tilde  \psi^r}_{\pm}(2r',2\rho') \, \\
    {\psi^b}{\raisebox{0.7ex}{\hspace{0.08mm}$'$}}_{\! \! \! \! \! \pm}(r',\rho') &\defeq {\tilde  \psi^b}_{\pm}(2r',2\rho') \, . 
\end{align}
\end{subequations}
Using the two previous equations, we can rewrite \eqref{eq: NEWEQ2} as
\begin{equation}
\label{eq: NEWEQ3}
\left(
\begin{array}{l}
    {\psi^r}{\raisebox{0.7ex}{\hspace{0.08mm}$'$}}_{\! \! \! \! \! +}(r',\rho') \\
    {\psi^b}{\raisebox{0.7ex}{\hspace{0.08mm}$'$}}_{\! \! \! \! \! +}(r',\rho') \\
     {\psi^r}{\raisebox{0.7ex}{\hspace{0.08mm}$'$}}_{\! \! \! \! \! -}(r',\rho') \\
    {\psi^b}{\raisebox{0.7ex}{\hspace{0.08mm}$'$}}_{\! \! \! \! \! -}(r',\rho')
\end{array} \right) =
\left(
\begin{array}{ll}
    c \, {\psi^r}{\raisebox{0.7ex}{\hspace{0.08mm}$'$}}_{\! \! \! \! \! +}(r',\rho'-1) & \hspace{-0.2cm} - \, \, is \, {\psi^b}{\raisebox{0.7ex}{\hspace{0.08mm}$'$}}_{\! \! \! \! \! -}(r',\rho'-1) \\
    c \, {\psi^b}{\raisebox{0.7ex}{\hspace{0.08mm}$'$}}_{\! \! \! \! \! +}(r',\rho'-1) & \hspace{-0.2cm} - \, \, is \, {\psi^r}{\raisebox{0.7ex}{\hspace{0.08mm}$'$}}_{\! \! \! \! \! -}(r'-1,\rho') \\
    c \, {\psi^r}{\raisebox{0.7ex}{\hspace{0.08mm}$'$}}_{\! \! \! \! \! -}(r'-1,\rho') & \hspace{-0.2cm} - \, \, is \, {\psi^b}{\raisebox{0.7ex}{\hspace{0.08mm}$'$}}_{\! \! \! \! \! +}(r',\rho'-1) \\
      c \, {\psi^b}{\raisebox{0.7ex}{\hspace{0.08mm}$'$}}_{\! \! \! \! \! -}(r'-1,\rho') & \hspace{-0.2cm} - \, \, is \, {\psi^r}{\raisebox{0.7ex}{\hspace{0.08mm}$'$}}_{\! \! \! \! \! +}(r'-1,\rho')
\end{array}  \right) \, .
\end{equation}

\subsubsection{Plain continuum limit}

We use the notations
\begin{subequations}
\begin{align}
t' &\defeq r' \epsilon' \\
x' &\defeq \rho' \epsilon' \, ,
\end{align}    
\end{subequations}
as well as
\begin{subequations}
\begin{align}
\phi^i{\raisebox{0.7ex}{\hspace{0.08mm}$'$}}_{\! \! \! \! \!  \pm}(t',x') \defeq \psi^i{\raisebox{0.7ex}{\hspace{0.08mm}$'$}}_{\! \! \! \! \!  \pm}(r',\rho') \, ,
\end{align}  
\end{subequations}
where $i=r,b$. With the previous notations, Eq.\ \eqref{eq: NEWEQ3} reads
\begin{equation}
\label{eq: NEWEQ3}
\left(
\begin{array}{l}
    {\phi^r}{\raisebox{0.7ex}{\hspace{0.08mm}$'$}}_{\! \! \! \! \! +}(t',x') \\
    {\phi^b}{\raisebox{0.7ex}{\hspace{0.08mm}$'$}}_{\! \! \! \! \! +}(t',x') \\
     {\phi^r}{\raisebox{0.7ex}{\hspace{0.08mm}$'$}}_{\! \! \! \! \! -}(t',x') \\
    {\phi^b}{\raisebox{0.7ex}{\hspace{0.08mm}$'$}}_{\! \! \! \! \! -}(t',x')
\end{array} \right) =
\left(
\begin{array}{ll}
    c \, {\phi^r}{\raisebox{0.7ex}{\hspace{0.08mm}$'$}}_{\! \! \! \! \! +}(t',x'-\epsilon') & \hspace{-0.2cm} - \, \, is \, {\phi^b}{\raisebox{0.7ex}{\hspace{0.08mm}$'$}}_{\! \! \! \! \! -}(t',x'-\epsilon') \\
    c \, {\phi^b}{\raisebox{0.7ex}{\hspace{0.08mm}$'$}}_{\! \! \! \! \! +}(t',x'-\epsilon') & \hspace{-0.2cm} - \, \, is \, {\phi^r}{\raisebox{0.7ex}{\hspace{0.08mm}$'$}}_{\! \! \! \! \! -}(t'-\epsilon',x') \\
    c \, {\phi^r}{\raisebox{0.7ex}{\hspace{0.08mm}$'$}}_{\! \! \! \! \! -}(t'-\epsilon',x') & \hspace{-0.2cm} - \, \, is \, {\phi^b}{\raisebox{0.7ex}{\hspace{0.08mm}$'$}}_{\! \! \! \! \! +}(t',x'-\epsilon') \\
      c \, {\phi^b}{\raisebox{0.7ex}{\hspace{0.08mm}$'$}}_{\! \! \! \! \! -}(t'-\epsilon',x') & \hspace{-0.2cm} - \, \, is \, {\phi^r}{\raisebox{0.7ex}{\hspace{0.08mm}$'$}}_{\! \! \! \! \! +}(t'-\epsilon',x')
\end{array}  \right) \, .
\end{equation}

We now Taylor expand the previous equation in $\epsilon'$ around $(t',x')$. The reader can check that the zeroth-order terms of the left- and right-hand sides cancel out, so that we can devide the whole by $\epsilon'$, and then take $\epsilon'=0$, which delivers the continuum limit equations in the rotated frame. After that, we go back to the original frame by using
\begin{subequations}
\begin{align}
i\partial_t' &\equiv \frac{1}{\sqrt 2} (i\partial_t - i \partial_x) \\
i\partial_x' &\equiv \frac{1}{\sqrt 2} (-i\partial_t - i \partial_x) \, ,
\end{align}
\end{subequations}
which of course demands to switch from $\phi^i{\raisebox{0.7ex}{\hspace{0.08mm}$'$}}_{\! \! \! \! \!  \pm}(t',x')$ to wavefunctions that we will denote by $\phi^i_{\pm}(t,x)$. The reader can check that this in the end delivers
\begin{subequations}
\begin{align}
i \partial_t \phi_+^r &= - (-\partial_x) \phi_+^r + m \phi_-^b \\
i \partial_t \phi_+^b &= - (-\partial_x) \phi_+^b + m \phi_-^r \\
i \partial_t \phi_-^r &= (-\partial_x) \phi_-^r + m \phi_+^b \\
i \partial_t \phi_-^b &= (-\partial_x) \phi_-^b + m \phi_+^r \, .
\end{align}    
\end{subequations}

\subsubsection{Rotation in the ``chirality tensor flavor'' space}

The previous equations are not yet what we were looking for initially, since they merge the two flavors, that is, they do not correspond to two versions of the same Dirac equation, one for each flavor. In order to achieve this last step, we must try linear combinations of the previous equations. After some trials, one realizes that the following linear combinations seem to work: $\tfrac{1}{\sqrt{2}}[(\mathrm{c})-i(\mathrm{d})]$, $\tfrac{1}{\sqrt{2}}[-i(\mathrm{c})+(\mathrm{d})]$, $\tfrac{1}{\sqrt{2}}[(\mathrm{a})-i(\mathrm{b})]$, $\tfrac{1}{\sqrt{2}}[-i(\mathrm{a})+(\mathrm{b})]$. They leads us to define new wavefunctions
\begin{subequations}
\begin{align}
 \phi^{(i)}_{\rightarrow} &\defeq \frac{1}{\sqrt{2}}(\phi^r_- - i \phi^b_-) \\
\phi^{(i')}_{\rightarrow} &\defeq \frac{1}{\sqrt{2}}(- i \phi^r_- + i \phi^b_-) \\
\phi^{(j)}_{\leftarrow} &\defeq \frac{1}{\sqrt{2}}(\phi^r_+ - i \phi^b_+) \\
\phi^{(j')}_{\leftarrow} &\defeq \frac{1}{\sqrt{2}}(- i \phi^r_+ + i \phi^b_+) \, ,  
\end{align}
\end{subequations}
and to finally choose $i = i'=0$,a nd $j=j'=1$, as the new flavor basis indices.

The reader can check that this means that we have performed the following rotation in the ``chirality tensor flavor'' basis:
\begin{equation}
\label{eq:rotation}
[\phi_d^{(i)}] \defeq (\sigma_x \otimes M) [\phi_c^f] \, ,    
\end{equation}
where  $f=r,b$ (``$f$'' for ``flavor'') and $i=0,1$ are the flavor indices in the original and in the new flavor basis, respectively, while $c=+,-$ (``$c$'' for ``chirality'') and $d=\rightarrow,\leftarrow$ (``$d$'' for ``direction'') are the chriality indices in the original and in the new chirality basis, respectively. Also, in the previous equation, we have introduced the flavor rotation matrix
\begin{equation}
M \defeq \frac{1}{\sqrt{2}} 
\begin{bmatrix}
    1 & -i \\
    -i & 1
\end{bmatrix} = e^{-i(\pi/4)\sigma_x} \, ,
\end{equation}
so this seems to be just a spinor representation, in flavor space, of the $-\pi/4$ rotation we have to perform in spacetime. Let us expand Eq.\ \eqref{eq:rotation} to give an explicit picture:
\begin{equation}
\label{eq:rotation}
\left( \begin{array}{l}
   \phi^{(0)}_\rightarrow  \\
   \phi^{(1)}_\rightarrow  \\
   \phi^{(0)}_\leftarrow  \\
   \phi^{(1)}_\leftarrow 
\end{array} \right)  = \frac{1}{\sqrt{2}}
\begin{bmatrix}
  0 & 0 & 1 & -i \\
  0 & 0 & -i & 1 \\
  1 & -i & 0 & 0 \\
  -i & 1 & 0 & 0
\end{bmatrix}
\left( \begin{array}{l}
   \phi^{r}_+ \\
   \phi^{b}_+  \\
   \phi^{r}_-  \\
   \phi^{b}_-
\end{array} \right) \, .   
\end{equation}

The reader can check that, after this rotation, we finally obtain what we were looking for, that is,
\begin{equation}
		i \mathbb{I}_{2} \partial_{t} \phi^{(i)}  = [\sigma_{3} (-i\partial_{x}) + m  \sigma_{1}] \phi^{(i)} \, ,
\end{equation}
where $i=0,1$, and where $\phi^{(i)}\defeq [\phi^{(i)}_{\rightarrow},\phi^{(i)}_{\leftarrow}]^\top$, where $\top$ denotes the transposition, which concludes our continuum-limit proof.

Up to the exchange of the chiralities via $\sigma_x$ above---about which some thought should be given{\mbox~---,} the new internal-degree-of-freedoms basis thus seems to correspond to having just rotated the flavor basis in agreement with the spacetime rotation needed to go back to the original frame.

\section{From Direct to Fourier lattice vectors}\label{sec:applattice}

In this article, we are interested in the BZ of two-dimensional and  four-dimensional direct spacetime lattices, which in particular are the integration domains of the Fourier expansion of functions over these lattices. The fact that the lattice describes spacetime and not just space will not play any role (because in this article we do not address the question of the metric of the spacetime lattices), hence, from now on in this Appendix we will use the word ``space'' for what is used as a spacetime. We will also sometimes even remove the word ``space'' whenever there is already the adjective ``direct'', since otherwise it would be a redundancy. We will also sometimes remove the word ``direct'', whenever not necessary.

A $2$D lattice can be embedded into a $3$D lattice. The $3$D lattice is constructed by selecting a discrete, additive subset of  $ \mathbb{R}^3$, such that any vector in  the 3D lattice  $\Lambda^{3} \subset \mathbb{R}^3 $ can be written
	\begin{equation}\label{eq:latticevector}
		 \vec v = n_1 \Vec{a}_1 +n_2 \Vec{a}_2 +n_3 \Vec{a}_3 \, ,
	\end{equation}
	where $ n_1 , n_2 , n_3 \in  \mathbb{Z} $ and $ \Vec{a}_1 , \Vec{a}_2 , \Vec{a}_3 $ are vectors in $ \mathbb{R}^3$ , called lattice vectors. In this article, we have translational invariance of the considered discrete schemes because these are the free-field schemes, i.e., it is at this free-fields level that we work to treat the FD problem---of course, the translational invariance of the free schemes is a consequence of the uniformity and isotropy of space, i.e., invariance of space under a translation (homogeneity) in any possible direction (isotropy). This translation invariance of the direct lattice creates a boundary on the Fourier space since momentum is defined to be the generator of the translation invariance.

 We define reciprocal lattice vectors, that can be seen as lattice vectors in Fourier space, as follows:
\begin{subequations}
\begin{align}
\Vec{k}_1 &\defeq 2 \pi \frac{\vec a_2 \wedge \vec a_3}{ \vec a_1 \cdot (\vec a_2 \wedge \vec a_3)}\\
\Vec{k}_2 &\defeq  2 \pi \frac{\vec a_3 \wedge \vec a_1}{ \vec a_1 \cdot (\vec a_2 \wedge \vec a_3)}\\
\Vec{k}_3 &\defeq  2 \pi \frac{\vec a_1 \wedge \vec a_2}{ \vec a_1 \cdot (\vec a_2 \wedge \vec a_3)} \, ,
\end{align}
\end{subequations}
	where $ \wedge$ is the usual exterior product on $ \mathbb{R}^3$ (in the case of $ \mathbb{R}^3$, it is also called ``vector product''). From all these definitions, one has $ \Vec{a}_i \cdot \Vec{k}_j = 2 \pi \delta_{ij}$. Note that $ -\Vec{k}_1,   \Vec{k}_2 ,   \Vec{k}_3  $, are also possible reciprocal lattice vectors. 

The  BZ is then defined by the Bragg equation~\cite{Kittel2004}, which can  actually  be written as 
	\begin{equation}\label{eq:bordereq}
		2 \Vec{p} \cdot \Vec{k} = \vec k\cdot \Vec{k}
	\end{equation}
	where $ \Vec{k} \defeq  n_{1} \Vec{k}_1  +  n_{2} \Vec{k}_2 +  n_{3}\Vec{k}_3$, $ n_1,n_2,n_3 \in \mathbb{Z} , |n_1+n_2+n_3| \leq 1$,---these $n_i$ are arbitrary, have nothing to to with the previous ones used for the direct lattice---, and $ \Vec{p}$ is the vector in Fourier space that characterizes the  BZ which has been bounded by the equation above, that is: the boundary of the BZ is the minimal compact subset of the set that contains all the possible $ \Vec{p}$s that solve Eq.\ \eqref{eq:bordereq}.  Note that conventionally a BZ is constructed by the intersection of planes that cut orthogonally each of the reciprocal lattice vectors and its additive inverses into two equal pieces.
    
	\subsection{A $2$D oblique lattice}
    
	Let $ \epsilon \times \{ \vec{x}, \vec{y}, \vec{z}\}$ be the orthonormal basis for $ \mathbb{R}^3$. The oblique lattice vectors are defined as
	\begin{subequations}
		\begin{align}
			\Vec{a}_1 &=\epsilon (\vec{y}-\vec{x}) \\ \Vec{a}_2 &= \epsilon (\vec{x} +\vec{y})\\ \Vec{a}_3 &=\epsilon \vec{z} \, ,
		\end{align}
	\end{subequations}
	and the reciprocal lattice vectors are
	\begin{subequations}\label{eq:rhombusreciprocal}
		\begin{align}
			\Vec{k}_1 &\defeq \frac{ \pi}{\epsilon}(\vec{x}  - \vec{y} ) \\
			\Vec{k}_2 &\defeq  -\frac{ \pi}{\epsilon} ( \vec{x} + \vec{y})\\
			\Vec{k}_3 &\defeq  \frac{2 \pi}{\epsilon}  \vec{z} \, .
		\end{align}
	\end{subequations}
	Since  the $z$ component of the three-dimensional lattice is independent, we disregard it, and we reduce the dimension of the lattice to two, with lattice vectors $ \Vec{a}_1, \Vec{a}_2$ and reciprocal lattice vectors $ \Vec{k}_1,  \Vec{k}_2 $. Hence, Eq.\ \eqref{eq:bordereq} for $ \Vec{p} = p_{x} \vec{x} + p_{y} \vec{y}$ becomes
	\begin{equation}
		\begin{split}
			(n_1-n_2)p_x - (n_1+n_2)p_y= \frac{\pi}{\epsilon}(n_1^2 + n_2^2) \, ,
		\end{split}
	\end{equation}
 with the $n_i$s being the coordinates of $\vec k$.
	We obtain $4$ possible solutions, yielding the minimal region,
	\begin{subequations}\label{eq:obliquelattice}
		\begin{align}
			p_x - p_y= \frac{\pi}{\epsilon} \label{eq:c7a} \\
			-p_x + p_y=  \frac{\pi}{\epsilon} \label{eq:c7b} \\
			-p_x - p_y= \frac{\pi}{\epsilon} \label{eq:c7c} \\
			p_x +p_y= \frac{\pi}{\epsilon} \label{eq:c7d} \, .
		\end{align}
	\end{subequations}
	
	\begin{center}
		\begin{tikzpicture}
			\draw[thick,->] (2,2) -- (4.5,2);
			\draw[thick,->] (2,2) -- (2,-0.5);
			\draw[thick,->] (2,2) -- (-0.5,2);
			\draw[thick,->] (2,2) -- (2,4.5);
			\filldraw[green, opacity=0.2] (4,2) -- (2,4) -- (0,2) -- (2,0) -- (4,2);
			\draw[red,ultra thick] (4,2) -- (2,4) -- (0,2) -- (2,0) -- (4,2);
			
			\draw[red, dashed] (5,3) --(1,-1);
			\draw[red, dashed] (5,1) --(1,5);
			\draw[red, dashed] (3,5) --(-1,1);
			\draw[red, dashed] (3,-1) --(-1,3);

			\draw node[rotate=-45] at (3.5,3.1) {$ (E.7d)$};
			\draw node[rotate=45]  at (0.5,3.1) {$ (E.7b)$};
			\draw node[rotate=-45]  at (0.5,1.1) {$ (E.7c)$};
			\draw node[rotate=45]  at (3.5,1.1) {$ (E.7a)$};
			\draw node at (4.7,2) {$ p_y$};
			\draw node at (2,4.7) {$ p_x$};
			\draw node at (4.04,2.4) { $ \frac{\pi}{\epsilon}$};
			\draw node at (2.4,4.06) { $ \frac{\pi}{\epsilon}$};
			\draw node at (-0.09,2.4) { $ -\frac{\pi}{\epsilon}$};
			\draw node at (2.4,-0.06) { $- \frac{\pi}{\epsilon}$};
			\draw[thick] (4.01,1.9) -- (4.01,2.1);
			\draw[thick] (1.9,4.01) -- (2.1,4.01);
			\draw[thick] (-0.01,1.9) -- (-0.01,2.1);
			\draw[thick] (1.9,-0.01) -- (2.1,-0.01);;
		\end{tikzpicture} 
		\captionof{figure}{ The  BZ of the oblique lattice. } 
        \label{Fig: bz of oblique}
		
	\end{center}
	The red lines in Fig.\ \ref{Fig: bz of oblique} are unions of the points that solve  Eqs.\ \eqref{eq:obliquelattice}. 
    Hence, the set of all possible solutions of Eq.\ \eqref{eq:bordereq} forms the boundaries of the BZ. Thus, the BZ of the oblique lattice, that has been defined by Eq.\ \eqref{eq:latticevector}, is the shaded area in Fig.\ \eqref{Fig: bz of oblique}. 
    Another way of constructing the BZ of the oblique lattice would be to take the intersection of the four planes that cut the vectors $\pm \Vec{k}_1,\pm \Vec{k}_2,$ orthogonally, into two pieces.  
    Observe that the line that was described by Eq.\ \eqref{eq:c7a} cuts orthogonally $ \Vec{k}_1 $ into two pieces, and Eq.\ \eqref{eq:c7b} cuts orthogonally $ -\Vec{k}_1 $ into two pieces. Eq.\ \eqref{eq:c7d} and Eq.\ \eqref{eq:c7c} cut orthogonally $ \Vec{k}_2, -\Vec{k}_2, $ into two pieces respectively. 

    Now, if the construction of a BZ is just the intersection of the normal planes of the corresponding reciprocal lattice vectors, why bother with the Bragg equation? The answer is that finding such an intersection is rather difficult in higher-dimensional situations, as we are going to see in the following subappendix.

	\subsection{A four-dimensional oblique lattice}\label{subsec:4DBZ}
	We are interested in simulations $(3+1)$D discrete-spacetime models, which can be realized with $4$D lattices. Let us provide useful facts about such $4$D lattices. In the article, we use the following $4$D oblique lattice:
	\begin{subequations}
		\begin{align}
			\vec a_0 &\defeq \epsilon (1,1,1,1) \\ 
            \vec a_1 &\defeq \epsilon (1,-1,1,1) \\
			\vec a_2 &\defeq \epsilon (1,1,-1,1) \\ 
            \vec a_3 &\defeq \epsilon (1,1,1,-1) \, .
		\end{align}
	\end{subequations}
	We want to find the reciprocal space of the this oblique lattice. It is constructed by means of the vectors below
	\begin{subequations}\label{eq:4dreciprocal}
		\begin{align}
			\vec{b}_{0} &\defeq  2\pi \frac{\vec a_1 \wedge \vec a_2 \wedge \vec a_3}{\vec a_0.(\vec a_1 \wedge \vec a_2 \wedge \vec a_3)}\\
			\vec{b}_{1} &\defeq -2\pi\frac{\vec a_0 \wedge \vec a_2 \wedge \vec a_3}{\vec a_0.(\vec a_1 \wedge \vec a_2 \wedge \vec a_3)}\\
			\vec{b}_{2} &\defeq 2\pi\frac{\vec a_0 \wedge \vec a_1 \wedge \vec a_3}{\vec a_0.(\vec a_1 \wedge \vec a_2 \wedge \vec a_3)}\\
			\vec{b}_{3} &\defeq -2\pi\frac{\vec a_0 \wedge \vec a_1 \wedge \vec a_2}{\vec a_0.(\vec a_1 \wedge \vec a_2 \wedge \vec a_3)} \, ,
		\end{align}
	\end{subequations}
	which yields
 \begin{subequations}
	\begin{align}
			\vec{b}_0 &= \frac{\pi}{\epsilon} (-1,1,1,1) \\\vec{b}_1 &= \frac{ \pi}{\epsilon}  (1,-1,0,0)\\
			\vec{b}_2 &= \frac{ \pi}{\epsilon}  (1,0,-1,0) \\\vec{b}_3 &= \frac{\pi}{\epsilon}  (1,0,0,-1) \, .
	\end{align}
\end{subequations}
	
Note that for each reciprocal vector above, and its additive inverse,  there exists a $3$D plane that cuts it orthogonally into two pieces. The intersection of these planes gives the BZ of the $4$D oblique lattice. However, it is not straightforward to study such an intersection. Instead, let use the Bragg equation. We let $  \vec b \defeq l_{0}\vec{b}_0  +l_{1}\vec{b}_1 +l_{2}\vec{b}_2 +l_{3}\vec{b}_3   $ and $  \vec p \defeq (p_{0}, p_{1}, p_{2} , p_{3})$. The BZ of the $4$D oblique lattice is given by the following Bragg equation:
\begin{equation}\label{eq:border4d}
\begin{split}
	2 \vec p \cdot \vec b- \vec b \cdot \vec b 
	&=-\frac{2 \pi}{\epsilon}( p_{0}( l_1 + l_2+ l_3-l_0 )+ p_{1} (l_0 -l_1 )+ p_{2} (l_0-l_2) + p_{3} (l_0 - l_3)\\
	& \ \ \ \, + \frac{\pi^{2}}{\epsilon^{2}} ( 2 l_{0}^2 + l_{1}^2 + l_{2}^2 + l_{2} l_{3} + l_{3}^2 + l_{1} (l_{2} + l_{3}) - 
	 2 l_{0} (l_{1} + l_{2} + l_{3}))) = 0 \, .
\end{split}
\end{equation}
	Let us find the constraints on the two-dimensional hyper-surfaces of the BZ. There exists $6$ distinct two-dimensional hyper-surfaces in the BZ. To describe the BZ projected on a chosen two-dimensional hyper-surface, one has to obtain $4$ equations as we had in Eq.\ \eqref{eq:obliquelattice}. When $ l_0 = l_1 = l_2 =0  $ and $ l_3 = \pm 1$, Eq.\ \eqref{eq:border4d} becomes
	\begin{equation}\label{eq:con4d1}
		p_{0} - p_{3} = \pm \frac{\pi}{\epsilon} \, .
	\end{equation}
  \noindent
	When instead  $ l_0 = l_1 = l_2 = \pm 1 $ and $ l_3 = 0$, we obtain
	\begin{equation}\label{eq:con4d2}
		p_{0} + p_{3} = \pm \frac{\pi}{\epsilon} \, .
	\end{equation}
 \noindent
We continue with the case where  $ l_0 = l_1 = l_3 =0  $,  $ l_2 = \pm 1$, and the case where $ l_0 = l_1 = l_3 =\pm 1 $,  $ l_2 =0  $, for which we respectively obtain
\begin{subequations}
\begin{align}\label{eq:con4d3}
	p_{0} - p_{2} &= \pm \frac{\pi}{\epsilon} \\
	p_{0} + p_{2} &= \pm \frac{\pi}{\epsilon} \, .
\end{align}
\end{subequations}
Similarly, for  $ l_0 = l_2 = l_3 =0  $,  $ l_1 = \pm 1$, and $ l_0 = l_2 = l_3 =\pm 1 $,  $ l_1 =0 $, we respectively obtain
\begin{subequations}
\begin{align}\label{eq:con4d4}
	p_{0} - p_{1} &= \pm \frac{\pi}{\epsilon}\\
	p_{0} + p_{1} &= \pm \frac{\pi}{\epsilon} \, .
\end{align}
\end{subequations}
We continue with  $ l_0 = \pm 1, l_1 = \mp 1,   l_2 =0  $,  $ l_3 = \pm 1$, and $ l_0 =0, l_1 =\mp 1,  l_2 =\pm 1 $,  $ l_3 =0 $, and we respectively obtain
\begin{subequations}
\begin{align}\label{eq:con4d5}
	p_{1} + p_{2} &= \pm \frac{\pi}{\epsilon}\\
	p_{1} - p_{2} &= \pm \frac{\pi}{\epsilon}\, .
\end{align}
\end{subequations}
For $ l_0 = \pm 1, l_1 = \mp 1,   l_2 = \pm 1  $,  $ l_3 =0$, and $ l_0 =0, l_1 =\mp 1,  l_2 =0 $,  $ l_3 =\pm 1 $,  we respectively obtain
\begin{subequations}
\begin{align}\label{eq:con4d6}
	p_{1} + p_{3} &= \pm \frac{\pi}{\epsilon}\\
	p_{1} - p_{3} &= \pm \frac{\pi}{\epsilon}\, .
\end{align}
\end{subequations}
Finally, when  $ l_0 = \pm 1, l_2 =  l_3 = l_1 =0$, and $ l_0 =0, l_1 =0,  l_2 =\mp 1 $,  $ l_3 =\pm 1 $, we respectively obtain
\begin{subequations}
\label{eq:con4d7}
\begin{align}
	p_{2} + p_{3} &= \pm \frac{\pi}{\epsilon},\\
	p_{2} - p_{3} &= \pm \frac{\pi}{\epsilon} \, .
\end{align}
\end{subequations}
Through the equations above, we observe that the projections of the BZ on each hyperplane of the $4$D reciprocal oblique, hyperplanes that we denote $p_0$\hspace{0.05cm}\textrm{---}\hspace{0.05cm}$p_1$,  $p_0$\hspace{0.05cm}\textrm{---}\hspace{0.05cm}$p_2$, $p_0$\hspace{0.05cm}\textrm{---}\hspace{0.05cm}$p_3$ and $p_1$\hspace{0.05cm}\textrm{---}\hspace{0.05cm}$p_2$,  $p_1$\hspace{0.05cm}\textrm{---}\hspace{0.05cm}$p_3$, $p_2$\hspace{0.05cm}\textrm{---}\hspace{0.05cm}$p_3$,  admit a rhombus.

\section{ A way of representing higher-dimensional BZs with two-dimensional BZs}\label{sec:applowerBZ}
We would like to convey a way of representing the higher-dimensional BZs that we are interested in, with the use of two-dimensional BZs that are easy to work with. Our approach will be based on embedding the higher-dimensional BZ into a space that is constructed by the Cartesian product of  two-dimensional BZs. Let us start with a simple example of three-dimensional BZ. We first describe a three-dimensional BZ in the following Sec.\ \ref{subsec:appbodycentred}. After this section, we continue with the four-dimensional oblique lattice in Sec.\ \ref{subsec:fouroblique}.

\subsection{Body-centered cubic lattice}\label{subsec:appbodycentred}

We give a study of the construction of the BZ of a body-centered cubic lattice which has the following lattice vectors (we sometimes use hats just to recall the unit-vector property, although before we used arrows),
\begin{subequations}
    \begin{align}
         \Vec{a}_1 &=\epsilon (\hat{y}-\hat{x}+ \hat{z}) \\ 
         \Vec{a}_2 &= \epsilon  (-\hat{y}+\hat{x}+ \hat{z}) \\ 
         \Vec{a}_3 &=\epsilon (\hat{y}+\hat{x}- \hat{z}) \, ,
    \end{align}
\end{subequations} 
and whose reciprocal lattice vectors are
\begin{subequations}
    \begin{align}
         \Vec{k}_1 &= \frac{ \pi}{\epsilon}(\hat{z}  + \hat{y} ) \\
        \Vec{k}_2 &=  \frac{ \pi}{\epsilon} (\hat{z} + \hat{x})\\
        \Vec{k}_3 &=  \frac{\pi}{\epsilon}  (\hat{x}  + \hat{y} ) \, .
    \end{align}
\end{subequations}
We observe that the body-centered cubic lattice cannot be reduced to two dimensions as we did for the BZ of the two-dimensional oblique lattice, where $ \vec{k}_{3} = \hat{z}$. However, in the present case, the first Brillouin zone is a volume whose boundary has been constructed by the possible $ \Vec{p} \defeq p_x \hat{x}+ p_y \hat{y}+p_z \hat{z}$ obeying Eq.\ \eqref{eq:bordereq}, that is,
\begin{equation}
    p_z(n_1 + n_2) + p_y (n_1 + n_3) + p_x(n_2 + n_3) = \frac{\pi}{2 \epsilon} ( (n_1 + n_2)^2 + (n_1 + n_3)^2 +(n_2 + n_3)^2).
\end{equation}
This gives $12$ possible equations for two dimensional surfaces:
\begin{subequations}\label{eq:bodycentred}
\begin{align}
  p_x - p_y &=  \pm \frac{\pi}{\epsilon} , \hspace{4mm} p_x + p_y= \pm \frac{\pi}{\epsilon} \label{subeq:1}\\
  p_x - p_z &=  \pm \frac{\pi}{\epsilon} , \hspace{4mm}  p_x + p_z= \pm \frac{\pi}{\epsilon} \label{subeq:2} \\
 p_y - p_z &=  \pm \frac{\pi}{\epsilon} , \hspace{4mm}   p_y + p_z= \pm \frac{\pi}{\epsilon} \, . \label{subeq:3}
\end{align}
\end{subequations}
Each subequation corresponds to a surface in three-dimensional Fourier space, and their intersection bounds a region which happens to be a regular rhombic dodecahedron. It has $12$ faces that correspond to the solutions of Eqs.\ \eqref{eq:bodycentred}. In three-dimensional Fourier space, we obtain Fig.\ \ref{fig:rhombicdodecahedron}.
\begin{center}
    \begin{tikzpicture}
    \coordinate ($p_x$) at (2,0,0);
     \coordinate ($p_y$) at (0,2,0);
     \coordinate ($p_z$) at (0,0,2);
     \draw[thick,->] (0,0,0) -- (3,0,0);
     \draw[thick,->] (0,0,0) -- (-3,0,0);
     \draw[thick,->] (0,0,0) -- (0,3,0);
      \draw[thick,->] (0,0,0) -- (0,-3,0);
     \draw[thick,->] (0,0,0) -- (0,0,5);
     \draw[thick,->] (0,0,0) -- (0,0,-5);
     \draw node at (3.2,0,0) {$p_x$};
      \draw node at (0,3.2,0) {$p_y$};
       \draw node at (0,0,5.2) {$p_z$};

      \begin{scope}[rotate around={0: (1,1,0)}]
     \filldraw[red, opacity= 0.7,thick] (0,2,0) -- (1,1,1)--(2,0,0) -- (1,1,-1)--(0,2,0);

     \filldraw[blue, opacity= 0.7,thick] (0,2,0)-- (1,1,1) -- (0,0,2)-- (-1,1,1) -- (0,2,0);

      \filldraw[red, opacity= 0.7,thick] (0,-2,0) -- (1.1,-1,0.9)--(2,0,0) -- (0.9,-1,-1.1)--(0,-2,0);
     
     \filldraw[blue, opacity= 0.7,thick] (0,-2,0)-- (1,-1,1) -- (0,0,2)-- (-1,-1,1) -- (0,-2,0);

      \filldraw[yellow, opacity= 0.7,thick] (0,0,2) -- (1,-1,1)--(2,0,0) -- (1,1,1)--(0,0,2);
   
     \filldraw[yellow, opacity= 0.7,thick] (0,0,2)-- (-1,-1,1) -- (-2,0,0)-- (-1,1,1) -- (0,0,2);
     
\end{scope}

      \draw[dashed] (2,0,2)--(2,0,-2)--(-2,0,-2)--(-2,0,2)--(2,0,2);
    
     \draw node at (0.7,2.1,0) { (0, $ \frac{\pi}{\epsilon}$, 0)};
     \filldraw ((0,2,0) circle (2pt) ;

     \draw node at (2.8,0.4,0) { ($ \frac{\pi}{\epsilon}$, 0, 0)};
     \filldraw ((2,0,0) circle (2pt) ;

       \draw node at (0.7,0,2) { (0, 0, $ \frac{\pi}{\epsilon}$)};
     \filldraw ((0,0,2) circle (2pt) ;

        \draw node at (-2.6,0.4,0) { ($ -\frac{\pi}{\epsilon}$, 0, 0)};
     \filldraw ((-2,0,0) circle (2pt) ;

       \draw node at (-2.9,0,2) { ($ -\frac{\pi}{\epsilon}$, 0, $ \frac{\pi}{\epsilon}$)};
     \filldraw ((-2,0,2) circle (2pt) ;

     \draw node at (2.9,0.3,-2) { ($ \frac{\pi}{\epsilon}$, 0, $ -\frac{\pi}{\epsilon}$)};
     \filldraw ((2,0,-2) circle (2pt) ;

       \draw node at (0.8,-2.1,0) { (0, $- \frac{\pi}{\epsilon}$, 0)};
     \filldraw ((0,-2,0) circle (2pt) ;
    \end{tikzpicture}

     \captionof{figure}{The first brillouin zone (BZ) of the body-centered cubic lattice \label{fig:rhombicdodecahedron}} 
  \label{fig:bodycentred}
\end{center}
Red, yellow and blue, indicate the solutions of Eqs.\ \eqref{subeq:1}, \eqref{subeq:2}, and \eqref{subeq:3}, respectively.

\subsubsection{Adding a degeneracy}

Let us now focus on the $p_x $\hspace{0.05cm}\textrm{---}\hspace{0.05cm}$p_y$ and $p_x$\hspace{0.05cm}\textrm{---}\hspace{0.05cm}$p_z$  cross-sections in Fig.\ \ref{fig:bodycentred}, and consider them as individual BZs. From Eqs.\ \eqref{eq:bodycentred}, we know that they are rhombus shaped. Then, using Eq.\ \eqref{eq:rhombusreciprocal}, we can define the reciprocal vectors for $p_x $\hspace{0.05cm}\textrm{---}\hspace{0.05cm}$p_y$ and $p_x$\hspace{0.05cm}\textrm{---}\hspace{0.05cm}$p_z$, in three dimensions, with the unit vectors $\hat{x}$, $\hat{y}$, and $\hat{z}$. We then let,
\begin{subequations}
    \begin{align}
        \Tilde{b}_{1} \defeq (\hat{x}+ \hat{y}) \frac{\pi}{\epsilon} \, , \hspace{2mm}\Tilde{b}_{2} \defeq (\hat{x}- \hat{y})\frac{\pi}{\epsilon} \, ,\\
        \Tilde{ \Tilde{b}}_{1} \defeq (\hat{x}+ \hat{z})\frac{\pi}{\epsilon} \, , \hspace{2mm}\Tilde{ \Tilde{b}}_{2} \defeq (\hat{x}- \hat{z})\frac{\pi}{\epsilon} \, ,
    \end{align}
\end{subequations}
where $ \Tilde{b}_{1}$ and $\Tilde{b}_{2}$  are the reciprocal vectors for $p_x $\hspace{0.05cm}\textrm{---}\hspace{0.05cm}$p_y$, and  $ \Tilde{\Tilde{b}}_{1}$ and $\Tilde{\Tilde{b}}_{2}$ are the reciprocal vectors for $p_x$\hspace{0.05cm}\textrm{---}\hspace{0.05cm}$p_z$. For the body-centered cubic lattice, we had three reciprocal vectors, but right now we have four, did we make a mistake? Do these four reciprocal vectors give the same BZ as the body-centered cubic lattice? There is only one way of understanding whether the structure that we propose yields the same BZ, as in Fig.\ \ref{fig:bodycentred}, namely, by imposing Eq.\ \eqref{eq:bordereq}.

We define $ \tilde{b} \defeq n  \Tilde{b}_{1} + m \Tilde{b}_{1} +  x \Tilde{ \Tilde{b}}_{1}+  l\Tilde{ \Tilde{b}}_{2}$, where $ n,m,x,l \in \mathbb{Z}$. Next, we find the minimal region that is determined by $ -2\vec{p} \cdot \tilde{b}+ \tilde{b} \cdot \tilde{b} = 0$. This is
\par\nobreak
{\small
\begin{equation}
        (m - n)^2 + (l - x)^2 + (l + m + n + x)^2   
 - 2 ((m - n) py
 + pz (-l + x) + px (l + m + n + x))=0 \, .
\end{equation}}
For (i) $ m = \pm 1, n=x=l=0$,  (ii) $ n = \pm 1, m=x=l=0$, (iii) $ x = \pm 1$, $m=n=l=0$, and~(iv)~$ l = \pm 1$, $m=n=x=0$, we obtain, respectively
\begin{subequations}
    \begin{align}
         p_{x}+p_{y} &= \pm \frac{\pi}{\epsilon}\\
          p_{x}-p_{y} &= \pm \frac{\pi}{\epsilon}\\
           p_{x}+p_{z} &= \pm \frac{\pi}{\epsilon}\\
            p_{x}-p_{z} &= \pm \frac{\pi}{\epsilon} \, .
    \end{align}
\end{subequations}
The previous equations lead to rhombuses in $p_x $\hspace{0.05cm}\textrm{---}\hspace{0.05cm}$p_y$ and $p_x$\hspace{0.05cm}\textrm{---}\hspace{0.05cm}$p_z$. But we have not yet obtained the relations between $p_y $\hspace{0.05cm}\textrm{---}\hspace{0.05cm}$p_z$ that were displayed in Eq.\ \eqref{eq:bodycentred}. Let us go further and try the cases (i) $ m= \pm 1 ,x= \mp1, n=l=0  $, and (ii) $ n= \mp 1 ,x= \pm 1, m=l=0  $; we obtain respectively
\begin{subequations}
    \begin{align}
         p_{z}-p_{y} &= \pm \frac{\pi}{\epsilon}\\
          p_{z}+p_{y} &= \pm \frac{\pi}{\epsilon} \, .
    \end{align}
\end{subequations}

The condition that is given in Eq.\ \eqref{eq:bordereq} does not see the difference between the reciprocal lattice vectors $ \{  \Tilde{b}_{1}, \Tilde{b}_{1}, \Tilde{ \Tilde{b}}_{1},\Tilde{ \Tilde{b}}_{2}\}$ and the reciprocal lattice vectors of the body-centered cubic lattice. Hence, the structure that is constructed by two individual BZs in the $p_x $\hspace{0.05cm}\textrm{---}\hspace{0.05cm}$p_y$ and $p_x$\hspace{0.05cm}\textrm{---}\hspace{0.05cm}$p_z$ planes, is a degenerate solution for the reciprocal lattice vectors of the body-centered cubic lattice. This degenerative representation is very useful, because now we do not need to study the standard reciprocal lattice vectors, rather we will focus on a finite number of two-dimensional BZs. 

\subsubsection{Embedding the BZ of the body-centered cubic lattice into the product of two BZs of two 2D oblique lattices, respectively}\label{subsec:fouroblique}
The BZ of the oblique lattice can be constructed by a quotient topology, $ \mathbb{C}/\Gamma^{2}$, where $ p_{0}+ip_{1} \in  \mathbb{C}$, and $\Gamma^{2} $ is spanned by $ (m_{0}+n_{0})/2 + i (m_{0}-n_{0})/2$ with $ n_{0},m_{0} \in \mathbb{Z}$. Let us consider $  \mathbb{C}/\Gamma^{2} \cross  \mathbb{C}/\Gamma^{2}$, which has coordinates $ ( p_{0}+ip_{1} ,p_{2}+ip_{3} )$. Note that $ ( p_{0}+ip_{1})$ has translational symmetry with respect to the lattice ``$ (m_{0}+n_{0})/2 + i (m_{0}-n_{0})/2$'', and $ ( p_{2}+ip_{3})$ has translational symmetry with respect to the second lattice ``$ (m_{1}+n_{1})/2 + i (m_{1}-n_{1})/2$''.

To obtain the BZ of the body-centered cubic lattice from $  \mathbb{C}/\Gamma^{2} \cross  \mathbb{C}/\Gamma^{2}$, we define a projection map, $ \rho : \mathbb{C}^{2} \to R^{3}  $, by the identification $ p_{0 } =  p_{2} $, which puts a constraint at the level of lattices, namely, $ m_{1} = m_{0}+n_{0}-n_{1} $. The  projection map $\rho$ induces a map $ \Phi^{*}$ on $\Gamma^{2} \cross \Gamma^{2}$, such that
\begin{equation}
\begin{split}
\Phi^{*}( (m_{0}+n_{0})/2 + i (m_{0}-n_{0})/2, (m_{1}+n_{1})/2 + i (m_{1}-n_{1})/2) \\
    = ( (m_{0}+n_{0})/2,(m_{0}-n_{0})/2, (m_{0}+n_{0})/2-n_{1}) \, .
\end{split}
\end{equation}
That final, fully relevant projection map, is defined to be 
\begin{equation}
    \pi:  \mathbb{C}/\Gamma^{2} \cross  \mathbb{C}/\Gamma^{2} \to R^{3}/ \Phi^*\hspace{-0.5mm}(\Gamma^{2} \cross\Gamma^{2}  ) \, .
\end{equation}

\subsection{Four-dimensional oblique lattice}\label{{subsec:fouroblique}}
\subsubsection{Adding a degeneracy}

Instead of the reciprocal lattice vectors in Eq.\ \eqref{eq:4dreciprocal}, we will use the following reciprocal lattice vectors for the four-dimensional oblique lattice,
\begin{subequations}
\label{eq:degenerate4d}
 \begin{align}
         \Tilde{b}^{x}_{0} =( \hat{E}+ \hat{p}_{x}) \frac{\pi}{\epsilon}\, , \hspace{2mm}         \Tilde{b}^{x}_{1} = (\hat{E}- \hat{p}_{x}) \frac{\pi}{\epsilon} \, , \\
          \Tilde{b}^{y}_{0} = (\hat{E}+ \hat{p}_{y}) \frac{\pi}{\epsilon}\, , \hspace{2mm}         \Tilde{b}^{y}_{1} = (\hat{E}- \hat{p}_{y}) \frac{\pi}{\epsilon} \, ,\\
           \Tilde{b}^{z}_{0} = (\hat{E}+ \hat{p}_{z}) \frac{\pi}{\epsilon}\, , \hspace{2mm}         \Tilde{b}^{z}_{1} = (\hat{E}- \hat{p}_{z}) \frac{\pi}{\epsilon} \, ,
 \end{align}
\end{subequations}
where $ \{  \hat{E}, \hat{p}_{x},\hat{p}_{y},\hat{p}_{z} \}$ is a choice of orthogonal coordinate system in $ \mathbb{R}^{4}$. We define $ \Tilde{\Tilde{b}} \defeq n^{x}_{0}  \Tilde{b}^{x}_{0} + n^{x}_{1}  \Tilde{b}^{x}_{1}+n^{y}_{0}  \Tilde{b}^{y}_{0}+n^{y}_{1}  \Tilde{b}^{y}_{1}+ n^{z}_{0} \Tilde{b}^{z}_{0}+ n^{z}_{1}  \Tilde{b}^{z}_{1}$, where $  n^{i}_{j} \in \mathbb{Z}$ for every $ i \in \{x,y,z\}$, $ j \in \{0,1\}$. We also define and $ \mathbf{p} \defeq (E,p_{x},p_{y},p_{z})$. Then, the solution of $ 2   \mathbf{p} \cdot\Tilde{\Tilde{b}} -\Tilde{\Tilde{b}} \cdot \Tilde{\Tilde{b}} = 0 $ for arbitrary $\mathbf{p} $ leads to a minimal volume which is $\mathcal{B}^{2'}$. 
Let us prove that this statement is correct.  
\par\nobreak
{\small
\begin{equation}\label{eq:border6d}
\begin{split}
2  \mathbf{p} \cdot\Tilde{\Tilde{b}} -\Tilde{\Tilde{b}} \cdot \Tilde{\Tilde{b}}&=  -\frac{\pi^{2}}{\epsilon^{2}}((n_{0}^{x} - n_{1}^{x})^2 - (n_{0}^{y} - n_{1}^{y})^2 - (n_{0}^{z} - n_{1}^{z})^2 - (n_{0}^{x} + n_{0}^{y} + n_{0}^{z} + n_{1}^{x} + n_{1}^{y} + n_{1}^{z})^2) \\
 & \ \ \ \, + \frac{ 2\pi}{\epsilon}  (E  (n_{0}^{x} + n_{0}^{y} + n_{0}^{z} + n_{1}^{x} + n_{1}^{y} + n_{1}^{z}) + (n_{0}^{x} - n_{1}^{x})  p_{x} + (n_{0}^{y} - n_{1}^{y})  p_{y} + (n_{0}^{z} - n_{1}^{z})  p_{z} ) = 0 \, . 
\end{split}
\end{equation}}
When comparing Eq.\ \eqref{eq:border4d} to Eq.\ \eqref{eq:border6d} (the `` $=0$ '' equation), we observe that if the following conditions are true,
\begin{subequations}\label{eq:cond6d}
  \begin{align}
          n_{0}^{z} &= l_{0}-n_{0}^{y}-n_{0}^{x} \, , \hspace{2mm} 
    \  n_{1}^{x} =  l_{1}+ n_{0}^{x} \, , \\
    n_{1}^{y} &=-l_{0}+ l_{2} + n_{0}^{y} \, , \hspace{2mm}
     n_{1}^{z} = l_{3}-n_{0}^{x}-n_{0}^{y} \, ,
  \end{align}  
\end{subequations}
then we find that Eq.\ \eqref{eq:border6d} is the same as Eq.\ \eqref{eq:border4d}. Since the $ n^{i}_{j} $s are arbitrary integers, the conditions above can be easily realized as they do not violate the integer nature of  the $ n^{i}_{j} $s. Equation \eqref{eq:border6d} describes the BZ of a four-dimensional oblique lattice: we can deduce this fact by producing constraints analog to Eqs.\ \eqref{eq:con4d1} and \eqref{eq:con4d7}, with the use of Eq.\ \eqref{eq:cond6d}. The table below shows that Eq.\ \eqref{eq:border6d} produces same the constraints as Eq.\ \eqref{eq:border4d}, for a chosen list $ (n_{0}^{x},n_{1}^{x}, n_{0}^{y},n_{1}^{y},n_{0}^{z},n_{1}^{z}  )$.

\vspace{2mm}
\begin{equation}
    \begin{tabularx}{0.8\textwidth} { 
  | >{\raggedright\arraybackslash}X 
  | >{\centering\arraybackslash}X 
  | >{\raggedleft\arraybackslash}X | }
 \hline
 $(n_{0}^{x},n_{1}^{x}, n_{0}^{y},n_{1}^{y},n_{0}^{z},n_{1}^{z}  )$ & The constraint \\
 \hline
$( \pm 1,0,0,0,0,0) $ & $E+p_{x}= \pm 1$  \\
\hline
$(0, \pm 1,0,0,0,0) $ & $E-p_{x}= \pm 1$  \\
\hline
$(0, 0, \pm 1,0,0,0) $ & $E+p_{y}= \pm 1$  \\
\hline
$(0, 0, 0, \pm 1,0,0) $ & $E-p_{y}= \pm 1$  \\
\hline
$(0, 0, 0, 0, \pm 1,0) $ & $E+p_{z}= \pm 1$  \\
\hline
$(0, 0, 0, 0,0, \pm 1) $ & $E-p_{z}= \pm 1$  \\
\hline
$(\pm 1, 0, 0, \mp1,0, 0) $ & $p_{x}+p_{y}= \pm 1$  \\
\hline
$(\pm 1, 0, \mp1, 0,0, 0) $ & $p_{x}-p_{y}= \pm 1$  \\
\hline
$(\pm 1, 0, 0, 0,0, \mp1) $ & $p_{x}+p_{z}= \pm 1$  \\
\hline
$(\pm 1, 0, 0, 0,\mp1, 0) $ & $p_{x}-p_{z}= \pm 1$  \\
\hline
$(0, 0, \pm 1, 0,0, \mp1) $ & $p_{y}+p_{z}= \pm 1$  \\
\hline
$(0, 0, \pm 1, 0,\mp1, 0) $ & $p_{y}-p_{z}= \pm 1$  \\
\hline
\end{tabularx}
\end{equation}
Hence, the reciprocal vectors in Eq.\ \ref{eq:degenerate4d} produce $\mathcal{B}^{2'}$.


\subsubsection{Embedding the BZ of the 4D oblique lattice into the product of three BZs of three  2D oblique lattices, respectively}

Let us embed the reciprocal lattice associated to Eqs. \eqref{eq:degenerate4d} into $ \mathbb{R}^{6}$, which has the following set of orthogonal unit vectors, $ \{  \hat{E}_{1}, \hat{p}_{x}, \hat{E}_{2},\hat{p}_{y}, \hat{E}_{3},\hat{p}_{z} \}$. Equations \eqref{eq:degenerate4d} become

 \begin{subequations}
\label{eq:degenerate6d}
\begin{align}
         \Tilde{b}^{x}_{0} &= (\hat{E}_{1}+ \hat{p}_{x}) \frac{\pi}{\epsilon}, \hspace{2mm}         \Tilde{b}^{x}_{1} = (\hat{E}_{1}- \hat{p}_{x}) \frac{\pi}{\epsilon} \, ,\\
          \Tilde{b}^{y}_{0} &= (\hat{E}_{2}+ \hat{p}_{y}) \frac{\pi}{\epsilon} \, , \hspace{2mm}         \Tilde{b}^{y}_{1} = (\hat{E}_{2}- \hat{p}_{y}) \frac{\pi}{\epsilon} \, ,\\
           \Tilde{b}^{z}_{0} &= (\hat{E}_{3}+ \hat{p}_{z}) \frac{\pi}{\epsilon} \, , \hspace{2mm}         \Tilde{b}^{z}_{1} = (\hat{E}_{3}- \hat{p}_{z}) \frac{\pi}{\epsilon} \, .
 \end{align}
 \end{subequations}
If we want to get back to the vectors of Eqs. \eqref{eq:degenerate4d}, we only need to apply a simple projection map $ \Pi$ that makes $ E_{1} = E_{2}= E_{3}$. Now we see that we only have two-dimensional reciprocal lattice vectors that are independent from each other. Because of this, we get three independent BZs on the planes $E_1 $\hspace{0.05cm}\textrm{---}\hspace{0.05cm}$p_x$, $ E_{2}-p_{y}$, and $ E_{3}-p_{y}$. This means that the reciprocal lattice vectors in Eqs.\ 
\eqref{eq:degenerate6d} generate a product of three rhombus-shaped BZs. Hence, this BZ can be written as $ \mathcal{B'}^{3} = \mathbb{C}/ \Gamma^{2} \cross \mathbb{C}/ \Gamma^{2} \cross \mathbb{C}/ \Gamma^{2} $. The projection map $ \Pi$  applied on $ \mathcal{B'}^{3}$ is defined as
\begin{equation}
    \Pi:  \mathbb{C}/\Gamma^{2} \cross  \mathbb{C}/\Gamma^{2}\cross  \mathbb{C}/\Gamma^{2} \to  \mathbb{C}^{2}/ \Phi^{*}(\Gamma^{2} \cross\Gamma^{2} \cross \Gamma^{2} ) \, ,
\end{equation}
where $\Phi^{*}: \Gamma^{2} \cross\Gamma^{2} \cross \Gamma^{2} \to \Theta^{2} $ is the induced map, which outputs a complex lattice $ \Theta^{2}$, that is, 

\begin{equation}
  \begin{split}
        \Phi^{*}(( n_{0}^{x}+ n_{1}^{x})/2 + i ( n_{0}^{x}-n_{1}^{x})/2, ( n_{0}^{y}+ n_{1}^{y})/2 + i ( n_{0}^{y}- n_{1}^{y})/2, ( n_{0}^{z}+ n_{1}^{z})/2 + i ( n_{0}^{z}- n_{1}^{z})/2  )\\
        = (( n_{0}^{x}+ n_{1}^{x})/2 + i ( n_{0}^{x}-n_{1}^{x})/2, ( n_{0}^{x}+ n_{1}^{x})/2,( 1+ i) - (n_{1}^{y} + i n_{1}^{z})) \, ,
  \end{split}
\end{equation}
From there we deduce that $\Theta^{2} =  \Gamma^{4}$, which proves that $ \Pi(\mathcal{B}^{3}) =  \mathcal{B}^{2'} $.

\end{document}